\documentclass[aps,twocolumn,prb,showpacs]{revtex4}

\usepackage{graphicx}
\begin{document}

%\preprint{}

\title[]{Spin-electric stripes: Electric voltage induced by spin currents}

\author{Y.~B.~Lyanda-Geller}
\email[]{yuli@purdue.edu}
%\homepage[]{}
%\thanks{}
\address{
Department of Physics and Birck Nanotechnology Center, Purdue University, West Lafayette, Indiana
47907 USA}

\date{July 5, 2011}

\begin{abstract}
At each of the boundaries of the two-dimensional (2D) rectangular conductor parallel to the electric current there arises a stripe with an electric field transverse to the current and a 100\% electron spin polarization. The two stripes have opposite spin orientations and opposite directions of electric fields. The magnitudes of the fields, directly related to the spin current if the spin relaxation is negligible, are the same. The periphery stripes are separated by a center-stripe, in which the magnitude and direction of the electric field depend on the ratio of the skew scattering and side jump spin currents. The spin polarization is zero on the centerline and reaches $\pm 1$ at the boundaries between the central and periphery stripes. Weak relaxation of the $z-$component of spin normal to the 2D plane modifies the magnitudes of the spin polarization and fields, with $\pm 1$ spin polarization persisting at the edges of the sample. Favorable experimental settings, in which the electron spin relaxation of the $z-$component of spin is suppressed but the spin current is not, are discussed.
\end{abstract}

\maketitle

{\it Introduction}
When an electric current is flowing in a conductor, the spin-orbit interactions
result in the asymmetric scattering\cite{mott}, an accumulation of spin polarization at
the sample boundary\cite{dyakonovperel},
or an average electron spin polarization\cite{aronov_lyandageller,edelstein}.
The accumulation of spin polarization at the boundaries in the presence of electric
current has been the subject of recent
experiments\cite{awschalom,Wunderlich,Sih1,Stern,Sih2,Gossard,nitta,Matsu,Jung} aimed at the observation of the spin Hall effects discussed
theoretically over
the years\cite{dyakonovperel,hirsch,Shuzhang,zhang,macdonald,shytov,Nikolic}. The conventional wisdom is that no ampermeter or voltmeter
exists
that can directly measure the spin Hall current or spin Hall voltage (the voltage and current
signals due to charge carriers of opposite spins cancel each other). 
Therefore these experiments
address
spatially-dependent electron spin polarization ${\bf S}({\bf r})$.
${\bf S}({\bf r})$ is generally defined by several effects 
besides the spin Hall current. The total spin density is determined by 
the spin polarization of electrons induced by the 
electric current, by the spin relaxation and spin diffusion,
and by the spin precession in the effective momentum-dependent fields 
(for example, the fields due to the Dresselhaus or Rashba coupling) 
and the external magnetic field. In general, there is no continuity equation for the spin current 
in contrast to the continuity equation for the charge current.     
Measurement of ${\bf S}({\bf r})$ (or time-dependent ${\bf S}({\bf r},t)$) 
is not capable of distinguishing between the several contributions to the net spin
density. Furthermore,  the inhomogeneous spin 
density arises due to the boundary effects \cite{halperin,Adagideli}. This happens even when the bulk 
spin Hall conductivity completely vanishes, e.g., in systems with only intrinsic spin-orbit interactions 
linear in momentum\cite{shytov,inoue,Rashba,Raimondi,ALGP}. 
Moreover, the charge current passing through the sample carries its own magnetic 
field, whose directions at two edges of the conducting film or the 2D plane are opposite to each other. The polarization of electrons due to the Zeemann interaction 
caused by this magnetic field leads to generation of opposite spins at two edges\cite{dyakonovperel,altshuler}. In semiconductor materials with a large g-factor, this contribution into spin polarization can exceed the contribution to the spin polarization caused by the spin current. Such spatially 
dependent polarization is not related to the spin Hall current 
but is indistinguishable from it experimentally when the spin density is studied.
Also, the average electron spin polarization in the presence of
electric current observed in\cite{awschalom1,silov,ganichev,rokhinson,ferro}, while contributing to the overall spin density, 
 is not related to the spin Hall effect in any direct way. The question arises: is it possible at all to find an unambiguous
signature of the spin Hall current showing no or negligible
contribution due to the other effects?

\begin{figure}[t]
\vspace{-6mm}
\includegraphics[scale=0.35]{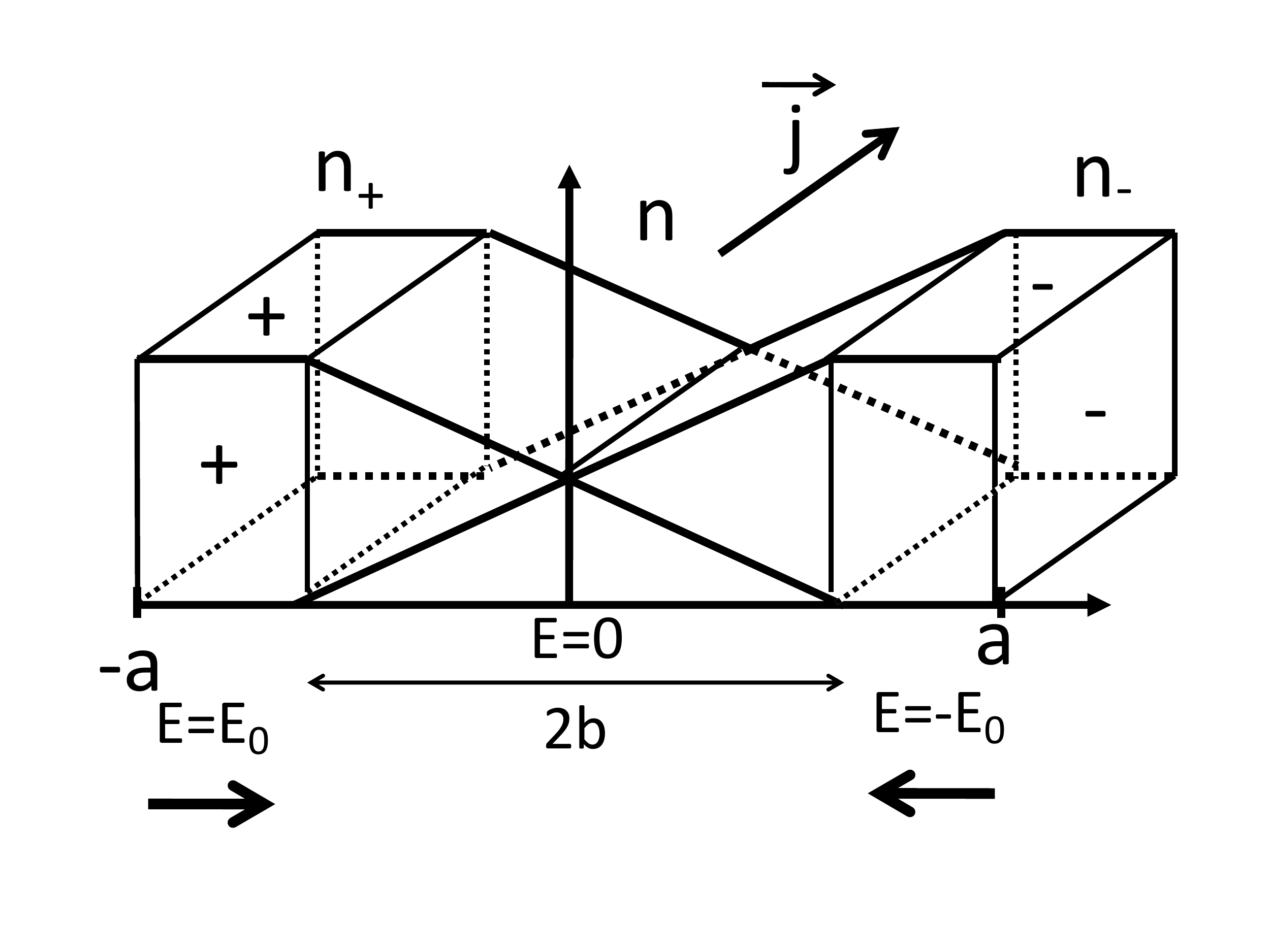}
\vspace{-13mm}
\caption{Schematic picture of electron density distribution in a 2D sample of width $2a$, 
with two periphery stripes with opposite electric fields transverse to a flowing current separated by a zero-field 
central stripe with width $2b$. The picture corresponds to case I of Sec. IA.  
In the periphery stripes, only electrons with one projection of spin (+ or -) contribute to the electric fields. In the center-stripe the gradients of density exist for both spin projections, and the transverse currents due to an applied external field  for each of the projections are cancelled by the corresponding diffusive currents. 
}
\vspace{-5mm}
\label{Fig1}
\end{figure}

In this paper, we predict that in two-dimensional (2D) systems with spin-orbit interactions that conserve an electron spin projection perpendicular to the
2D plane, the electric current results in two stripes with uniform electric
fields of opposite signs and equal magnitudes at two opposite boundaries of the sample. The electric current flows along the stripes, and the electric fields are perpendicular to the flowing current, and are directly related to the spin Hall current.  In the stripes, electrons are fully spin polarized, with opposite spin orientations at opposite sides of the sample. The two stripes are separated by a third stripe in 
the central region of the sample. On the center-line, the electron spin polarization is zero. The picture of the spin polarization in the central stripe is almost like in the conventional spin accumulation, except that when the spin polarization grows from the center-line to the
boundaries between the central and the periphery stripes (or falls in the case of the opposite spin direction), it reaches maximal $\pm 1$ value at that boundary, and cannot grow (fall) further into the periphery stripes. The magnitude and direction of the electric field in the central stripe depends on the relation between microscopic mechanisms of the bulk spin Hall current. For example, if the only spin current is due to 
side-jump-like effects \cite{Luttinger,Berger}, then the electric field in the central stripe vanishes, as in Fig.~\ref{Fig1}. In contrast, the skew scattering spin current results in the electric field in a central stripe, which is linear in transverse coordinate, vanishes at the center-line and is quadratic in the current flowing through the sample. 
The symmetry with respect to the center line characterizes the system in the absence of an external magnetic field only. 
In the presence of the external magnetic field, the stripes become asymmetric. We will discuss 
various experimental situations and their influence on the transverse electric field profile in stripes. 

The origin of the different roles of side-jump-like and skew scattering spin currents in the stripes is their different dependence on the density of contributing electrons and on the spin density and polarization. Side-jump-like spin currents are independent of the spin density and depend only on the total density, in contrast  to 
the skew scattering spin current, which depends on both densities. If the spin polarization varies in space, the spin current can change its magnitude and even its direction. This may result in a non-trivial spatial disribution of electric field in stripes.

\begin{figure}[t]
\vspace{-5mm}
\includegraphics[scale=0.35]{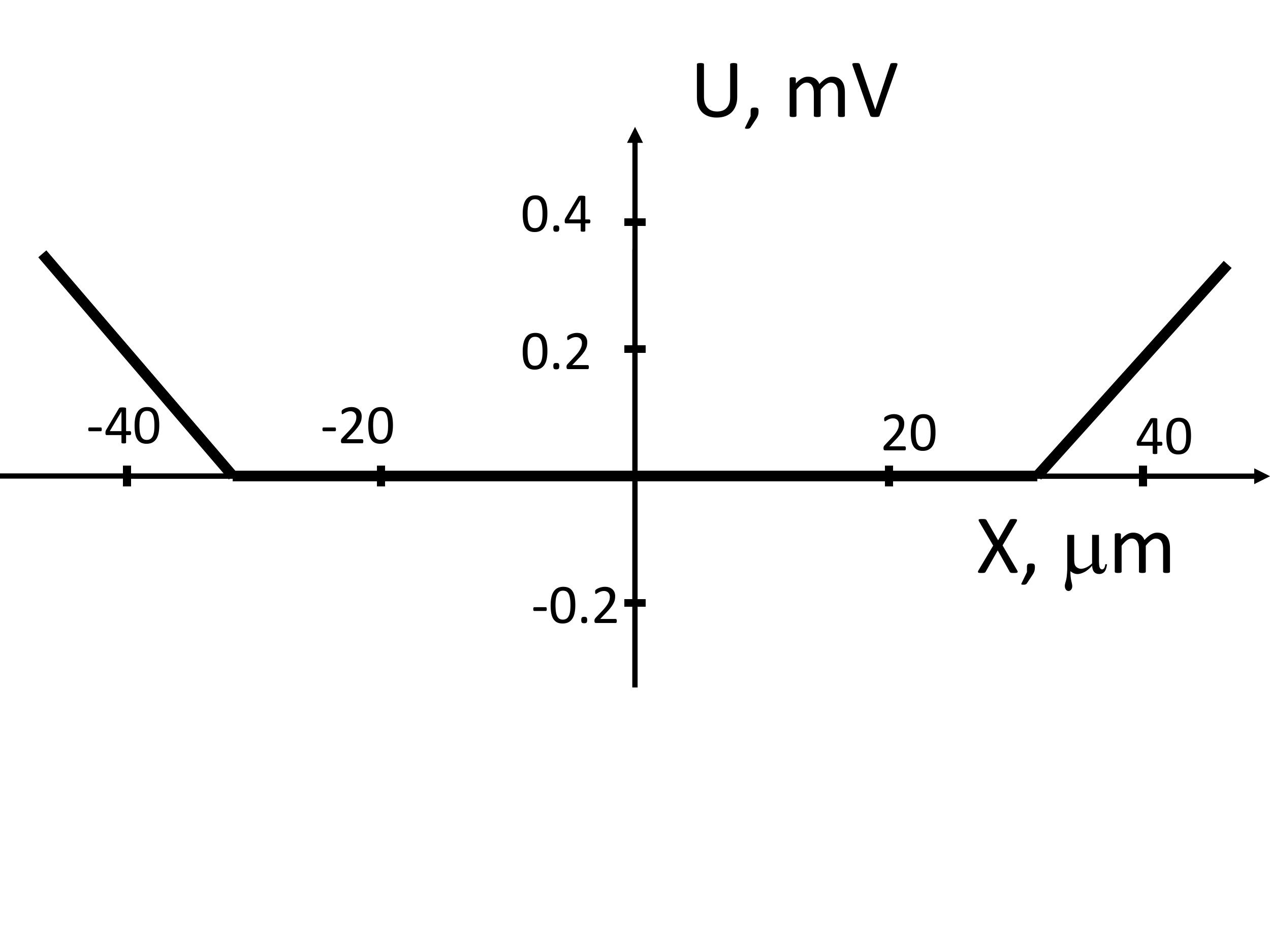}
\vspace{-23mm}
\caption{The Hall direction potential distribution across the sample (case I Sec.~1A). Slopes of potential curves on the right and on the left correspond to opposite electric fields in two periphery stripes, and are directly related to the spin Hall effect and are linear in the flowing electric current. The underlying parameters are discussed in Sec. IV. }
\vspace{7mm}
\label{Fig2}
\end{figure}

The transverse electric fields result in a potential difference between the central stripe and the edges of the sample. The example of a potential profile that corresponds to Fig.~\ref{Fig1} is shown in  Fig.~\ref{Fig2}. 
Measuring the voltage in an experiment similar to \cite{willett} is a straightforward method of observation of 
the spin-electric state. Positioning probes to measure voltage caused by the electric fields in the periphery stripes would constitute a direct electric measurement of the spin currents. Furthermore, if a single periphery stripe becomes a part of a separate electric circuit with the current flowing, it is possible to transfer its giant spin polarization for the purpose of applications.

\begin{figure}[t]
\vspace{-10mm}
\includegraphics[scale=0.4]{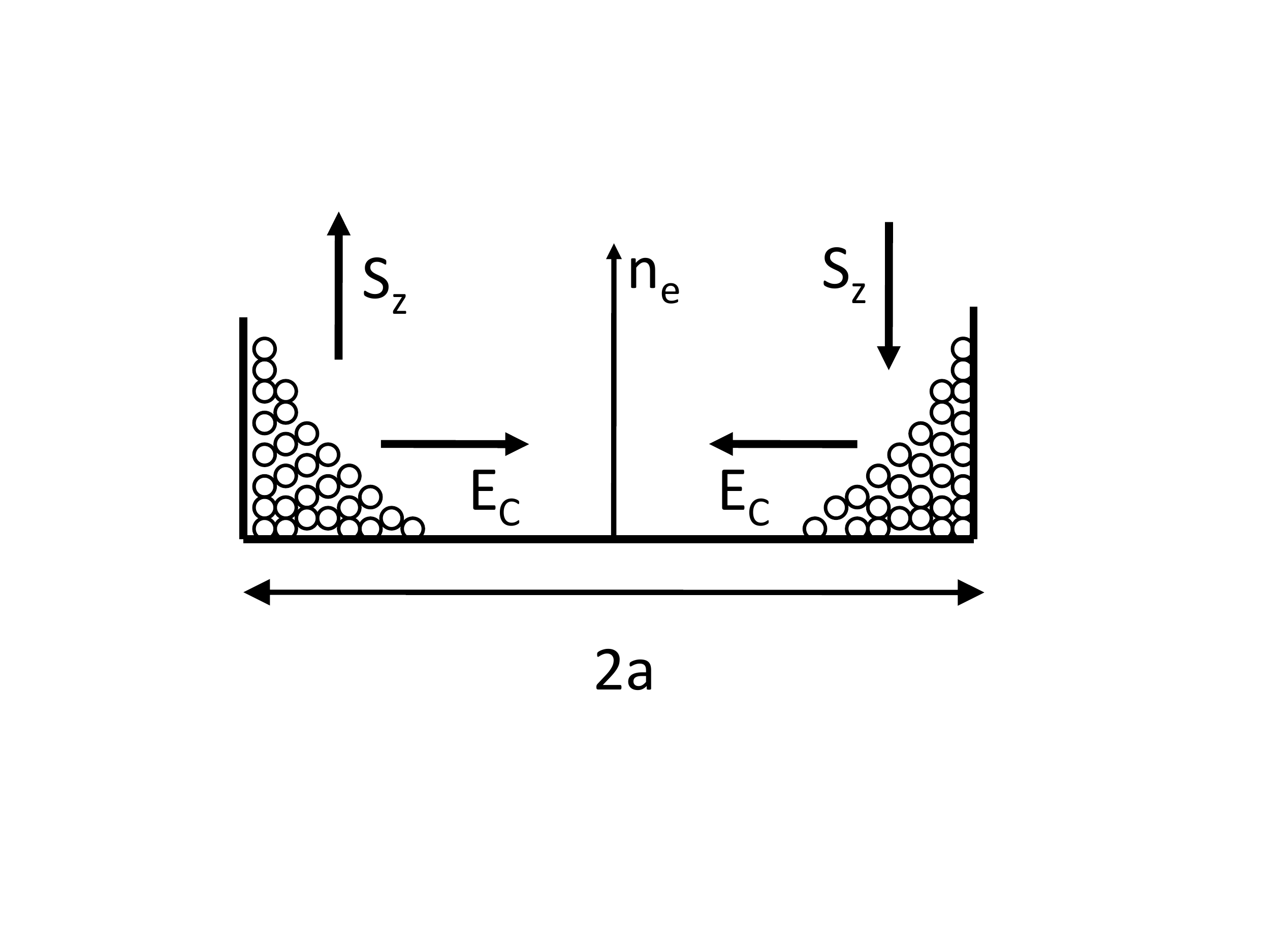}
\vspace{-2.5cm}
\caption{Cartoon showing the physical origin of the periphery stripes with opposite electric fields due to the spin current in the absence of $z$-projection spin relaxation.
When spin polarization reaches $\pm 1$ at the boundaries of the periphery stripes, this spin polarization makes the spin current equal to the Anomalous Hall current, which leads to charge accumulation near boundaries of the stripes. The Coulomb interactions prevent electrons from mounting walls of the sample and results in electric fields.}
\vspace{-3mm}
\label{cartoon}
\end{figure}

Spin-electric stripes are accompanied by the charge redistribution in the sample.
The electric field in the periphery stripes
is linear in the externally applied electric field, and the flowing current. These are "true" electric fields and charge distribution, which do not require 
nonlinearities\cite{finkler,stanescu} or pre-set, artificially created macroscopic inhomogeneity of charge density in 
the sample \cite{pershin_diventra}.
Charge carriers
with both spin projections contribute to the macroscopic electric field and charge density distributions.

Furthermore, we find
that spin-electric stripes arise even in the presence of spin relaxation, if the 
latter is sufficiently weak. As we demonstrate, macroscopic consequences of the spin current differ 
in systems with strong and weak spin relaxation.
In systems with strong spin
relaxation, it is the latter that balances
the spin current, keeps the total concentration of charge carriers constant
across the sample, and regulates spin accumulation at the
boundaries. The spin current induced by an external electric field in the presence of spin relaxation results in a spatial dependence of the spin density. Correspondinly, a diffusive spin current arises, which balances the externally induced spin current, particularly canceling the total spin current at the surface. At strong spin relaxation, stripes do not arise. Once spin relaxation is not present, or is
weak (the criteria for the spin relaxation strength will be discussed
below), it is no longer capable
of balancing the spin current. Spin polarization and density are limited only by their maximal (minimal) values, which are 
reached
in the periphery stripes. Here the spin polarization is a sourse of the Anomalous Hall effect (AHE), which results in a charge build up.
This buildup is opposed by the Coulomb interactions and electric fields, somewhat similar to the usual Hall effect, but, 
of course, with opposite electric fields in two stripes, in contrast to the usual Hall effect.  
For example in the case shown in Fig.~\ref{Fig2} that corresponds to side-jump-like currents only contributing to the spin Hall effect, the potential profile caused by 
transverse electric fields is aimed at keeping electrons in the sample.
Were there no Coulomb interactions, charges with 
opposite spins as a result of multiple scattering and spin current would stream towards the opposite boundaries and mount the walls of the sample, leaving the middle of 
it free of charge. The Coulomb interaction prevents this (Fig.~\ref{cartoon}) and
results in the electric field and spin-electric stripes. We find that because the sample breaks 
in the periphery stripes and a central stripe, and thus becomes non-uniform, other curious spatial distributions of the electric field and 
potential in the stripes are possible, depending on the ratio of the side-jump-like and skew scattering 
spin currents, and their variation across the sample. 

The ideal picture in the absence of relaxation of the $z-$component of spin is not much affected by weak spin relaxation. We show that in this case, spin polarization and the electric field in the periphery stripes experience a small change from the periphery towards the center region. 100\% spin polarization is still present at the edges of the sample. The division of the sample into three stripes remains meaningful in the presence of weak spin relaxation: at the boundaries between the periphery stripes and the central stripe, the electric field is a steep function of the transverse coordinate. 

 The state of the matter that we describe is different from that discussed by Dyakonov and Perel
\cite{dyakonovperel} or Hirsch \cite{hirsch}. 
The distinction is both on macroscopic and microscopic levels. 
On the macroscopic level, no voltage 
due to the spin Hall effect between the 
center of the sample and its sides arises in these studies (the voltage probe suggested by Hirsch is in the longitudinal direction,
is quadratic in the anomalous Hall resistance and constitutes a signal that is three orders of magnitude smaller 
than the transverse voltage we discuss). Furthermore, the phenomenon we predict here arises 
due to the spin Hall current in 
the presence of significant spatial separation of electrons with opposite spin projections.  
Moreover, discussing the spin-electric state in the presence of spin relaxation, we will see where the traditional spin accumulation physical picture fails and the spin-electric stripes emerge.
On the microscopic level, the state we suggest arises 
due to elimination or suppression of the part of the spin-orbit interaction responsible for the relaxation of the $z-$projection of spin, while the spin-orbit interactions that produce the spin current are kept intact. This physics is described by the Hamiltonian different from those studied in 
\cite{dyakonovperel,hirsch}. 

We consider macroscopic properties of the spin Hall system in terms of an ideal picture of the 2D metal.
We also discuss the role of the effects of inhomogeneous distribution of impurity centers in semiconductor heterostructures, and show that they do not alter the principal macroscopic observables in any essential way.  
 
Spin-orbit interactions that conserve the electron spin projection
transverse
to the 2D plane arise for generic spin-orbit interactions\cite{note2}
\begin{equation}
H=\frac{\hbar^2{\mathbf k}^2}{2m}+ U({\mathbf r}) +\alpha {\mathbf 
\sigma}_z\cdot [{\mathbf k}\times\nabla_{\bf r} U({\mathbf
r})]_z,
\label{H}
\end{equation}
where ${\mathbf k}$ and ${\mathbf r}$ are the two-dimensional wavevector and
coordinate, $m$ is the effective mass,
${\mathbf \sigma}_z$ is the $z-$Pauli matrix, 
$U({\mathbf r})$ includes potentials due to impurities and due to an external
electric field, and $\alpha$ is the spin-orbit constant (see Appendices B and C). 
The remarkable properties of this Hamiltonian in 2D metals, containing
just one Pauli matrix,  because of the presence of only in-plane electron
momenta,
have been noticed in studies of
weak localization effects\cite{hikami,lgm}. Spin accumulation
in this system has been also calculated \cite{vignale1}. A similar
situation with conserved spin can
 arise in materials with intrinsic spin-orbit interactions
of the Rashba and Dresselhaus type, but only for special crystallographic
orientations or for equal magnitudes of the Rashba and Dresselhaus
constants\cite{knap,aleiner,zhang1,awschalom2009,maslov}.
For spin-orbit interactions described by Eq.(\ref{H}), no spin precession,
no generation of average spin by the electric field and no spin relaxation 
of the $z$-component of spin
takes place. Therefore,
the spin current $J^{(z)}_s$ is unambigously defined.

Even the simplest spin-orbit interactions Eq.(\ref{H}) lead to several
microscopic
contributions to the
anomalous Hall effect and the spin Hall current.
In this work, we use a universally accepted 
approach \cite{review,halperin,Vignale,dasSarma,YLG,culcer} for the spin current, and calculate it 
in cases favorable for observation of the spin-electric stripes. The role of the remote doping by donors and doping in the quantum well by donors and acceptors is studied. By using various degrees of compensation of donors by acceptors in the quantum well, it is possible to eliminate skew scattering altogether. In this respect 2D structures with remote doping are a more favorable experimental setting compared to 3D samples\cite{Chazalviel}.

Simple 2D metal systems, in which the electron spin-orbit scattering is described by the Hamiltonian (\ref{H}), have not been acheived technologically.  We therefore turn our attention to semiconductor heterostrutures. 
In order to find a feasible experimental setting, we embark on a comprehensive 
study of the spin-orbit interactions in quantum wells. 
We shall demonstrate that there can exist a class of material systems, in which spin-orbit interaction leading to the spin current $J^{(z)}_s$ exists, 
but spin-orbit interactions resulting in spin relaxation of the $z-$projection of spin vanish. In these systems, 
the intrinsic spin-orbit coupling of the Dresselhaus type and strain-induced intrinsic spin-orbit interactions are removed by using a special crystallographic orientation of the 2D plane. Intrinsic spin-orbit 
interactions of the Rashba type are removed using symmetric potential confinement of the quantum well, leaving only the residual Rashba interactions 
associated with large scale fluctuations of density of impurities located in symmetrically positioned (with respect to the quantum well) 
$\delta$-doping layers\cite{sherman}.
 In our consideration of the spin-orbit effects for 2D electrons, 3D short- and long-range random potentials
 are taken into account. The role of the boundary conditions and the role of the difference of parameters in the barriers and the quantum well is investigated. 
 We demonstrate that by selecting heterostructure materials, it is possible to eliminate this residual Rashba interaction, 
as well as spin-flip scattering due to impurities of the quantum well.
The only source of spin relaxation in these systems 
is then extremely small interactions of the free electron spins with the spins of nuclei.
It turns out that not only are such materials theoretically possible but also there already exists an electronic system in InAlAs/InP/InAlAs double heterostructure, in which it looks feasible to observe the spin-electric stripes caused by the spin Hall effect. 

The manuscript is organized as follows: In Sec. I, we present the discussion of the stripe structure due to spin Hall current in the 2D metal.
Sec. IA deals with a macroscopic equations and properties of the state in the absence of spin relaxation, IB describes the charge distribution corresponding to the stripes, 1C describes spin-electric stripes in  the presence of magnetic dield perpendicular to the 2D plane,  I.D discusses 
a macroscopic picture in the presence of non-uniform doping, and Sec. IE discusses microscopic mechanisms of the spin current.  
In Sec. II  we discuss the spin-electric state the presence of spin relaxation. Sec. IIA gives macroscopic equations in the presence of spin relaxation, IIB shows how the traditional spin accumulation picture stops working and a spin-electric state emerges, IIC gives solutions of the macroscopic equations for the spin polarization and electric fields and presents a criterion of weak relaxation of the $z-$component of spin. Sec. III discusses a strength of spin-orbit interactions and spin relaxation in various electronic systems and feasibility of spin-electric stripes. In particular, IIIB discusses spin-orbit interactions in the quantum wells, and how they can be eliminated or strongly suppressed in order to observe the spin-electric state,  IIIB.3 describes impurity spin-orbit interactions in 2D systems, IIIB.4 discusses the
residual Rashba intrinsic spin-orbit interactions due to large-scale fluctuations of doping potential in heterostructures, IIIB.6 is devoted to the Dyakonov-Perel spin relaxation for non-uniform spin-orbit interactions, which is the case for the residual Rashba interaction, IIIB.7 analyses the role of intrinsic spin-orbit coupling for the non-equilibrum and equilibrium spin currents in [110] grown quantum well. Sec. IV, which discusses feasible experimental setting for observation of the spin-electric state in InAlAs/InP/InAlAs double heterostructure, is followed by a conclusion. There are three appendices in the paper: Appendix A discusses details of the solution of the 2D electrostatic problem that defines self-consistently the distribution of charge density and electric fields in the sample. Appendix B discusses 
spin-orbit constants in various systems, including the role of boundary conditions in heterostructures, and treats spin-orbit constants in quantum wells in the III-V materials. Appendix C discusses the effect of different materials parameters in the quantum well and the barriers on spin-orbit constants responsible  for relaxation of the $z$-component of electron spin.

\section{\it Spin-electric stripes in the absence of spin relaxation} 

The principal goal of this section is to solve macroscopic constituitive equations, a macroscopic electrostatic problem in the 2D electron system, and
to find the electric field and the electron density profile in the presence of the
spin current. We will also discuss the microscopic theory of spin current aiming at experimentally relevant situations, 
and describe the stripe state in the presence of an external magnetic field 
perpendicular to the plane of the 2D gas.

\subsection{\it Constitutive equations of the system}

\begin{figure}[t]
\vspace{-12mm}
\includegraphics[scale=0.35]{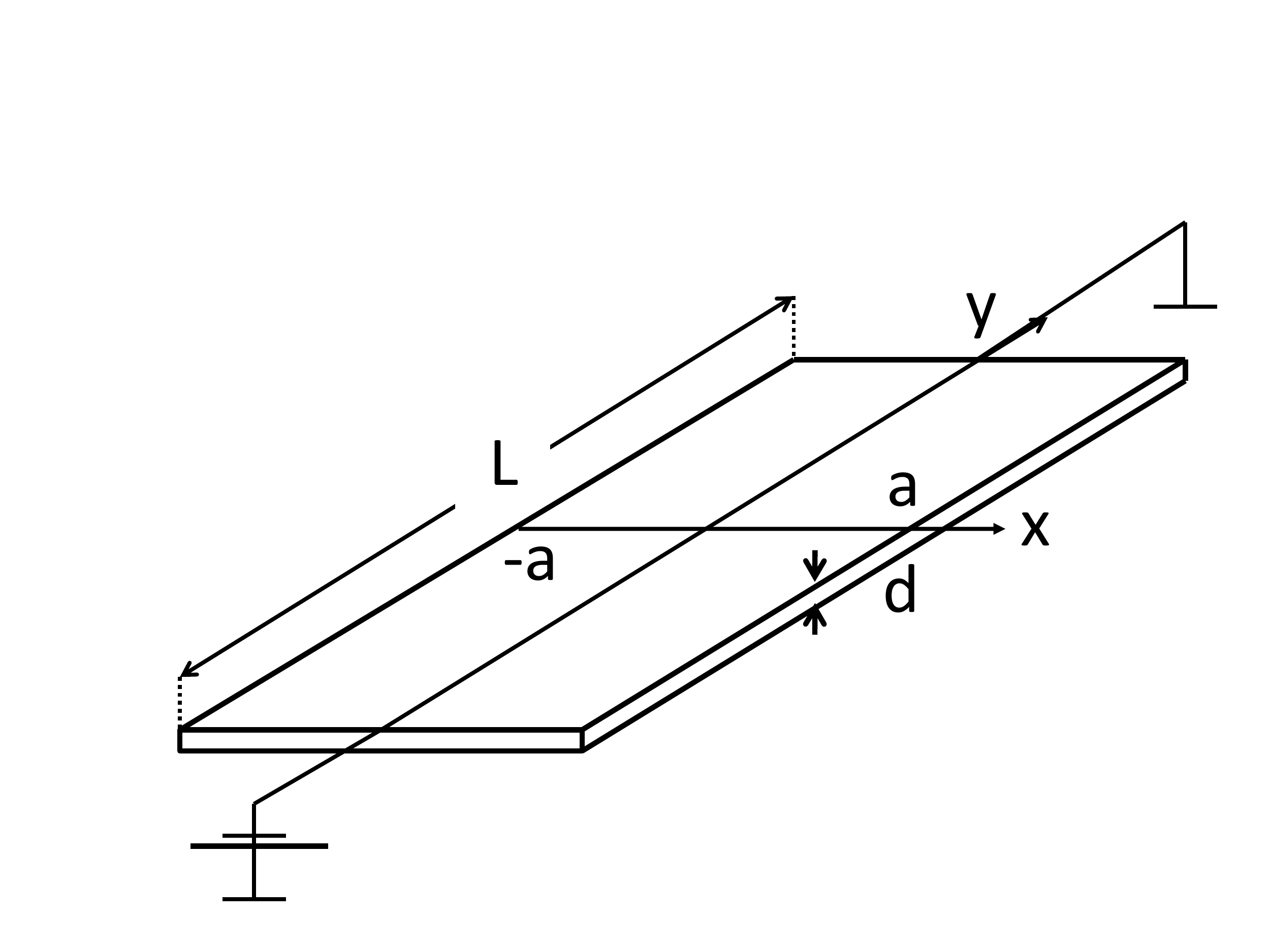}
\vspace{-5mm}
\caption{Schematic picture of the spin Hall sample. 2D electron is gas confined to a quantum well of width d.}
\vspace{-5mm}
\label{Device}
\end{figure}

The system is assumed to be a rectangle in the $xy$ plane, as shown in Fig.~\ref{Device}, 
with a width 2a
in $x$-direction and the length of the sample between the electrodes in $y$-direction 
$L\gg a$, considered infinite. The boundary conditions for this
configuration
is that there be no current flow in $x$-direction out of the sides of the
device,
and no electric field parallel to the contacts in $y$-direction.
The electric field ${\bf E}=-\nabla \phi$, where $\phi$ is the electric
potential, and the current density is $j(n)$, where $n=en_e$ is the charge density and $n_e$ is the electron 
density,
are assumed to be independent of the $y$-direction, i.e., we are dealing with the strictly a 2D electrostatic problem. 

The fundamental constitutive equations for the system are
\begin{eqnarray}
{\bf j}_{\pm}= \sigma (n_{\pm}){\bf E}+D(n_{\pm})\nabla_{\bf r} n_{\pm} +\nonumber \\
\sigma_{\perp}(n_{\pm}) ({\bf E}\times {\bf z}) +
D_{\perp}(n_{\pm}) (\nabla_{\bf r} n_{\pm}\times{\bf z}),
\label{j}
\end{eqnarray}
where signs $\pm$ correspond to the spin $z$-projection, which is
conserved
for systems described by the Hamiltonian Eq.(\ref{H}),
$n_{\pm}$ is the concentration of carriers with the corresponding spin
projection,
$\sigma (n_{\pm})$ is the longitudinal conductivity due to carriers with
$\pm$ spin projection, $\sigma_{\perp}(n_{\pm})$ is the transverse
conductivity,
$D(n_{\pm})$ is the longitudinal diffusion constant describing the current
directed along the gradient of the carrier concentration, and
$D_{\perp}(n_{\pm})$ describes diffusion in the direction transverse to the concentration 
gradient. The longitudinal component of these equations expresses Ohm's law and corrections to 
the current due to spin polarization. The transverse component vanishes, because of vanishing currents at the boundary of 
the 2D plane and their continuity in the absence of spin-relaxation of the $z-$component of spin. 
Expressions for all the coefficients in these equations are obtained from
the microscopic description. The Einstein relations hold for both
longitudinal and transverse transport, and give relations between  
$\sigma (n_{\pm})$ and $D(n_{\pm})$, and between $\sigma_{\perp}(n_{\pm})$ and
$D_{\perp}(n_{\pm})$.  

Eqs.(\ref{j}) for the current of charge carriers with spin projections
"$+$" ("$-$")
contain carrier density, conductivity, and longitudinal and transverse
diffusion
constants  corresponding to the same sign of the spin projection\cite{remark}. 
The electric field 
${\bf E}(x,y)$, which enters the equations for both spin projections, is defined,
apart from the external electric field, by all charges in the system,
including carriers of both spin projections and
impurities supplying the charge carriers (for doped semiconductors)
or ions (for metals).

We now describe the physics and find the solutions of these equations. 
As a starting point, we take a uniform sample.
Assuming a certain sign of $\alpha$ (the sign of $\alpha$ can differ from system to system\cite{sign,Abakumov}), 
charge carriers
with
spin projection "$+$" leave the boundary $x=a$, 
and carriers with spin "$-$" leave the boundary $x=-a$ because of the spin current (Figs.~\ref{Fig1} and 
\ref{cartoon}). Furthermore, charge carriers 
with any transverse coordinate and spin projection "$+$" attempt to move in the negative direction of $x$, towards $-a$,
and carriers with the spin projection "$-$" attempt to move in the positive direction of $x$, towards $a$.
We now consider an ideal case with positive charges (due to impurities or ions) uniformly 
distributed over the sample, 
in the conditions of the overall electrical neutrality.
Then the sum $n_+ + n_-$, which is the total charge carrier density $n$, in the first approximation can be 
considered constant with transverse
coordinate $x$,
\begin{equation}
n(x)=n_+(x) + n_-(x)=Const
\label{const}
\end{equation}
because in such situation the Coulomb interactions generally favor homogeneous distribution of
carriers. However, the densities of electrons with each component of spin $n_+(x)$ and $n_-(x)$ are not uniform, and therefore coefficients in Eq.(\ref{j}) are coordinate-dependent. Furthermore, the spin polarization 
\begin{equation}
S_z(x)=n_+(x)-n_-(x)
\label{Sz}
\end{equation}
 is not uniform. At the periphery of the sample a non-zero spin polarization 
arises defined by the sign of the spin-orbit constant defining the spin current. In conditions of a uniform total density, the spin polarization at the centerline (centerplane) of the sample, on the symmetry grounds, 
\begin{equation}
S_z(0)=0. 
\label{zero} 
\end{equation}

We note that by our definition, $S_z(x)$ has the same dimensionality as the charge density $n$.
We will primarily use $S_z(x)$ and call it, for simplicity, the spin density. The polarization is given by $p=S_z(x)/n$, and electron spin polarization is $\hbar p/2$. 
The spin density in the strict sense is $\hbar S_z(x)/2e$, where $e$ is the value of the electron charge.  

In order to solve the constituitive equations (\ref{j}), we need the dependences of the coefficients in these equations on 
density of charge carriers (with a given projection of spin). We assume that the conductivity of the 2D gas is defined by its Drude value
\begin{equation}
\sigma (n_{\pm})=e\tau n_{\pm}/m=\mu n_{\pm},
\label{Drude}
\end{equation}
where $\tau$ is the transport relaxation time, and $\mu$ is the electron mobility. 
We note here that for two independent electron species with up and down projections of spin, spin polarization effectively 
leads to two quasi-Fermi energies\cite{Giuliani}. 
In terms of the degree of electron polarization $p$, these energies in two dimensions are given by
\begin{equation}
\epsilon_F^{\pm}=\epsilon_F^{0}(1\pm p),
\end{equation}
where $\epsilon_F^{0}$ is the Fermi energy for unpolarized electrons $(p=0)$. 
The charge densities of electron species with up and down spin projections are
\begin{equation}
n_{\pm}=\frac{em}{2\pi\hbar^2}\epsilon_F^{\pm}.
\end{equation}
In a degenerate electron gas, generation of electron-hole pairs (an electron above quasi-Fermi-energy, and a hole below it) for minority species takes place at a different energy compared to majority electrons. This happens because the Pauli principle allows for unoccupied states for minority species in the vicinity of the lower quasi-Fermi level. Correspondingly, the Drude conductivity is a sum of conductivities of up and down species, with each partial low-temperature conductivity defined by the quantities taken at the corresponding quasi-Fermi level. Other kinetic coefficients defined by 
the quantities at the Fermi level for unpolarized electrons, for two independent electron species are defined similar to conductivity.
The longitudinal diffusion coefficient for each of the electron species is given by 
\begin{equation}
D(n_{\pm})=\pi\hbar^2n_{\pm}\tau/em^2=\zeta n_{\pm} .
\label{diffusion}
\end{equation}
Here we assume that $\tau$ is energy independent, and has the same value for both spin projections. 
We also note that the expression for the longitudinal diffusion coefficient explicitly assumes the 2D case. The coefficients $\zeta$ and $\mu$ are 
independent of densities. The generalizations taking  into account the spin projection dependence of $\tau$, $\zeta$ and $\mu$ are straightforward.  

We now observe that in the first approximation, there is no gradient of densities for up and down spin projections in 
the longitudinal direction $y$. Such gradients appear only in the second order in the spin-orbit constant (one order 
comes from the transverse spin current, and the other is required to generate the coordinate dependence of density in the $y$-direction, when spin-orbit scattering transforms the transverse spin current into the longitudinal one). The effects of the term proportional to $D_{\perp}(n_{\pm})$ are cubic in the spin-orbit constants, because $D_{\perp}(n_{\pm})$ is connected by the Einstein relation to $\sigma_{\perp}$ and is linear in the spin-orbit constant. Therefore, such effect are small and from now on we neglect terms with $D_{\perp}(n_{\pm})$.

 For the solution of Eqs.(\ref{j}) we also need the coordinate dependence of the transverse 
conductivity. As we shall see in Sec. IE, the transverse spin conductivity in relevant systems
is defined by two principal contributions: one is linear in the electron density (this contribution originates from the effects of the side-jump group) and the other is quadratic in density (and comes from skew scattering):
\begin{equation}
\sigma_{\perp}(n_{\pm})=\pm( \eta_1 n_{\pm} +\eta_2 n_{\pm}^2.)
\label{source}
\end{equation}

To make observations on the physics of the phenomenon, it is instructive to write down the identities for transverse spin and charge currents by adding and subtracting Eqs.(\ref{j}) and taking into account 
that these currents vanish, $j_x^{\pm}=0$ because of the boundary conditions and continuity in the 
absence of spin relaxation:
\begin{eqnarray}
&\mu  S_z  E_x+\frac{\zeta n }{2}\frac{d}{dx}S_z + \big(2\eta_1 n+\eta_2 (n^2+S_z^2)\big)\frac{E_y}{2}=0
\label{sc1}\\
&\mu  n E_x +\frac{\zeta}{4}\frac{d}{dx}S_z^2+(\eta_1+\eta_2 n) S_z E_y=0,
\label{cc1}
\end{eqnarray}
where $n_{+}^2-n_{-}^2= S_z \cdot n$,
and the strusture of the $\eta_2-$term comes from $(n_-^2+n_+^2)$.
  In the diffusion term, $\nabla_x S_z^2=2 S_z\nabla_x S_z$  comes from
\begin{equation}
\zeta (n_+\nabla_x n_++ n_-\nabla_x n_-)=\zeta \nabla_x ( n_+^2+ n_-^2)/2
\label{struct}
\end{equation} 
and assumption of uniform $n$. The combination of transverse conductivity parameters $\eta_1$ and $\eta_2$ in Eq.(\ref{sc1}) for the spin current and in Eq.(\ref{cc1}) for the charge current are different in general. Only in the case of fully polarized electrons $(S_z=\pm n)$, when the values of the spin current and the AHE current are the same, the combinations of skew scattering and side-jump-like terms in both equations become $\pm (\eta_1+ \eta_2 n)\cdot n$ and are the same. At small $S_z$, the spin current and the AHE current differ, and that will have important consequences. We underscore that 
the spin current depends on spin density via the skew scattering contribution only. In the case of varying spin polarization, the value and even the direction of the spin current can vary across the sample. 

Vanishing transverse spin current Eq.(\ref{sc1}) is a result of the balance of the spin mobility current, which arises if $S_z\ne 0$ and $E_x\ne 0$ are generated due to spin-orbit interactions, a diffusive spin current, which arises if $S_z$ is not uniform, and the bulk (sourse) spin Hall current. Vanishing transverse charge current (\ref{cc1}) is a balance of mobility current steming from $E_x$, diffusive current arising due to non-uniform $n_+^2+n_-^2=(S_z^2 + n^2)/2$, which is equivalent to non-uniform $S_z$ at unifom $n$ , and the AHE current. These identities are non-linear equations for $S_z$ and $E_x$, and solving them is not straightforward. Instead, to find a solution of the problem, we return to the constituitive equations (\ref{j}) for $j_x^{\pm}=0$.

We first solve these equations for the case when both $n_+(x)$ and $n_-(x)$ are non-zero:
\begin{eqnarray}
n_+(x) >0; & n_-(x)>0.
\label{positive}
\end{eqnarray}
We then divide Eq.(\ref{j}) for $ j^+=0$ by $n_+$ and Eq.(\ref{j}) for $ j^-=0$ by $n_-$ , and obtain
\begin{eqnarray}
& \mu E_x +\zeta\frac{d}{dx} n_+ + ( \eta_1 +\eta_2 n_+)E_y=0
\label{eq+}\\
& \mu E_x +\zeta\frac{d}{dx} n_- -  (\eta_1 +\eta_2 n_-)E_y=0.
\label{eq-}
\end{eqnarray}
Since we neglected terms with $D_{\perp}(n_{\pm})$, in these equations $E_y$ is an externally applied 
electric field that generates a current in the longitudinal direction, and $E_x(x)$ is the transverse electric field that we have to find alongside with the charge density $n_+$ $(n_-= n- n_+)$. Adding Eqs.(\ref{eq+}) and (\ref{eq-}), and taking into account 
Eq.(\ref{const}), we obtain the equation for the transverse electric field 
\begin{equation}
E_x=\frac{-\eta_2 S_z(x) E_y}{\mu}.
\label{sum}
\end{equation}
Thus, the transverse electric field is defined by $S_z$ and the skew scattering constant $\eta_2$, if Eqs.(\ref{positive})  are satisfied. The skew scattering term in Eq.
(\ref{sum}) arises as a specific linear combination of the AHE and spin Hall currents.
Subtracting  Eq.(\ref{eq-}) from Eq.(\ref{eq+}), we obtain the equation for of spin polarization
\begin{equation}
\zeta \frac{dS_z}{dx}+ ( 2\eta_1 +\eta_2 n)E_y=0.
\label{subtract}
\end{equation}
It is easy to recognize that this equation describing the gradient of spin density reflects the balance of the diffusive spin current and the spin 
current generated by $E_y$ at $S_z=0$, i.e., in accord with Eq.(\ref{zero}), at $x=0$. This is what defines the spin density, and, via Eq.(\ref{sum}), the transverse electric field in the whole central stripe, which is geometrically restricted by Eq.(\ref{positive}). Indeed,
solving Eq.(\ref{subtract}) with the boundary condition Eq. (\ref{zero}), we have 
\begin{equation}
S_z(x)=-\frac{2\eta_1 +\eta_2 n}{\zeta}E_y x 
\label{polcenter}
\end{equation}
for the spin density. The spin density is the first order in $E_y$, and the first order in spin-orbit constants.  
The principal story is that the spin density grows or falls with $x$ and inevitably reaches $\pm n$. If it reaches $n$ then $n_-=0$.If it reaches $(-n)$ then $n_+=0$. In both instances Eqs. (\ref{positive}) are no longer valid. The coordinates $-a<x<a$ at which this happens are the boundaries $\pm b$ of the central stripe. 

For the transverse electric field in the central stripe we have
\begin{equation}
E_x(x)=\frac{\eta_2(2\eta_1+\eta_2 n)}{2\mu\zeta}E_y^2 x
\label{fieldcenter}
\end{equation}
 We see that the electric field is quadratic in $E_y$ and is independent of the polarity of the external field. Besides the skew scattering constant, it is defined by the combination of skew scattering and side jump contributions that describe the spin Hall current at $x=0$. The electric 
field in the central stripe does not play an exclusive role, because transverse currents of electrons with each of the spin projections are also balanced by the diffusive spin currents of electrons 
with the corrresponding spin projection, due to spatially-dependent $S_z(x)$ given by Eq.(\ref{polcenter}) . 

The boundaries of the central stripe are given by
\begin{equation}
b=\pm \frac{n\zeta}{(2\eta_1 +\eta_2 n)E_y},
\label{boundcp}
\end{equation}
 The boundaries $\pm b$ are defined by skew scatering and side jump currents via the combination that describes spin current  at $x=0$. 

Once the maximal $S_z=n$ or minimal $S_z=-n$ spin density is reached, it cannot grow (or decrease for opposite $x$) any further, and $S_z=\pm n$ outside the central stripe, i.e., in the periphery stripes.
Because the spin density is constant in each of the periphery stripes, and $n_+=0$, $n_-=n$ or $n_-=0$, $n_+=n$, the diffusive terms in Eqs.(\ref{j}) vanish. 
Spin currents in these two periphery stripes $ j_{\pm}=\sigma_{\perp}(n_{\pm})E_y$ are compensated by the corresponding electric currents  $j=\sigma(n_{\pm}) E_x$, and this gives two opposite electric fields in the two periphery stripes.
 The magnitude of this transverse electric field is given by
\begin{equation}
|E_x^{(p)}|=|\frac{\sigma^\pm_{\perp}}{\sigma}E_y|=|\frac{(\eta_1 +\eta_2 n)E_y}{\mu}|.
\label{valE}
\end{equation}
The combination of side-jump $\eta_1$ term and skew scattering $\eta_2$ term here characterizes 
the magnitude of the AHE and spin Hall currents, which coinside at maximal spin polarization.
On the formal grounds, the sign of the electic field is opposite for the two periphery stripes due its dependence on the sign 
of the spin density, because $S_z$ defines whether $E_x^{(p)}$ is determined by the equation for $j_+$ or $j_-$ :
\begin{equation}
sign(E_x^{(p)})= -sign(S_z)\cdot sign(\eta_1 +\eta_2 n)\cdot sign(E_y).
\label{signEi}
\end{equation}
We note that a convenient vantage point for the consideration of the electric field in the periphery stripes is afforded by Eq.(\ref{cc1}). Indeed, because the diffusive term vanishes, it has a form
\begin{equation}
\mu  n E_x^{(p)} +(\eta_1+\eta_2 n) S_z E_y=0,
\label{vantage}
\end{equation}
and represents a balance of Ohm's current in the presence of the transverse electric field and 
the AHE current that arises in periphery stripes due to the full spin polarization. At $|S_z|=n$, we arrive to Eq. (\ref{valE}) for the magnitude of the electric field.  This picture looks in much the same way as in the conventional "charge" Hall effect \cite{rendell}, but here two opposite 
electric fields arise in the two periphery stripes. In order to find the direction of 
$E_x^{(p)}$ at positive and negative $x$, we determine
the sign of the spin density on the boundaries of the central and periphery stripes, which 
determines the spin polarization in the periphery stripes: 
\begin{equation}
sign(S_z)=-sign(2\eta_1 +\eta_2 n)\cdot sign(E_y)\cdot sign(x).
\label{signSz} 
\end{equation}
It is noteworthy that the sign of spin density in the periphery stripes depends on the sign of the spin current  at $x=0$ (the centerline), governed by the combination $(2\eta_1 +\eta_2 n)$. Combining Eq.(\ref{signEi}) and Eq.(\ref{signSz}), we find the sign of $(E_x^{(p)})$ defining the direction of the field in the periphery stripes:
\begin{eqnarray}
& sign(E_x^{(p)})= [sign(E_y)]^2\cdot sign(\eta_1 +\eta_2 n) \cdot\nonumber\\
& sign(2\eta_1 +\eta_2 n) \cdot sign(x).
\label{signE}
\end{eqnarray}
Remarkably, althouh the magnitude of the transverse electric field in the periphery stripes is linear in 
the externally applied electric field (and thus linear in the longitudinal flowing electric current), the 
direction of $E_x^{(p)}$ does not depend on the polarity of the external field. We underscore the cause 
of this effect: $E_x^{(p)}$ depends on $E_y$ directly, and, independently of this, also on the sign of the spin density in the stripes. Having 
maximal  (minimal) value, the spin density in the stripes does not depend on the magnitude of the external field but depends, as Eq.(\ref{signSz}) shows, on the direction of the field. Thus,
the $sign(E_y)$ appears twice in Eq.(\ref{signE}), and upon the change of polarity of the external field,
the direction of  $E_x^{(p)}$ does not change. However, the electric field is an odd function of the transverse coordinate $x$.

The direction of the electric field in the periphery stripes turns out to depend on the product of the
two combinations of side jump-like and skew scattering currents. One combination defines 
the magnitude of AHE and spin Hall currents at maximal spin polarization, 
the other, coming from $S_z$ on the boundaries between the periphery and central stripes, defines the 
spin Hall current on the center-line of the sample. The transverse electric field across the sample is defined by Eqs. (\ref{fieldcenter}),(\ref{valE}) and (\ref{signE}). A general feature of the transverse electric field is a 
steep dependence on coordinate at the boundary between the central and periphery stripes. In zeroth approximation to charge-electric field distribution,  at central-periphery stripe boundaries the electric field magnitude experiences "jumps" 
\begin{equation}
\Delta E_x= \pm \frac{2\eta_1 +\eta_2 n}{2\mu }E_y.
\end{equation}
The numerator in this equation is defined by the value of the spin current at $x=0$.
From the discussion of the electrostatic picture in the next subsection and Appendix A, we will see that in fact there are two extremely narrow transitional 
ranges of $x$ between the central and periphery stripes, in which the electric field has a finite derivative as a function of $x$. Still, if spin relaxation is negligible, from comparison of the value of electric field and its variation with $x$ in the central stripe with the value of the electric field in the periphery stripes, one can separate the side jump-like currents from the skew scattering. 

In a real electronic transport experiment, a voltage rather than the electric field is measured. 
For uniform distribution of impurities, the electric potentials are 
the same at the two boundaries of the sample. However, voltage, which is directly related to spin current, can be measured between a probe at the edge of the sample and a probe at $x$ in the periphery stripe. The magnitude of this voltage is
\begin{equation}
U_{sc}=\frac{(\eta_1+\eta_2 n)  E_y}{\mu}|x|.
\label{uper}
\end{equation}
The magnitude of potential difference between the centerline 
and a probe at a point $x$ within the central stripe is given by
\begin{equation}
 U_c(x)= \frac{\eta_2(2\eta_1+\eta_2 n)}{4\mu\zeta}E_y^2 x^2. 
\label{ucenter}
\end{equation}
At $x=0$ the potential is chosen to be zero. This equation holds even if the sample is narrow, and the spin polarization does not reach its maximal possible magnitude.
The magnitude of potential difference between the centerline and a probe within a periphery stripe is given
\begin{equation}
U_p(x)=- \frac{\zeta n(8\eta_1+3\eta_2 n)}{4\mu(2\eta_1+\eta_2 n)}+
\frac{(\eta_1 +\eta_2 n)E_y}{\mu} |x|
\label{uside}
\end{equation}
The sign of these potential differences depends on electric field configurations. In particular,
if $E_x^{(p)}(x<0) <0$, the potential is higher at the edge than at the centerline. If $E_x^{(p)}(x<0) >0$ the potential at the edge is lower than that at the centerline. We discuss examples of these situations below. If measured, potential differences between an edge of the sample and several probes within it would provide
information about the magnitudes of side jump group currents or the skew scattering contribution. 

The overall shape of the transverse electric fields and potentials depend on 
the magnitudes and signs of the $\eta_2$ skew scattering and $\eta_1$ side-jump-like contributions. 
We consider four situations that will be described microscopically in Sec. IE 
with a justification of their physical parameters in Sec. IV.  We mention here that the 
effect of $\eta_1$ currents has the same sign as the 
Drude conductivity, and the $\eta_2$ current has the sign opposite to $\eta_1$ contributions in the case of dominant scattering by the attractive impurity potential, see, e.g., \cite{Chazalviel,Vignale,dasSarma}. In the case of repulsive impurity potential the sign of the $\eta_2$ contribution is the same as the sign of the $\eta_1$ currents. Then, in the case of remote doping with large setbacks, when the principal scattering mechanism is due to impurities in the quantum well, while the carriers are primarily due to remote donors, the skew scattering current can be eliminated by introducing full compensation of quantum well donors by acceptors. 

\begin{figure}[t]
\vspace{-5mm}
\includegraphics[scale=0.35]{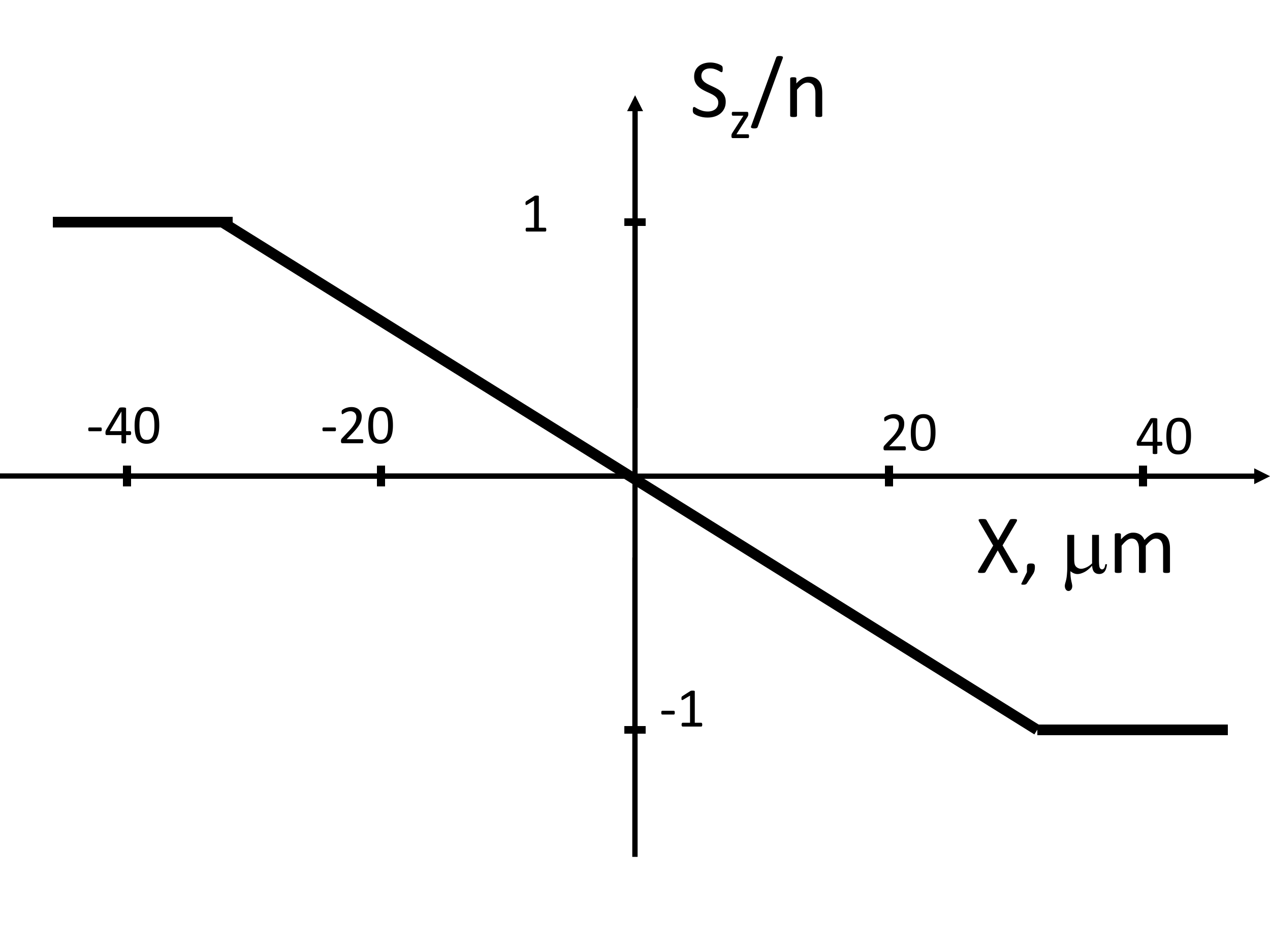}
\vspace{-10mm}
\caption{Dependence of spin polarization of electrons on coordinate transverse to passing electric current in case I. The slope of the dependence in the central stripe is calculated for parameters dicsussed in Sec.IV. 
Similar spatial dependences characterize cases II,III and V, but with the corresponding slopes and widths of the central stripes.}
\vspace{-5mm}
\label{Fig5}
\end{figure}

\begin{figure}[t]
\vspace{2mm}
\includegraphics[scale=0.35]{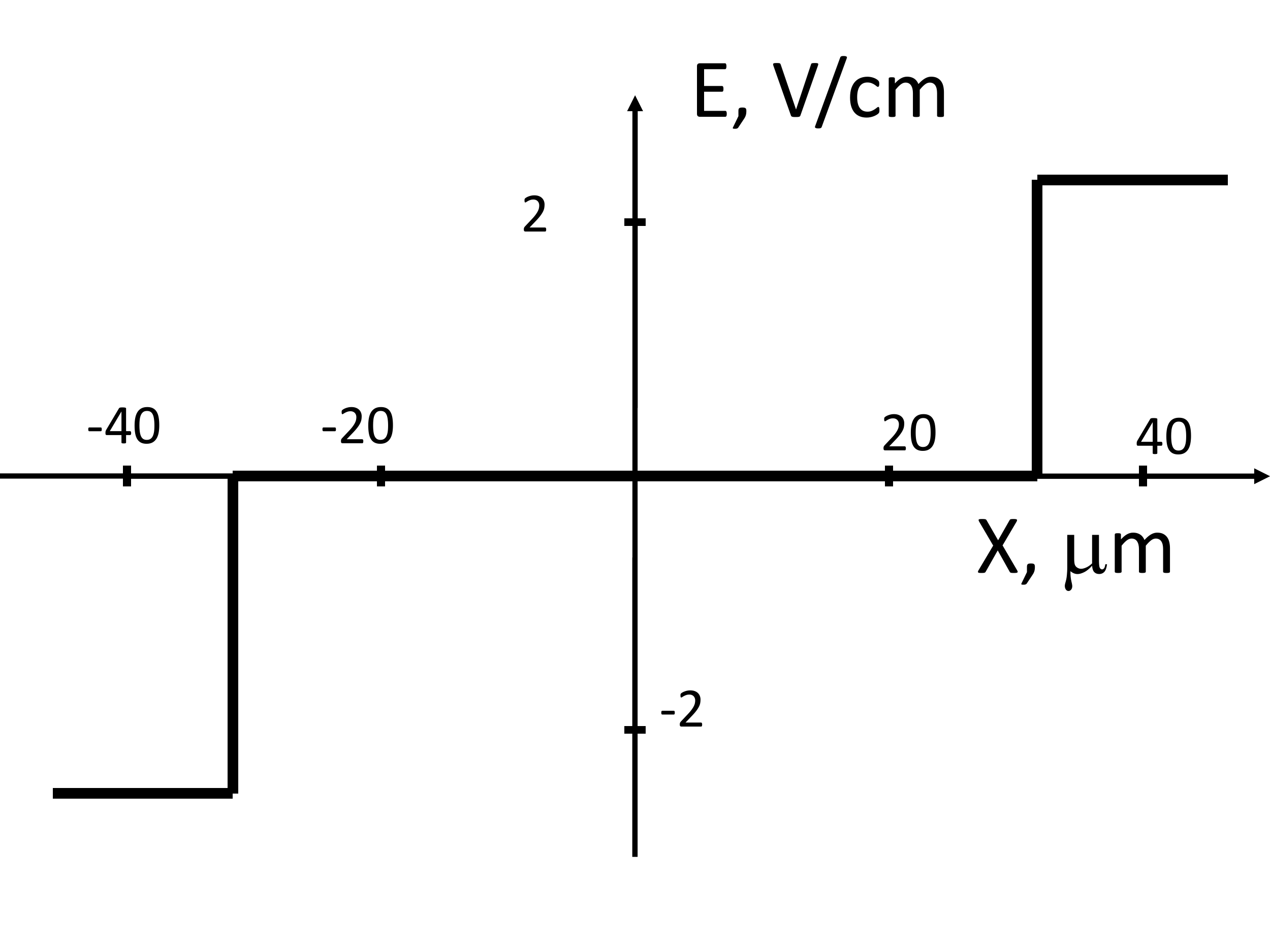}
\vspace{-9mm}
\caption{Spatial dependence of transverse electric field for case I. The underlying parameters are discussed in Sec. IV. }
\vspace{-2mm}
\label{Fig6}
\end{figure}

The qualitatively distinct cases are: 

 $\bullet$ Case I. Skew scattering is absent: $(\eta_2=0)$. The electric field $E_x^{(c)}$ 
in the central stripe vanishes. In the periphery stripes,  $E_x^{(p)}(x<0) <0$ and  $E_x^{(p)}(x>0) >0$.
The spin polarization is $p=1$ at $x<0$ and $p=-1$ at $x>0$. The spin polarization is shown in 
Fig.~\ref{Fig5}, the transverse electric field is shown in Fig.~\ref{Fig6}, the potential distribution is given by Fig.~\ref{Fig2}.

\begin{figure}[t]
\vspace{-4mm}
\includegraphics[scale=0.35]{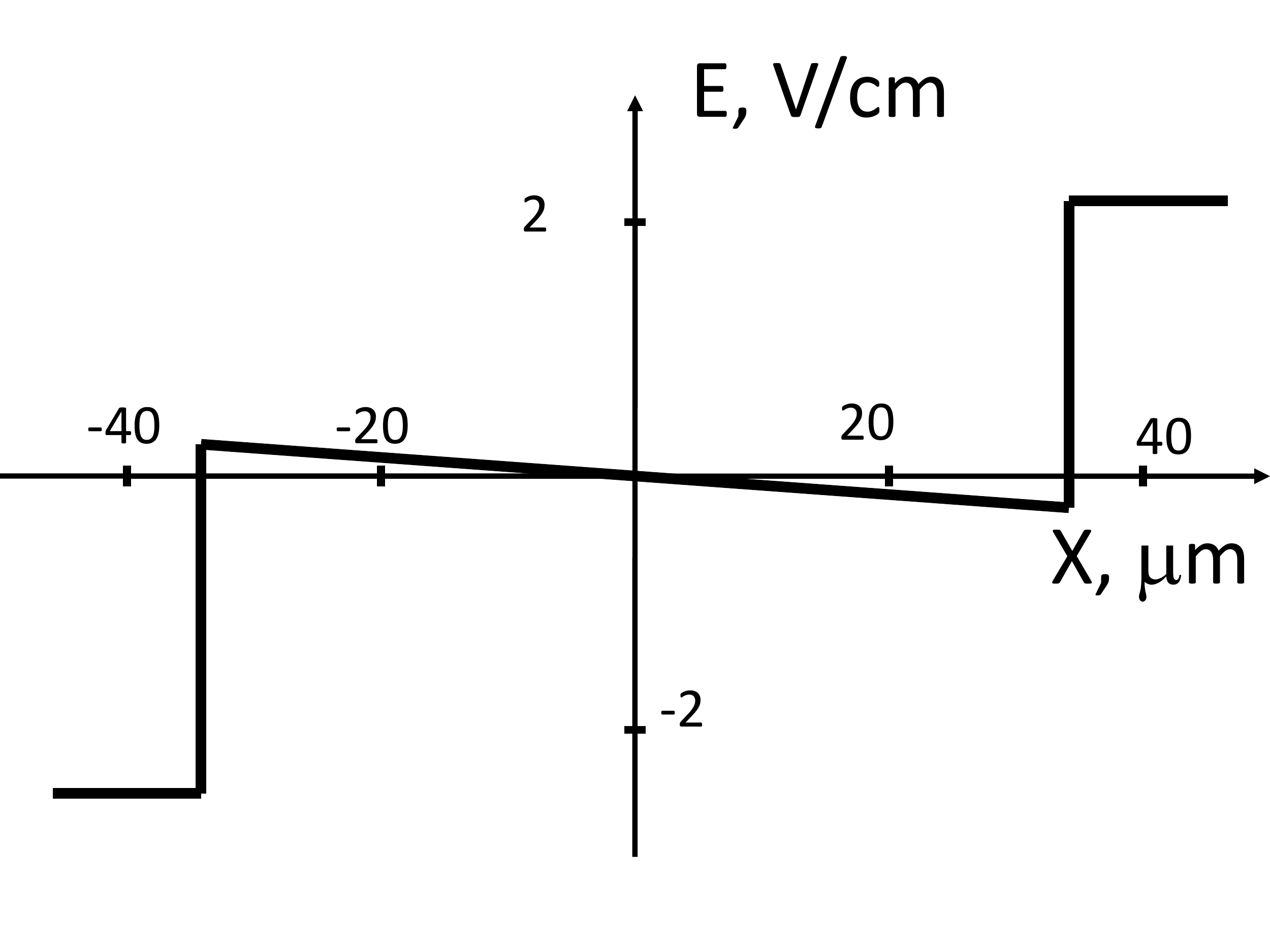}
\vspace{-8mm}
\caption{Dependence of transverse electric field on coordinate transverse to the flowing current for case II. The underlying parameters are discussed in Sec. IV.}
\vspace{-3mm}
\label{Fig7}
\end{figure}

\begin{figure}[t]
\vspace{2mm}
\includegraphics[scale=0.7]{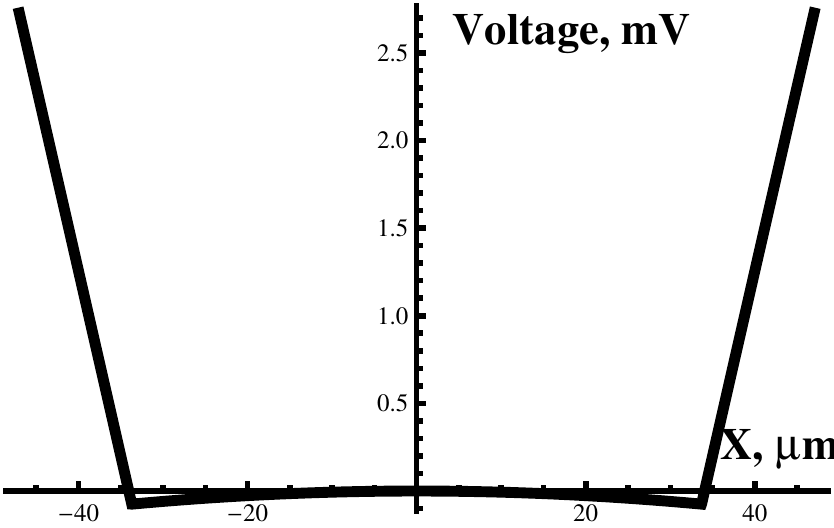}
\vspace{-1mm}
\caption{Electron potential energy profile along the direction transverse to the flowing current for case II. The underlying parameters are discussed in Sec. IV.}
\vspace{1mm}
\label{Fig8}
\end{figure}

$\bullet$ Case II. Side-jump-like term dominates both the AHE and the spin currents at $x=0$: $\eta_1+\eta_2n>0$, $2\eta_1+\eta_2n>0$  Skew scattering is characterized by $\eta_2<0$. 
The slope of the electric field in the central stripe $dE_x^{(c)}/dx>0$ defines the camel back potential profile. In the periphery stripes,  the situation is similar to case I: $E_x^{(p)}(x<0) <0$ and  $E_x^{(p)}(x>0) >0$. The distribution of the spin polarization is also similar to case I:  $p=1$ at $x>0$ and $p=-1$ at $x<0$. In comparison with case I, the central stripe is wider and the periphery stripes are narrower.  The transverse electric field is shown in Fig.~\ref{Fig7}, and the potential profile 
is given by Fig.~\ref{Fig8}.

\begin{figure}[t]
\vspace{-4mm}
\includegraphics[scale=0.35]{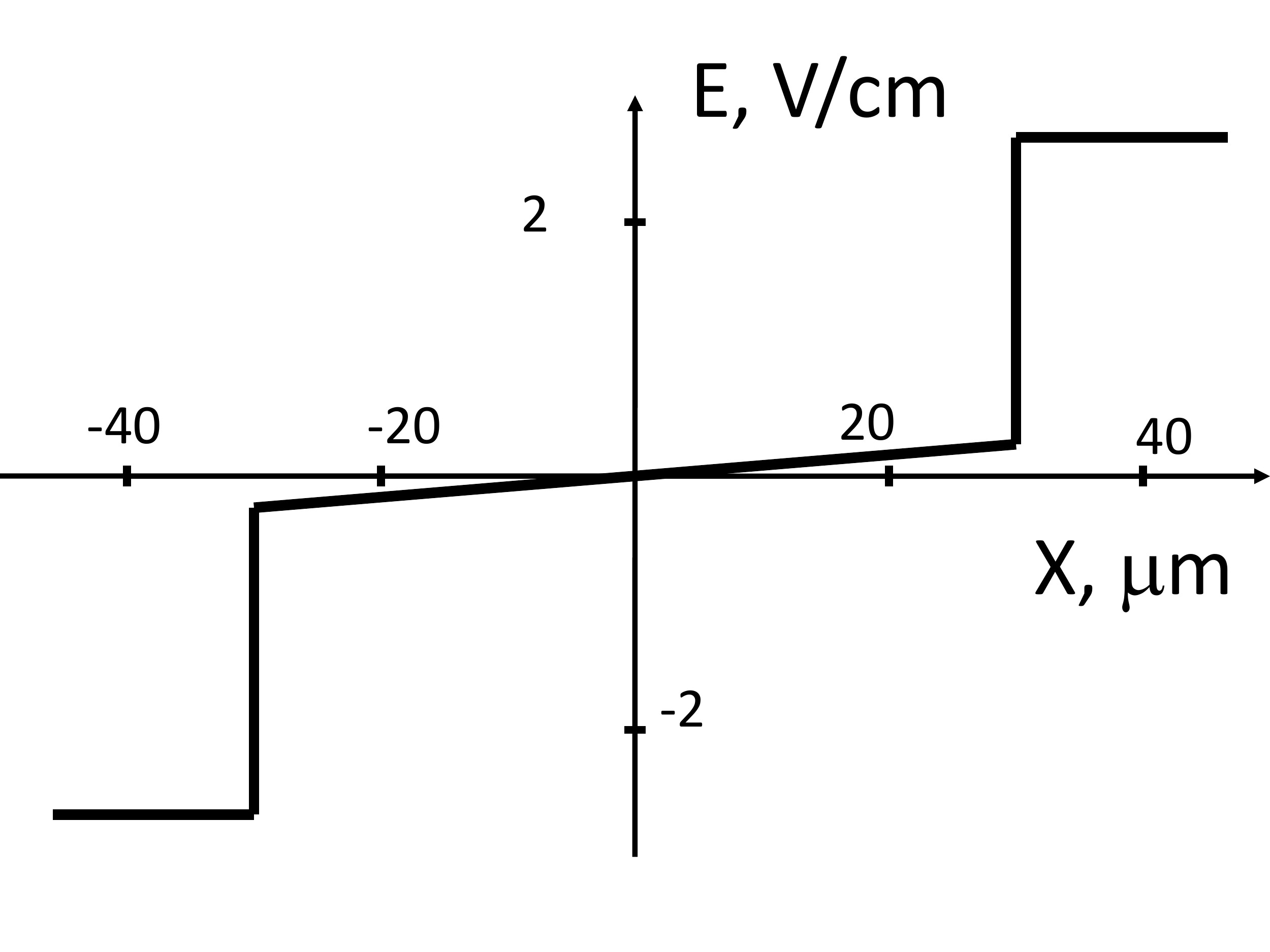}
\vspace{-8mm}
\caption{Spatial dependence of transverse electric field for case III. The underlying parameters are discussed in Sec. IV.}
\vspace{-3mm}
\label{Fig9}
\end{figure}

\begin{figure}[t]
\vspace{2mm}
\includegraphics[scale=0.7]{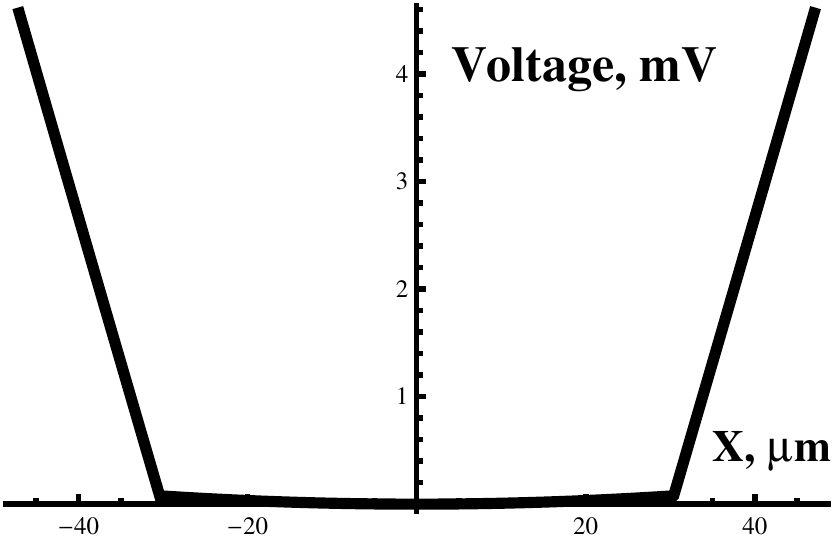}
\vspace{0mm}
\caption{Electron potential energy profile along the direction transverse to the flowing current for case III. The underlying parameters are discussed in Sec. IV.}
\vspace{-1mm}
\label{Fig10}
\end{figure}

$\bullet$ Case III. Side jump-like current is dominant both in the spin current at $x=0$,  $2\eta_1+\eta_2n>0$, and in the AHE,  $\eta_1+\eta_2n>0$, and skew scattering  contribution with $\eta_2>0$ is present. The qualitative behavior of electric field and spin polarization in the periphery stripes is the same as in cases I,II. In the central stripe, the transverse electric field has a slope  $dE_x^{(c)}/dx<0$, which results in a parabolic shape of the potential profile. The spin polarization across the sample is similar to that in Fig.~\ref{Fig5} for the case I. The central stripe is narrower, and the periphery stripes are wider for the case III compared to cases I,II. The transverse electric field is shown in Fig.~\ref{Fig9}, and the potential profile 
is given by Fig.~\ref{Fig10}.

We will see in Sec. IV that cases I,II and III, describe the proposed experimental settings, although spin relaxation modifies the spin density, electric field and potential energy shapes, as discussed in Sec. II. There are other shapes of spin density, 
electric field and potential that may characterize the spin-electric stripes in the absence of spin relaxation. It turns out, however, that in the InAlAs/InP/InAlAs system, which is a possible existing experimental setting for observation of the stripes,  the  ratio of skew scattering and side-jump-like contributions required  for these shapes make spin relaxation strong. Still, we briefly discuss these shapes.

\begin{figure}[t]
\vspace{-2mm}
\includegraphics[scale=0.35]{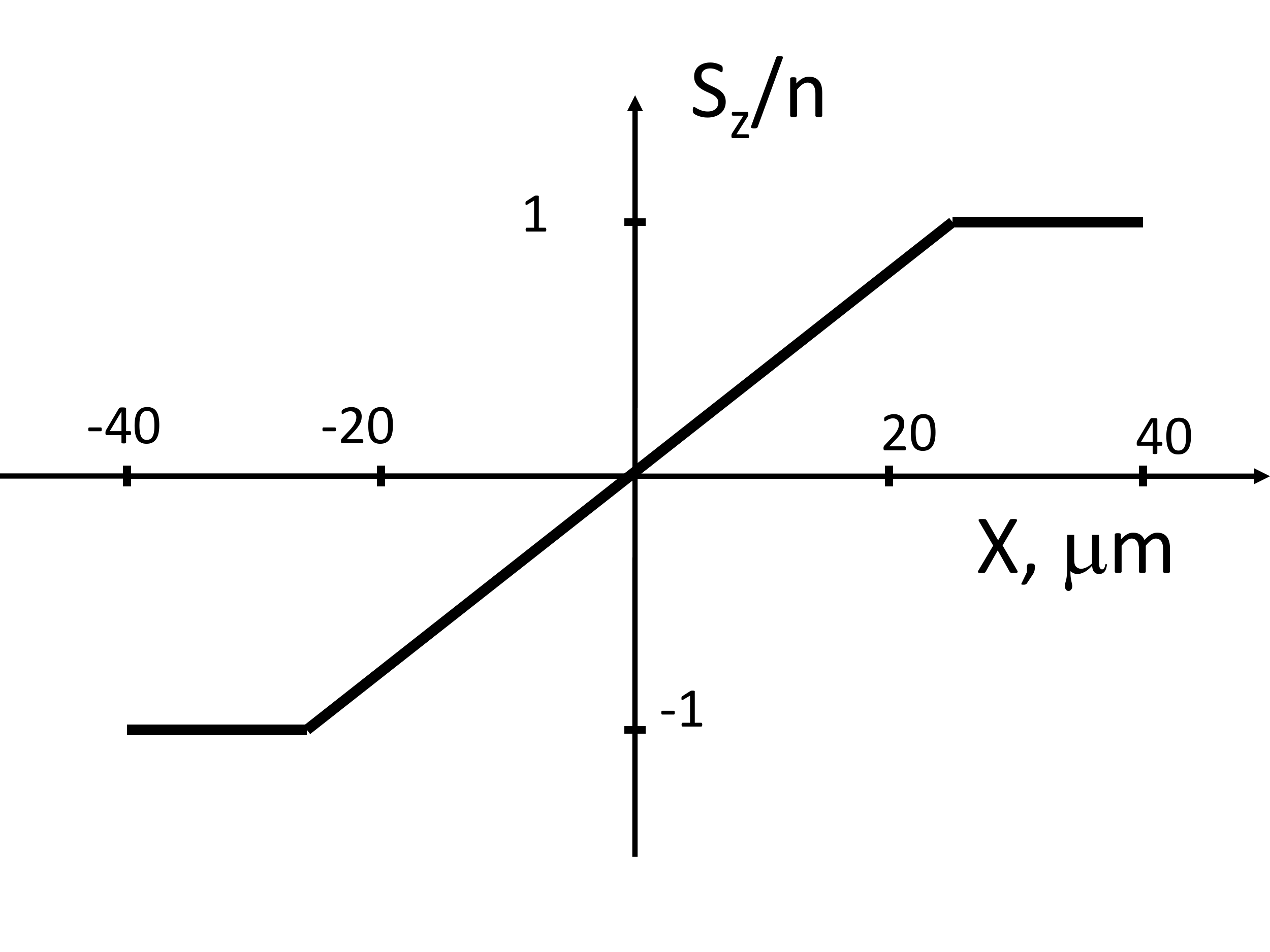}
\vspace{-9mm}
\caption{Qualitative picture of spin polarization profile along the direction transverse to the flowing current for case IV.}
\vspace{-2mm}
\label{Fig11}
\end{figure}

$\bullet$ Case IV.   Skew scattering $\eta_2<0$ dominates both the AHE and the spin currents at $x=0$:$\eta_1+\eta_2n<0$, $2\eta_1+\eta_2n<0$. The profile of electric field and potential in this case is similar to 
case I. The spin polarization, however, reverses its sign compared to case I,  as shown in Fig.~\ref{Fig11}.

\begin{figure}[t]
\vspace{1mm}
\includegraphics[scale=0.35]{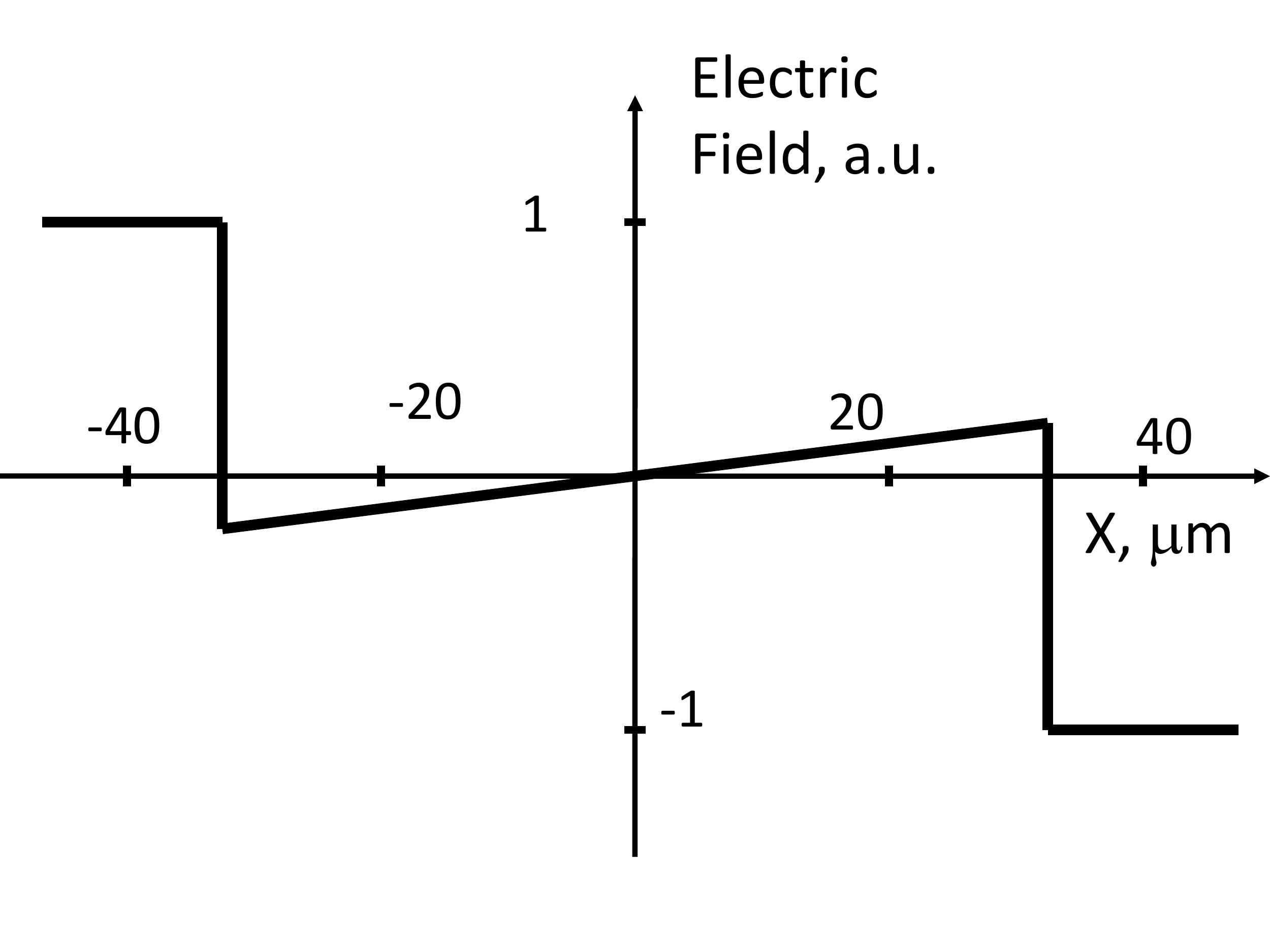}
\vspace{-8mm}
\caption{Qualitative picture of electric field profile along the direction transverse to the flowing current for case V.}
\vspace{-2mm}
\label{Fig12}
\end{figure}

\begin{figure}[t]
\vspace{1mm}
\includegraphics[scale=0.35]{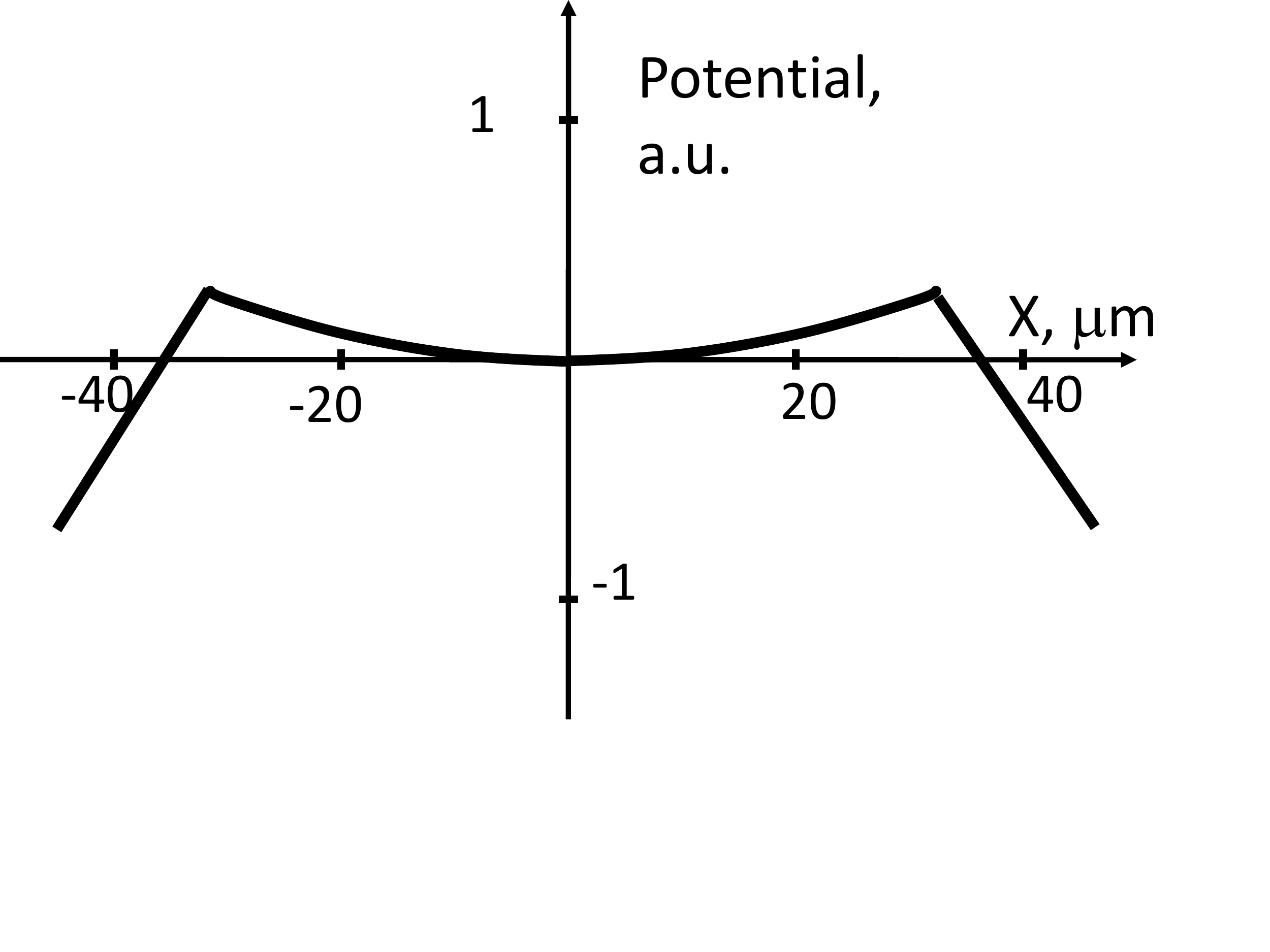}
\vspace{-15mm}
\caption{Qualitative picture of electron potential energy profile along the direction transverse to the flowing current for case V.}
\vspace{-2mm}
\label{Fig13}
\end{figure}

$\bullet$ Case V. 
Skew scattering $(\eta_2<0)$ dominates the AHE  $\eta_1+\eta_2n<0$, but the side jump-like contributions dominate the spin current at $x=0$, $2\eta_1+\eta_2n>0$. This is a surpprising situation. 
 In the central stripe, the transverse electric field has a slope  $dE_x^{(c)}/dx<0$. 
The spin polarization is qualitatively the same as in the case I. However, the electric field in the periphery stripes is opposite compared to the cases I-IV. The physical picture explaining this field distribution and the corresponding potential energy shape is as follows. In the central stripe
electrons with spin up travel towards negative $x$ and reach the boundary of the periphery stripe with $S_z=1$. In the periphery stripe with these negative $x$, the transverse current of particles with spin up, instead of continuing to flow towards the edge of the sample, as in Figs.~(\ref{Fig2},\ref{cartoon}), reverses its direction and flows towards the boundary of the periphery and the central stripe. This is because the spin current at $S_z=1$ and the spin current at $S_z=0$ have opposite signs for the set  of parameters in the case V. Still, the sum of all contributions to the currents satisfies conservation laws of Eqs. (\ref{sc1},\ref{cc1}) in both the center-stripe and the periphery stripes. The corresponding profile of the electric field that compensates the reversed flow of spin (and AHE) current in the periphery stripes is shown in Fig.~\ref{Fig12}, and the potential profile is shown in Fig.~\ref{Fig13}.

Measurement of potential difference between the side contact to periphery region, and probes in the periphery and the central stripes is the most straightforward experiment.
In this regard we note that the results presented above implicitly assume the dependence of the spin current on density given by Eq.(\ref{source}).
While the term proportional to $\eta_1$ is universal (independent of impurity scattering mechanism), we shall see that the skew scattering $\eta_2$ term 
generally depends on the ratio of the square of the transport time and the power  $3/2$ of the collision time (Sec. IE). For short-range scattering, when all 
scattering times are energy-independent, the  Eq.(\ref{source}) holds. In quantum wells with ionized impurity scattering mechanism being dominant, the skew scattering term in the absence of screening could depend on density as 
$n_{\pm}^{5/2}$ rather than $n_{\pm}^2$.
 Screening will bring back the power in the dependence closer to 2. A meaningfull experiment is then to measure the dependence of the transport time on density,
and use this dependence to solve Eqs. (\ref{eq+}, \ref{eq-}) with a modified density dependence of $\eta_2$ term. This will result in modification of the solution in the central stripe, and a different functional dependence of the electric field and spin polarization there. The transverse electric field in the periphery stripes will remain linear in $E_y$,
and will be defined precisely by Eqs.(\ref{valE})  with the only modification of density dependence of $\eta_2$ term in brackets.  It will be then possible to separate 
the modified first and second terms in Eq.(\ref{uside}) and determine $\eta_1$ and $\eta_2$.

Other experiments aimed at observing spin-electric stripes are also feasible.
For example, the electric field spatial distribution can potentially be measured optically via birefringence 
(more precisely, linear birefringence - the Pockels effect\cite{Landau}).
The spin structure also can allow testing via traditional measurement of spin polarization distribution. 
The periphery of the sample is $100\%$ polarized, the two domains have opposite spin polarization, and the polarization in each of the 
domains is constant in the absence of spin relaxation. We shall see in Section II that this picture is only slightly modified by the presence of weak spin relaxation: the polarization reaches maximal possible 
magnitude only at the edge and decreases within periphery stripes towards boundaries with the center stripe.  We should also remark that only in zeroth order in charge-electric field distribution, the total carrier density is constant: in the next approximation, discussed in the following subsection, the total density becomes non-uniform. Interestingly, while 
the circular polarization of the optical signal for the preiphery domains does not depend on the density distribution, the intensity of the signal can sense the modified charge density.
 
We underscore that maximal spin polarization caused by passing current through the sample is not limited to current-carrying electrons. All electrons in the periphery stripes are characterized by one $z-$projection of spin. The spin polarization at the centerline is zero, and is described by Eq.(\ref{polcenter}) in the central stripe. 
We reiterate that, strictly speaking, the above picture is valid when the $z$-component of charge
carrier spins is conserved, so that spin relaxation plays no role in limiting
accumulation of spins
near the boundary, as well as in the balance of spin densities in the bulk. It is then 
the Coulomb interaction and the maximal possible spin polarization that set the limit on charge accumulation near boundaries
and leads to formation of the electric field stripes. 
Our solution for spatial distributions of electric fields and densities of each of the spin species 
takes into account the dependence of the electron energy and charge densities on the spin polarization.  

\subsection{\it Electric field and charge density in the spin-electric state}

We now discuss the disctribution of electric fields and charge density using 
considerations of 2D electrostatics. 
So far, we have assumed carriers with both spins independent of each
other and found that
passing electric current results in transverse electric field periphery stripes with uniform and opposite electric fields. The electric field in the 
central stripe is linear in transverse coordinate,  and depends on microscopic mechanism of the spin current. If only the side jump type currents are present then the electric field in the central stripe vanishes.  
We treat the electron density as constant across the sample, similarly to the conventional Hall effect.
However, the electric field and the total charge density, which includes electron charge density $n=n_++n_-$ 
and a background charge density due to ions/impurities, 
   define each other self-consistently.  
Indeed, the distribution of the electric field
in the system is given by the Poisson equation
\begin{equation}
div {\bf E}= 4\pi\rho,
\label{P}
\end{equation}
where $\rho$ is the charge density. Normal components of the field
$E_z$
are discontinuous
at the 2D plane:
\begin{equation}
E_z(z+0)-E_z(z-0)=4\pi (N_0-n_+-n_-),
\label{norm}
\end{equation}
where $N_0$ is the 2D positive charge density (due to ions in metals or ionized
impurities in semiconductors). 
The electron charge density $n$, depending on the experimental situation, 
comes either from electrons in metals, or from doping in semiconductors. In semiconductor heterostructures, electrons come from doping of the 2D quantum well itself and from remote doping layers. 

We now demonstrate that in the next approximation to electric-field-charge distribution, the charge distribution corresponding to the electric field stripe structure is non-uniform.  For this demonstration, we consider the situation 
when electrons come only from the quantum well impurities,  so that the total charge of electrons and impurities in the 2D plane is zero (total electroneutrality). To simplify our task, we will also restrict the self-consistent solution of the electrostatic problem to the case I with only side jump group spin currents present. We note that while the electric fields and the voltages between the center-region and the periphery of the sample are directly observable, the distribution of charge density is much more difficult to identify.  True charge density depends on fluctuations in distribution of impurities. It would be of interest to address this and other aspects of the electrostatic problem, such as charge-field distribution in the presence of symmetric remote doping with two additional layers of positive charge, and to include all microscopic mechanisms of the spin current, for specific experimental configurations. 

We now solve Eq.(\ref{P},\ref{norm}) for case I, by using the boundary conditions for
the electric field in the stripes and constraints on electric charge density. The electric field in stripes is described by Eq.(\ref{valE},\ref{signE}) at $\eta_2=0$, 
and is zero in the central stripe. The distribution of charge in a sample corresponds to the electric field obtained 
in the initial approximation to charge-electric field distribution, and is subject to constraint on the minimal value of possible electron charge. The 
origin of the constraint is as follows. If the distribution of the electric field consistent with the spin current results in a discontinuity 
according to electrostatics, and the charge at discontinuity must be positive, then this charge density
cannot exceed the density of positively charged ions 
(or ionized impurities located in a quantum well) $N_0$. We assume that each donor or ions supplies one electron, and correspondingly, as soon as 
the $N_0$ limit is reached, no electron charge can further be removed from a given location in order to reach values of positive charge bigger than 
 $N_0$.
This restriction matters because, similarly to the case of charge
profile of the 2D charged strip\cite{halle,smythe}, we can expect 
divergencies of charge density at the boundaries of the stripes, in which the electric field is non-zero.   
For example, in a 2D charged metal film, 
divergencies of charge density at the edges result in a finite
electric potential $\phi$. The presence of excess electrons at the boundaries in our case is aided by the spin current. 
If electrostatics requires some ranges of
$x$ to be abandoned by electrons with both spin projections, there is a constraint on the
remaining positive charge at these $x$.

In the long 2D strip configuration we only need to find the distribution of electric
field in $x-$direction, and the dependence of charge density on $x$,
because the system is uniform in the direction $y$ of the electric current
flow. 
Consider a complex function of complex
variable $E(t)$.
The real part of this function $E_x(x,z)$, $Re E(x+iz)$ defines the
electric
field in the sample:
\begin{equation}
 \frac{1}{2\pi} [ E(x+i0) + E(x-i0)]=E_x (x),
\end{equation}
and the complex part of $-Im E(x+iz)$, as follows from Eq.(\ref{norm}),
defines the distribution of charge:
\begin{equation}
[ E(x+i0) - E(x-i0)]=4\pi (N_0-n_+(x)-n_-(x)),
\end{equation}
 We now find the function $E(t)$, using the following conditions:
\begin{eqnarray}
Re E(x)  =  E_0 & \mbox{if $b_1<|x|<a$}
\label{a}\\
Re E(x) =  0   & \mbox{ if $|x|< b$}
\label{b}\\  
Im [E(x+i0)-E(x-i0)]  & = -4\pi N_0 \nonumber \\   \mbox{if $b<|x|<b_1$}
\label{c1}\\
Im [E(x+i0)-E(x-i0)] & >  -4\pi N_0  \nonumber \\   \mbox{if $|x|<b$ or $b_1<|x|<a$ }
\label{d},
\end{eqnarray}
where $b_1$ defines the range $b<|x|<b_1$ of minimal positive charge. 
These ranges are transitional regions between the central and the periphery stripes.
The value of $b_1$ is defined from known $E_0$,  $N_0$,  
by using the asymptotics of $E(t)$ as $|t|\rightarrow
\infty$, 
which for electrically neutral sample gives $E(t)\rightarrow 0 (1/t^2)$ as $|t|\rightarrow
\infty$.
The condition Eq.(\ref{c1}) is the
constraint
on the positive charge density. Furthermore, this constraint also affects the boundaries of ranges in Eqs (\ref{a},\ref{b},\ref{d}).  
However, as we shall see, $b_1$ is exponentially
close
to $b$, and in reality the constraint is relevant only 
at the points $b$ (positive and negative) merged with $b_1$\cite{but}, which coinside with the boundaries between the central and the periphery stripes.    
 By introducing Eqs.(\ref{a},\ref{b},\ref{c1},\ref{d}) as above, we
describe the electric field and charge distribution in the stripes 
self-consistently. 

\begin{figure}[t]
\vspace{-1mm}
\includegraphics[scale=0.8]{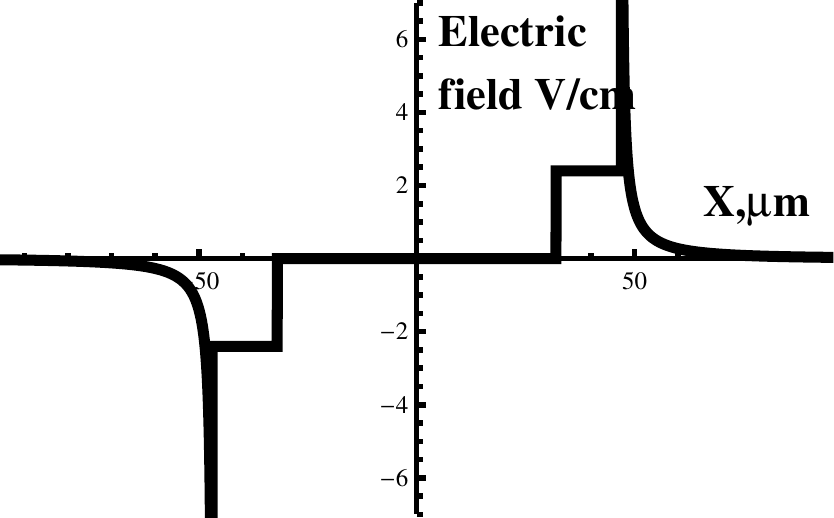}
\vspace{2mm}
\caption{Distribution of the transverse electric field due to the spin current across the 2D sample. $N_0=5\times 10^{11}cm^{-2}$, $E_0=2.4 V/cm.$}
\vspace{-2mm}
\label{Fig14}
\end{figure}

The details of solution of Eqs.(\ref{P},\ref{norm}) using methods of 2D electrostatics are given in Appendix A. 
The electric field in the system is given by 
Eqs.(\ref{electricfield},\ref{y},\ref{b1}) and the charge density is defined by Eqs. (\ref{n1},\ref{n0},\ref{n2}).
The electric field profile is shown in Fig.~\ref{Fig14} (electroneutral quantum well). 

\begin{figure}[t]
\vspace{-1mm}
\includegraphics[scale=0.8]{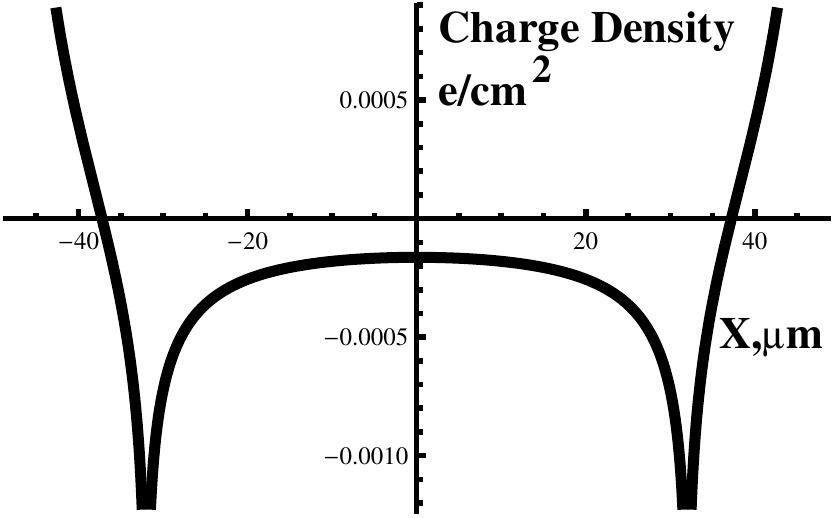}
\vspace{2mm}
\caption{Distribution of the charge accross the 2D sample, caused by the spin current. 2D charge carriers are assumed to originate from donors in the quantum well only, 
$N_0=5\times 10^{11}cm^{-2}$, $E_0=2.4 V/cm.$}
\vspace{-1mm}
\label{Fig15}
\end{figure}

The charge density is shown in Fig.~\ref{Fig15}. We note that the values $b$ and $b_1$ that define intervals of $x$  abandoned by carriers of both spin polarizations (this indeed occurs) are exponentially close, so that such intervals are essentially two singular points symmetric with 
respect to the centerline $x=0$ of the sample.
However, in the vicinity of these points, there is a visible logarithmic dependence of charge density,  
which potentially can be detected. The logarithmic character of the position dependence of the charge density 
is defined by Eqs(\ref{n1},\ref{n2}), and Eq. (\ref{b1}) at $n_{rd}=0$. 
We note that measurements of charge distribution and spin polarization can potentially demonstrate 
the logarithmic behavior of density in the vicinity of $|x|=b$,
and the square root features in the vicinity of the edges at $|x|=a$.

\subsection{\it Spin-electric stripes in the presence of a perpendicular magnetic field}

Magnetic field $ H$ in z-direction (perpendicular to the 2D plane) results in the two principal effects:
(i) usual (orbital) Hall effect resulting in a uniform electric field, and (ii) in a magnetic field induced spin polarization of charge carriers. This spin polarization, even if all other sources of spin polarization are ignored, results in the Anomalous Hall effect of 
spin-polarized carriers, i.e. the transverse charge current. In the constitutive equations (\ref{j}) for 
currents due to electrons with both spin projections
we must add the effect of charge Hall current, which at small magnetic fields is given by 
\begin{equation}
{\mathbf J}_{\pm}^{cH}= \mu  n_{\pm}  \theta\omega_c\tau({\mathbf z}\times {\mathbf E}),
\end{equation}
where $\omega_c = eH/mc$, is the cyclotron frequency. The coefficient $\theta$ is defined by the energy dependence of momentum relaxation 
time $\tau$, and for $\tau$ independent of energy $\theta=1$. In contrast to the spin Hall current, this contribution comes with the same sign in equations for the currents due to electrons with "$+$" and "$-$" 
$z-$projections of spins. Furthermore, in  equations (\ref{j}), the densities $n_+$ and $n_-$ are now magnetic-field dependent. 
We note that taking into account 
this dependence of density on magnetic  field automatically accounts for the AHE 
current, and no further modifications of constituitive equations in magnetic field are necessary. 

In a degenerate 2D gas at low temperatures, in the absence of sources of polarization other than magnetic field, the electron spin polarization
$S_z = p_H n$, where $p_H=g\mu_B H/\epsilon_F$,  $g$ is the electron g-factor and $\mu_B$ is the 
Bohr magneton. Then on symmetry grounds, such spin polarization is the only source of the spin polarization at $x=0$, and the boundary condition for $S_z$ now is $S_z(0)=p_H n$.
Solving the constituitive equations, for the spin polarization in the central stripe we get
\begin{equation}
S_z(x)=p_H  n -\frac{2\eta_1 +\eta_2 n}{\zeta}E_y x. 
\end{equation}
For the electric field in the central region we have
\begin{equation}
E_x(x)= -\frac{\eta_2}{\mu }(p_H  n -\frac{2\eta_1 +\eta_2 n}{\zeta}E_y x)E_y +\omega_c\tau E_y.
\end{equation}
The second term in this expression for the electric field comes from the charge Hall effect. 
It is interesting to note that similarly to the case $H=0$, the side jump and skew scattering terms 
contribute to electric field in the central domain in a non-trivial way, which potentially gives a path 
for additional experiments in magnetic field that would allow their separation.

\begin{figure}[t]
\vspace{0mm}
\includegraphics[scale=0.35]{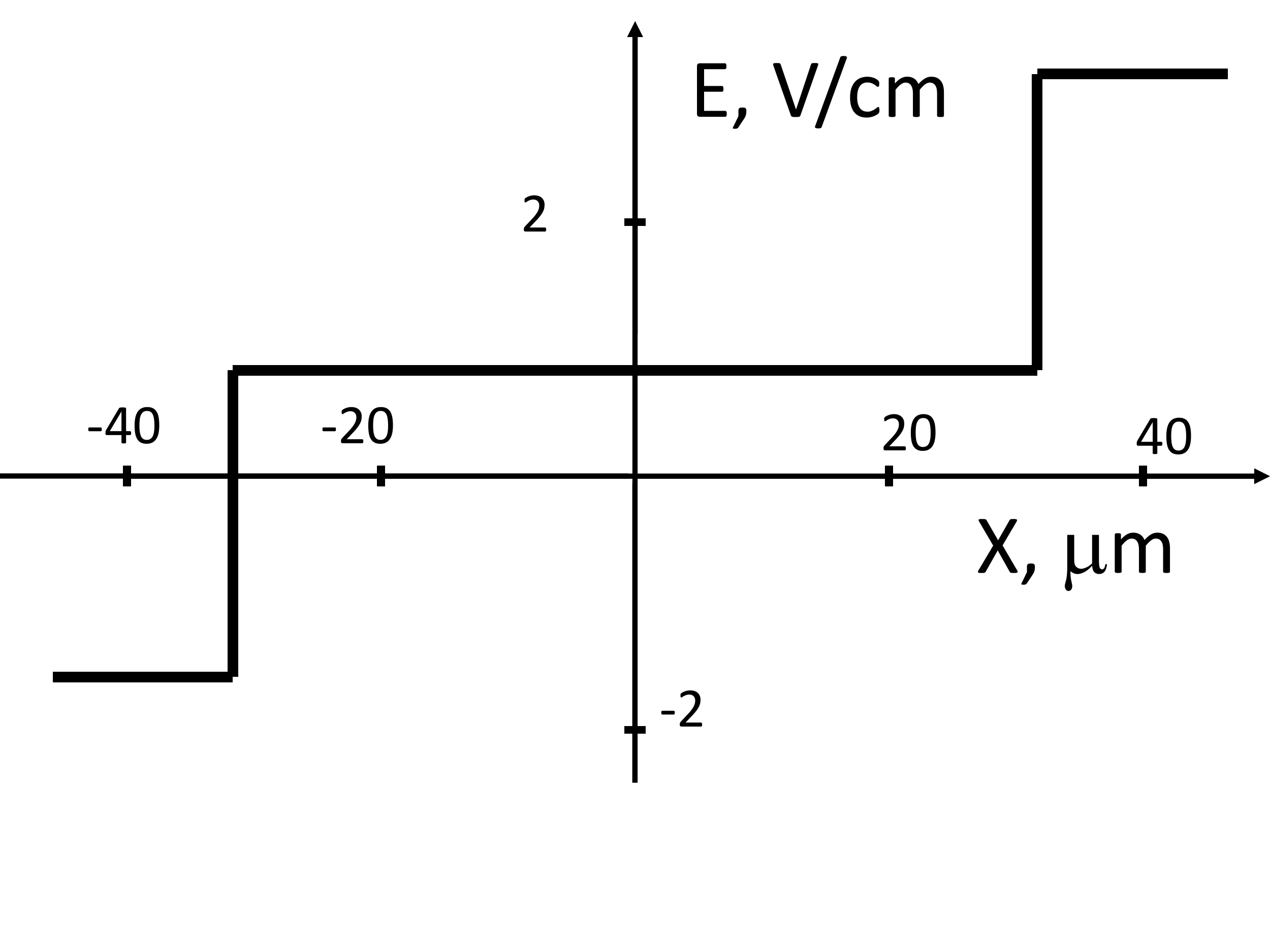}
\vspace{-13mm}
\caption{Dependence of transverse electric field on coordinate transverse to the flowing current in the presence of magnetic field perpendicular to the 2D plane. Case I.}
\vspace{-2mm}
\label{Fig16}
\end{figure}

We now consider the properties of the periphery stripes in the presence of magnetic field.
The uniform Zeemann spin polarization results in the shifts of the boundaries of the stripes, and makes periphery domains asymmetrically positioned with respect to the centerline of the sample. Indeed, the boundaries $\pm b$ between the periphery stripes and the central 
stripe are defined as $x$ at which the electron spin polarization reaches maximal possible values $S_z/n=\pm 1$. In the presence of 
magnetic field induced spin polarization, $p_H$, depending on its sign, the boundary is closer to the centerline on one side and further from the centerline on the other side.  The two boundaries are now given by  
\begin{equation}
b_{\pm}=\frac{(p_H \pm 1)n\zeta}{(2\eta_1 +\eta_2 n)E_y}.
\end{equation}
Thus, application of magnetic field is capable of moving the stripe boundaries.
The electric fields in the two periphery stripes are given by
\begin{equation}
E_x^{\pm}= \omega_c\tau E_y \mp \frac{(\eta_1 +\eta_2 n)E_y}{\mu},
\end{equation}
where the upper/lower sign corresponds to the spin polarization in stripes, $S_z/n=\pm 1$. 
As expected, the electric field due to charge Hall effect leads to electric fields in two stripes 
having different magnitudes. If the charge Hall effect field exceeds the magnitude of its spin Hall effect 
counterpart, electric fields in domains may become of the same signs, in contrast to their opposite 
signs at $H=0$. For the charge Hall current having half value of the spin Hall current in the case I, the distriibution of electric fields in the sample is shown in Fig.\ref{Fig16}.
The potential difference between 
the centerline $x=0$ of the sample and the right edge $x=a$, $U(0)-U(a)= V_H/2 -V$ in the presence of magnetic field is different  from the potential difference between $x=0$ and the left edge $x=-a$,  $U(0)-U(-a)=-V_H/2 -V$ by the amount of the 
charge Hall voltage, i.e. the potential difference between the left and right edges $U(-a)-U(a)= V_H$ . The half-sum of these two potential differences $[2U(0)-U(a)-U(-a)]/2=-V $ defines the voltage drop between the edges and the centerline due to spin Hall current. 

\subsection{Effects of non-uniform doping}

The ideal picture we considered so far is a metallic electrically neutral 2D plane. However, we shall see that a feasible experimental setting is a semiconductor heterostructure with remote doping. Remote doping leads to conditions of electrical neutrality fulfilled in the whole sample, so that the 2D plane itself may actually be charged. Were the 2D plane a metal, in the absence of an external current and therefore spin current, it would be an equipotential irrespective whether it is charged or not. Correspondingly, no fields in the Hall direction would exist in the absence of external voltage, and, once the voltage is applied, it would generate transverse electric field domains, in much the same way as we have discussed already. Thus, in the case of the charged metal  the observable voltage between the center of the sample and the edges clearly remains the same voltage shown in Fig.~\ref{Fig2}, Fig.~\ref{Fig8} and \ref{Fig10}. The concentration profile in a charged metallic plane is going to be modified compared to the electrically neutral plane. We will not consider this issue here, and rather turn our attention to the semiconductor setting.

In semiconductors, the quantum well is inevitably doped by impurities. We shall see below 
that in order to achieve the best conditions of experimental observation of the stripes, it is useful to dope the quantum well itself, and in addition to use the remote doping. Doping profile is never uniform, both in-plane 
and along the growth direction $z$. In particular, one of the effects of this non-uniformity (as well as non-uniformity of the remote doping), which we analyze below, is an additional channel for the the spin relaxation of charge carriers. The question arises whether impurities can in any way alter the observable potentials (fields) in the proposed experiment, and obscure the observation of the electric field stripes. 

Non-uniform distribution of donors results in a built-in electric fields and the corresponding non-uniform electron charge density. Nevertheless, in the 
absence of an external voltage, and therefore absence of the spin current and transverse electric fields that it induces, measuring of voltage between the center of the sample and the edges results in zero signal in the steady state regime. The situation is similar to measuring a voltage drop in a Schottky barrier or p-n junction. Without a battery in
 an external circuit, when  the 
circuit is closed, one has two Schottky barriers (or p-n junctions)
compensating each other, and zero signal. However, if the external voltage in 
longitudinal direction is switched on, the spin current is generated, and the transverse electric fields are induced in the periphery stripes and the central stripe. Then the external circuit 
with a voltmeter will show the voltage drop precizely as in Figs \ref{Fig2}, \ref{Fig8} and \ref{Fig10}. Thus, while a non-uniform distribution 
of impurities in a semiconductor quantum well may alter the electrons charge density distribution compared to the ideal metallic case, the measured voltage between a probe within the sample and a probe at the edge remains the same. 

\subsection{\it Contributions to the spin current and their dependence on electron density.}
The microscopic picture of the spin current of electrons described by the Hamiltonian (\ref{H}) includes two primary 
mechanisms: skew scattering assymetry and the effects associated with the 
change of the center of gravity of the 
wavepacket of electrons subject to electric field and impurity potential (one of the effects 
in the latter group is the side jump current). It is important for us here that the sign of skew scattering current depends on the donor or acceptor character of impurities doping the quantum well, and that the magnitude of the skew 
scattering contribution is 
parametrically different from the magnitude of the effects in the
side jump group, and has a different dependence on density of charge carriers. 
Skew scattering and side jump contributions have been calculated in numerous works, see, e.g., \cite{Abakumov,YLG,Vignale,dasSarma,review}. 
In order to explain conditions, in which spin-electric stripes arise or are absent, we need here 
the expressions for these currents in both 3D and 2D conductors in certain experimental setups and 
spin relaxation times, in the same microscopic model. We therefore present the calculation of 
these quantities.

The skew scattering contribution to the spin current density in the ideal case of Eq.(\ref{H}) 
can be formulated in terms of 
distribution functions of electrons with spin up (or spin down), because there are no spin 
flip processses, and 
two spin species of electrons can be considered independently. 
We note that the effect of electron-electron interaction of skew scattering leading to the spin 
Coulomb drag  can be also taken account by this approach. As was shown in \cite{Vignale}, the 
spin Coulomb drug contribution ultimately introduces a correction to the skew scattering contribution.
Our estimates will show that the resulting reduction of the skew scattering effects amounts to just a 
few percent in the relevant range of parameters.
 
The contribution to the spin current by skew scattering effect is given by 
\begin{equation}
j_{sk}= -e\int d^2{\mathbf k}
{\mathbf v}_{\mathbf k} f^{as}_{\mathbf k},
\end{equation}
 where the spin-dependent (proportional to ${\mathbf \sigma}_z$) 
distribution function $f^{as}_{\mathbf k}$ is obtained from the kinetic (rate) 
equation for the electron distribution function 
$f_k$
\begin{equation}
-\frac{e}{\hbar}{\mathbf E}\cdot\frac{\partial f_{\mathbf k}}{\partial {\mathbf k}}=
\sum_{{\mathbf k}^{\prime}}(W^{(0)}_{{\mathbf k}, {\mathbf k}^{\prime}}+
W^{(1)}_{{\mathbf k}, {\mathbf k}^{\prime}})(f_{\mathbf k} - f_{{\mathbf k}^{\prime}}),
\label{kineq}
\end{equation}
where for 2D electrons
\begin{equation}
W^{(0)}_{{\mathbf k}, {\mathbf k}^{\prime}}=\frac{2\pi}{\hbar S}n_i
|V_{{\mathbf k}, {\mathbf k}^{\prime}}|^2
\delta(\epsilon_{\mathbf k}-\epsilon_{\mathbf k}^{\prime})
\label{sym} ,
\end{equation}
$V_{{\mathbf k}, {\mathbf k}^{\prime}}$ is the Fourier component of the impurity potential, 
$n_i$ is the impurity 
density,  and $S$ is the area of the 2D sample. For the calculation of longitudinal electric current and 
momentum relaxation time, it is sufficient to calculate the 
probability of scattering in the Born approximation. 

The probablity of asymmetric (skew) scattering in the framework of perturbation theory arises in the 
approxination next to the Born approximation:
\begin{eqnarray}
& W^{(1)}_{{\mathbf k}, {\mathbf k}^{\prime}}= -\frac{(2\pi )^2}{\hbar S^2}n_i
\sum_{{\mathbf k}^{\prime\prime}}Im [V_{{\mathbf k}{\mathbf k}^{\prime\prime}}
V_{{\mathbf k}^{\prime\prime}{\mathbf k}^{\prime}}V_{{\mathbf k}^{\prime}{\mathbf k}}]\nonumber\\
& \delta(\epsilon_{\mathbf k}-\epsilon_{\mathbf k}^{\prime})
\delta(\epsilon_{\mathbf k}-\epsilon_{\mathbf k}^{\prime\prime}).
\label{asym}
\end{eqnarray}
We underscore that the asymmetric skew scattering probability that we consider arises from generic 2D spin-orbit interactions given by Eq.(\ref{H}) and does not 
rely on the presence of intrinsic spin-orbit interactions, e.g., of Rashba type, as in the case discussed in \cite{borunda,Nunner}.  We will see that such contributions 
for various symmetries of intrinsic spin-orbit interactions do not play any role in the context of spin-electric stripes induced by the nonequilibrium spin current in 
section IIIB.7.

We will assume the model, in which the principal mechanism of electron scattering is from point-like 
impurities doping 
the quantum well, 
\begin{equation}
V({\mathbf r})=\sum_i V_0\delta ({\mathbf r}-{\mathbf R}_i), 
\label{potential}
\end{equation} 
where ${\mathbf R}_i)$ gives the impurity 
location. The amplitude $V_0$ is negative when Eq.(\ref{potential}) models attraction of electrons to ionized donors. 
The amplitude $V_0$ is positive when Eq.(\ref{potential}) describes charged acceptors.
Point-like scattering is a reasonable approach if all charge carriers come from impurities in the quantum well itself. 
This implies that if charge carriers in 2D sample are supplied, apart from impurities doping the well, 
by the remote $\delta$-doping, then the remote doping 
layers should be 
separated from the 2D electron gas by sufficiently large setbacks. Then a smooth random potential from 
remote doping is 
negligibly small, and scattering 
by impurities of the quantum well are dominant. 
In this case from Eq. (\ref{H}) the scattering amplitude, including spin-dependent part, is 
\begin{equation}
V_{{\mathbf k}{\mathbf k}^{\prime}}=V_0(1+i\alpha{\mathbf \sigma}_z \cdot({\mathbf k}\times{\mathbf k}^{\prime})_z ). 
\label{matrix}
\end{equation}
From Eq.(\ref{kineq}), the asymmetric distribution function describing skew scattering is given by
\begin{equation}
f^{as}_{\mathbf k}= \sum_{{\mathbf k}^{\prime}} W^{(1)}_{{\mathbf k}, {\mathbf k}^{\prime}}\tau 
(f^{({\mathbf E})}_{\mathbf k} - f^{({\mathbf E})}_{{\mathbf k}^{\prime}}),
\label{asym_dist}
\end{equation}
where the momentum relaxation time $\tau$ is calculated from Eqs.(\ref{kineq},\ref{sym},\ref{matrix}) 
and in the leading approximation neglecting small spin-orbit term is defined by
\begin{equation}
\frac{1}{\tau}= 2\pi n_i V_0^2 \nu,
\label{time}
\end{equation}
with $\nu=\frac{m}{2\pi\hbar^2}$ being the electron density of states per spin; the distribution 
function $f^{({\mathbf E})}_{\mathbf k}$ describing electron flow in a 
conductor in the presence of electric field is given by
\begin{equation}
f^{({\mathbf E})}_{\mathbf k}= e\frac{\hbar}{m} {\mathbf E}\cdot {\mathbf k} \tau 
\frac{\partial f_0(\epsilon_{\mathbf k})}
{\partial \epsilon_{\mathbf k}},
\label{flow}
\end{equation}
where $f_0(\epsilon_{\mathbf k})$ is the Fermi distribution function.
Using Eqs.(\ref{asym},\ref{matrix},\ref{asym_dist},\ref{time}), 
we obtain the expression for skew scattering contribution to the
spin current
\begin{equation}
j_{sk}=-2 \alpha n \sigma\beta E_y,
\label{skr}
\end{equation}
where $\sigma=ne^2\tau/m$ is the longitudinal conductivity, and $\beta$ is the parameter distinguishing successive orders of the Born approximations:
\begin{equation}
\beta= \frac{V_0m}{\hbar^2}=\pm \sqrt{\frac{\pi \hbar n}{\epsilon_F \tau n_i}},
\end{equation}
where the upper sign corresponds to scattering by acceptors and the lower by donors. The expression on the right of this equation can be used for estimates of $\beta$, which can reach $\sim$1 \cite{Born,abrikosov}. 

When both donor and acceptor impurities are present, it is possible 
to suppress skew scattering at large compensation. Provided both donors and acceptors are shallow hydrogen-like impurities,
at full compensation the skew scattering vanishes, but electrons coming from remote doping contribute to longitudinal mobility and side-jump group currents. Assuming the concentration of quantum well donors $n_d$ and quantum well acceptors $n_a$, the total skew scattering current is given by
\begin{equation}
j^t_{sk}=2 \alpha n \sigma\beta K E_y,
\end{equation}
where $K= (n_d-n_a)/(n_d+n_a)$ is the degree of compensation. Defined this way, $K=0$ corresponds to full compensation, and $K=\mp1$ correspond to only acceptors or only donors present, correspondingly. We note that because of remote doping, the electron density 
in the quantum well $n$, and partial densities of electron species $n_\pm$ are the parameters of the system that can be tuned independently 
of $K$.

Provided that $\tau$ is independent of energy, the skew scattering current of carriers with a certain projection of spin polarization  is quadratic in density of these carriers (being proportional to $\sigma_\pm n_\pm$).
The parameter $\eta_2$ that determines the characteristics of the stripe state (Sec. IA) is given by
\begin{equation}
\eta_2=\frac{2\pi}{m}\alpha \beta \tau K .
\end{equation} 
 
We now consider the parametric dependence of the effects like side jump. 
These effects can be considered separately 
for the two spin species of electrons in much the same way as the skew scattering contribution to the current.
The side jump current is given by
\begin{equation}
J_{sj}= e\sum_{{\mathbf k}, {\mathbf k}^{\prime}} W^0_{{\mathbf k}, {\mathbf k}^{\prime}}
R_{{\mathbf k}, {\mathbf k}^{\prime}}f_{\mathbf k},
\label{sj}
\end{equation}
where $W_{{\mathbf k}, {\mathbf k}^{\prime}}$ is the scattering probability 
(it is sufficient to take into account symmetric part $W_0$ only),
$f_{\mathbf k}$ is the electron distribution function 
(it is necessary to account for longitudinal flow of electrons in the presence of 
electric field described by $f^{{\mathbf E}}$),  and
\begin{equation}
R_{{\mathbf k}, {\mathbf k}^{\prime}}= i\frac{\partial}{\partial {\mathbf k}}+ 
 i\frac{\partial}{\partial {\mathbf k}^{\prime}} \Phi_{{\mathbf k}, {\mathbf k}^{\prime}} 
+ {\mathbf \Omega}_{\mathbf k}- {\mathbf \Omega}_{\mathbf k}^{\prime}
\label{jump}
\end{equation} 
is the "side jump", i.e., 
the displacement of center of gravity of the electron wavepacket in the proceess of scattering \cite{Berger}. 
Furthermore, $\Phi_{{\mathbf k}, {\mathbf k}^{\prime}}$ is the 
phase of the matrix element of scattering, and
\begin{equation}
{\mathbf \Omega}_{\mathbf k}=i\int u^*_{\mathbf k}\frac{\partial}{\partial {\mathbf k}}u_{\mathbf k}dV
\label{Omega}
\end{equation}
is the diagonal part of the matrix element of 
the electron coordinate associated with the Bloch amplitudes $u_{\mathbf k}$ in the wavefunction. 
In the presence of spin-orbit interactions, ${\mathbf \Omega}_{\mathbf k}= \alpha\sigma_z (k_x, -k_y, 0)$ .
This additional contribution to coordinate 
is often viewed as a dipole associated with the spin-orbit interaction, 
so that the spin-dependent part of the potential in the Hamiltonian 
(\ref{H}) is considered to be due to this dipole moment. 

We would like to note here that the side-jump contribution expression (\ref{sj}) can be equaivalently derived by 
the three methods:
(i) by taking into account terms that are second-order in scattering in off-diagonal density matrix and 
summing over all semiconductor bands\cite{Luttinger}; (ii) by taking into account a single band 
and corrections of the first order in scattering to the current operator together with 
first order corrections to the density matrix 
\cite{ivchenko1,culcer}, and (iii) by considering a semiclassical picture of the 
contribution to the current arising 
from the shift of the center of gravity of the wavepacket in the course of electron scattering \cite{Berger}. 
Apart from the direct contribution of the side jump to the current given by Eq.(\ref{sj}), 
there are other similar 
contributions, which we collectively refer to as the side jump group effects. 
 The side jump group effects remain under debate since 1954 \cite{Luttinger1954}, 
and the common understanding is 
that several contributions to those effects partially compensate one another, while the difference of opinions 
is to what extent this compensation occurs and what is the physical meaning of such compensation and of 
various contributions themselves. There is an agreement, however, what is the overall sum of 
side jump contributions in the simple case of elastic scattering by impurities, as well as on the sign of 
the side jump contributions compared to skew scattering. 
Our goal here is to calculate the magnitude of the constant $\eta_1$.

Inserting  (\ref{matrix}), (\ref{time}), (\ref{flow}) into Eq.(\ref{sj}), we obtain
\begin{equation}
j_{sj}= \frac{ne^2E_y\alpha}{\hbar}= \pi \sigma\frac{\hbar}{\epsilon_F\tau}\alpha nE_y. 
\label{sjr}
\end{equation}
As was shown in \cite{YLG,Vignale,review,dasSarma,Nozieres,culcer}, this turns out to be a correct expression for the sum of all side-jump-group effects. 
Side jump current of species with certain spin projection is linear in 
  density of these species as opposed to quadratic dependence on electron density of the skew scattering contribution. A signature of the difference between the two is the factor of $\epsilon_F$ in the denominator 
of the expression describing $j_{sj}$ . The linear dependence on density for side jump currents holds even if the momentum relaxation time depends on energy. 
The parameter $\eta_1$ that describes the characteristics of the stripes is given by:
\begin{equation}
\eta_1=\frac{2e\alpha}{\hbar}.
\end{equation} 

The side-jump-group currents can be made dominant over skew scattering contribution in a situation when compensation of impurities is close to 100\% (i.e. almost equal number of donors and acceptors). While in 3D samples \cite{Chazalviel} equal compensation is not possible because no current carrying electrons present in a sample, in 2D case the charge carriers can be supplied by the remote doping, and even full compensation of quantum well donors by quantum well acceptors is feasible. At certain compensation, the skew scattering effect can vanish. If the strenth of the donor and acceptor potentials are the same, this happens at full compensation. As we show in Sec.III, in low mobility samples, the magnitude of spin relaxation rate is minimized, and that results in feasible experimental setting.  Results of calculation of contributions to the spin current are presented in Sec. IV.

\section{\it Spin-electric stripes in the presence of spin relaxation}
 
Some spin relaxation will always be present in the sample
(even if $L_{so} > 2a$). In numerous contributions since the celebrated paper by M.I. Dyakonov and 
V.I. Perel \cite{dyakonovperel}, the effects of asymmetric skew scattering, and, particularly, spin Hall conductivity, have been discussed in terms of spin accumulation near the boundaries of the sample, which is limited by spin relaxation.

The physical picture of electric field domains in the absence of spin relaxation is predicted in Sec. I.
However the question arises whether this effect can survive the presence of spin relaxation.
I will now show that when spin relaxation is sufficiently strong, it is the
spin relaxation that limits the
accumulation of spins closer to the boundaries of the planar sample. However, at electric
fields larger than
the certain field, 
the spin current, properly defined, reaches such a magnitude that the 
spin relaxation, and the associated spin diffusive current are not sufficient to balance the spin current induced y the external field, and limit spin and charge accumulation. Then the Coulomb
interaction takes on the leading role and
results in the transverse electric field and the spin-electric stripes.
I will first demonstrate why spin relaxation can be insufficient to limit the spin accumulation,
so that electric field and the periphery stripes with $100\%
 $ sample edge spin polarization must be a necessary ingredient of the physical picture. I will then derive the properties of domains in the presence of spin relaxation.

\subsection{Constituitive equations in the presence of spin relaxation.}

In the presence of spin relaxation, there is no longer a conservation of the currents caused by 
charge carriers with each spin projection, and instead, the spin density satisfies the continuity 
equation
\begin{equation}
\frac{d}{dx} ((j_x^{+}-j_x^{-})=\frac{2S_z}{\tau_s}.
\label{spindensity}
\end{equation}
A factor of two and the sign in the right hand side of Eq.(\ref{sc}) is due to our definitions of currents in Eq.(\ref{j}) and spin density 
$S_z$ in Eq.(\ref{Sz}), and $\tau_s$ is the microscopic spin relaxation time for the $z-$projection of spin. We assume the steady state and , for simplicity, no external magnetic field. 
The charge current is conserved: 
\begin{equation}
\frac{d}{dx}(j_x^{+}+j_x^{-})=0.
\label{chargecurrent}
\end{equation} 
We will assume first that $j_x^{+}$ and $j_x^{-}$  are described by the same equations (\ref{j},\ref{Drude},\ref{diffusion}, \ref{source}) as in section I.1, and thus consider 
the non-conservation of the spin currrent (\ref{spindensity}) to be the principal effect due to spin relaxation. This approach is along the lines of \cite{dyakonovperel}. We will discuss the other possible roles of spin relaxation on a microscopic level in the end of this subsection. 
In order to pinpoint where the physical picture of spin accumulation gets replaced by the 
physical picture of spin-electric stripes, we will now use the same model for the source spin current 
as in \cite{dyakonovperel}, i.e., assume linear in density spin Hall conductivity and set $\eta_2=0$ in Eq.(\ref{source}).
 
Inserting Eqs. (\ref{j},\ref{Drude},\ref{diffusion},\ref{source}) 
into Eqs. (\ref{spindensity},\ref{chargecurrent}), and assuming the total density $n(x)=const$ and $\eta_2=0$, we have the following equations for the transverse currents: 
\begin{eqnarray}
&\frac{d}{dx}[\mu  S_z  E_x+\frac{\zeta n }{2}\frac{d}{dx}S_z] =\frac{2S_z}{\tau_s}
\label{sc}\\
&\frac{d}{dx}[\mu  n E_x +\frac{\zeta}{4}\frac{d}{dx}S_z^2+\eta_1 S_z E_y]=0,
\label{cc}
\end{eqnarray}
The nonlinear terms in these  equations were neglected in \cite{dyakonovperel}, where no density or coordinate dependence of the 
coefficients in continuity equations were taken into account.  In this work, these factors are important. 

\subsection{Why spin accumulation picture is insufficient and how spin-electric stripes arise}

The usual assumption for solving Eqs.(\ref{sc},\ref{cc}) is that $S_z$ is proportional 
to the electric current, so that the first term in (\ref{sc}) becomes non-linear in electric 
field and therefore can be dropped. Let us follow this assumption for a moment.
Then the same procedure applies to the $S_z^2$ term, and all terms containing the electric field, and we arrive 
to the standard spin diffusion equation 
\begin{equation}
\frac{d^2}{dx^2}S_z= \frac{S_z}{L_s^2},
\end{equation}
where $L_s^2=\zeta n \tau_s$. This spin diffusion equation is solved with the boundary conditions at the edges $x=\pm a$
\begin{equation}
\frac{\zeta n}{2}\frac{d}{dx}S_z +\eta_1 nE_y=0,
\end{equation}
which reflects the absence of flow of the spin current out of  the sample. We underscore that 
in the diffusion equation itself, the gradient of the source term in this model vanishes, $\nabla_x  \eta_1 nE_y=0$, but nevertheless the spin polarization arises due to the spin current affecting boundary conditions.

What these equations describe is a traditional spin accumulation picture. Let us now see whether this 
picture always applies.
The solution of the spin diffusion equation is
\begin{equation}
S_z(x)=\frac{-2\eta_1E_y L_s}{\zeta \cosh{(a/L_s)}}\sinh{\frac{x}{L_s}}.
\label{diffsol}
\end{equation}
This solution for a planar sample, see , e.g., \cite{Zutic}, is an analog of the 3D solution for a cylindrical sample in\cite{dyakonovperel}.
At small $x$ (the center region) there is linear in $x$, $E_y$ and $\eta_1$ spin polarization, due to the spin current. If the spin relaxation length is small, there is linear in electric 
field spin polarization at the edges, as very well known.

However, let us now pose a question what happens if the spin relaxation length and time are sufficiently long, and the magnitude of electric field is sufficiently large, to the extent that
\begin{equation}
|S_z(b, |b|<a)| =\pm n,
\end{equation}
where
\begin{equation}
b= L_s arsh\Big (\frac{\zeta n\cosh{(a/L_s)} }{\eta_1 E_y L_s} \Big ).
\end{equation} 
This amounts to the spin polarization reaching its maximal possible value at some $x$ such that $0<x< a$ (and minimal possible value for the opposite spin projection at $0>-x> -a$).
Then at $b<|x|  \le a$, the solution given by Eq.(\ref{diffsol})
not only results in spin polarization exceeding its maximal possible value, but in contradiction to this value being maximal possible, the solution continues to grow with $x$ approaching the boundaries. For negative polarization, 
at the opposite coordinate, $S_z/ n < -1$, and the spin polarization according to the solution decreases with $x$ approaching the boundary despite having already reached the minimal possible value. Therefore, the solution 
(\ref{diffsol}) fails at sufficiently long spin relaxation times and large electric fields. 

The reason is, in these conditions one cannot neglect the electric field in Eqs.(\ref{cc},\ref{sc}). Spin relaxation of the z-component of spin alone cannot limit the accumulation of spin caused by the spin current, and the electric field takes over this role. We are going to show now that the range of coordinates $-b<x<b$ defines the central stripe and 
the ranges $-a<x<-b$ and $b<x<a$ define the spin-electric periphery stripes.

\subsection{The solution of constituitive equations for spin-electric domains}

To find the elecric field and the spin density, we must solve the system of the non-linear equations 
(\ref{sc},\ref{cc}) supplemented by the boundary conditions of the absence of flow of the spin current and charge current out of the sample at $x=\pm a$:
\begin{eqnarray}
&\mu S_z  E_x+\frac{\zeta n }{2}\frac{d}{dx}S_z+\eta_1nE_y=0\\
&\mu  n E_x +\frac{\zeta}{4}\frac{d}{dx}S_z^2+\eta_1 S_z E_y=0
\label{boundarynl}
\end{eqnarray}
We will assume that spin relaxation is weak. In the model we discuss now, the corresponding 
condition is 
\begin{equation}
\eta_1E_y n \gg \frac{\zeta n^2}{L_s}.
\label{conditiondr}
\end{equation}
The more general condition at $\eta_2\ne 0$ is 
\begin{equation}
\sigma^{\pm}_{\perp}E_y  \gg \frac{D_{\pm}n_{\pm}}{L_s}.
\label{conditiongen}
\end{equation}
The physics of this condition is that the spin current caused by the electric field is much stronger than the diffusive spin current caused by non-uniformity of spin density on the scale of spin relaxation length. We immediately note  that the possibility to reach this regime is 
non-trivial. Indeed, if to assume that the spin current and spin relaxation are governed by the same spin-orbit constant $\alpha$, then $\eta_1\propto \alpha$, $1/L_s\propto \alpha$,
and the strength of spin interaction does not enter this condition. If that were true, too strong electric fields, in which electrons gain energy $eE_y\ell\sim \epsilon_F$ on the scale of the mean free path $\ell$, are required to fulfill (\ref{conditiondr}).  However, we shall see that in 2D gas 
different spin constants govern the spin current and spin relaxation.

We now will solve the system of equations (\ref{sc},\ref{cc}) by perturbation theory using 
the small parameter of Eq. (\ref{conditiondr}). In the zeroth approximation, i.e. 
$1/\tau_s =0$, the stripe solution of subsection I.A. holds. We next find first order corrections 
in $1/\tau_s$ to this solution. In the periphery region, say, around $x=-a$ 
we write down the spin polarization in the periphery stripe in the form
\begin{equation}
S_z^p(x)=n-s(x),
\label{polcor}
\end{equation}
and the electric field in the form
\begin{equation}
E_x(x)=- \frac{\eta_1 E_y}{\mu} + {\tilde e}(x)= E_0 + {\tilde e}(x) ,
\label{ecor}
\end{equation}
where $s(x)$ and ${\tilde e}(x)$ are first order in  $1/\tau_s$, and $E_0$ is the electric field in the stripe at the 
boundary $x=-a$ 
in the absence of spin relaxation.
Inserting Eqs.(\ref{polcor},\ref{ecor}) into Eqs.(\ref{sc},\ref{cc}), and keeping only 
linear in $1/\tau_s$ terms, we have
\begin{eqnarray}
& \nabla_x\Big (-\mu E_0 s(x) + n\mu {\tilde e}(x) -\frac{\zeta n}{2}\nabla_x s(x)\Big )= \frac{2n}{\tau_s}\nonumber \\
&\nabla_x\Big (\mu n {\tilde e}(x) - \frac{\zeta n}{2}\nabla_x s(x) -\eta_1s(x)E_y\Big )=0.
\label{linearized}
\end{eqnarray}
This system must be solved with boundary conditions on spin and charge current for ${\tilde e}(x)$ and $s(x)$, which amounts to the two expressions in round brackets  in (\ref{linearized}) vanishing at $x=-a$. This turns out to be equaivalent to 
\begin{equation}
s(-a)=0. 
\end{equation}
We note that because both the electric field and spin polarization are the odd functions of $x$, it is sufficient to solve the system of equations at $x<0$ in order to obtain the whole solution. At negative 
$x$ for the periphery region, we have
\begin{eqnarray}
&s(x)= \frac{n(x+a)}{\eta_1 E_y \tau_s},\\
& {\tilde e}(x)= \frac{x+a}{\mu\tau_s}+\frac{\zeta n}{2\mu \eta_1E_y\tau_s}.
\end{eqnarray}
The resulting spin polarization and the electric field in the periphery stripe are given by
\begin{eqnarray}
& S_z^p= n\Big (1+\frac{2(x+a)b}{L^2_s}\Big ),
\label{spinp}\\
& E_x(x)=E_0\Big (1+\frac{2(x+a)b}{L^2_s}-\frac{2b^2}{L^2_s}\Big ),
\end{eqnarray}
where $b$ is the $x$-coordinate of the boundary of the central and the periphery stripe in the absence of spin relaxation, given by Eq.(\ref{boundcp}) at $\eta_2=0$.   
 The absolute value $|b|$ is the half-width of the central stripe, and at $x<0$ $b$ is negative.
Compared to the picture with spin relaxation absent, the magnitude of the electric field is slightly smaller
 at $x=-a$, and it decreases further as $x$ arroaches the center region, as 
opposed to being constant. The spin polarization has its maximal possible value at the edge, 
and instead of keeping this value throughout the periphery stripe, decreases towards 
the center region. 

We now find the solution of Eqs. (\ref{sc},\ref{cc}) in the center region, in the linear in 
$1/\tau_s$ approximation. For $\eta_2=0$, we have found $E_x=0$  
in the absence of spin relaxation, as seen from Eqs.(\ref{sum},\ref{fieldcenter}). We assume that  
the electric field can arise in the first order in $1/\tau_s$:
\begin{equation}
E_x(x)= {\tilde e}_c(x).
\label{ecenter}
\end{equation}
For the spin polarization, we assume
\begin{equation}
S_z^c(x)= -\frac{2\eta_1E_y}{\zeta}x + s_c(x),
\label{scenter}
\end{equation} 
where the first term is the spin polarization in the center region in the absence of spin relaxation, and $s_c(x)$ is the first order correction in spin relaxation rate.
Then, inserting Eqs.(\ref{ecenter},\ref{scenter}) into Eqs. (\ref{sc},\ref{cc}), we obtain the following equations for  ${\tilde e}_c(x)$ and  $s_c(x)$:
\begin{eqnarray}
&\nabla_x\Big (-\frac{2\eta_1E_y \mu}{\zeta}x\cdot {\tilde e}_c(x)+
\frac{\zeta n}{2}\nabla_xs_c(x)\Big )= -\frac{4\eta_1E_y}{\zeta \tau_s}x\\
&\mu n {\tilde e}_c(x) - \eta_1E_y \nabla_x s_c(x) =0.
\end{eqnarray}
These equations must be solved with boundary conditions reflecting that the spin polarization and the electric field vanish at $x=0$:
\begin{eqnarray}
 & s_c(0) =0,\\
& {\tilde e}_c(0)=0.
\end{eqnarray}
The solution of these equations for  $s_c(x)$ is :
\begin{equation}
s_c(x)= \frac{1}{\tau_s}\frac{nx}{\eta_1 E_y},
\end{equation}
and therefore the spin polarization in the center region is
\begin{equation}
S_z^c(x)= -\frac{n}{|b|}x\big(1 + \frac{2b^2}{L^2_s}\big).
\label{spinc}
\end{equation}
The solution for the electric field is
\begin{equation}
E_x(x)=  \frac{2E_0 b x}{L^2_s }.
\label{ind}
\end{equation}
Here the product of $E_0$ and $b$ is positive for in the whole range of $x$. We note that this result (as our whole procedure) applies in the range of electric fields compatible with Eq.(\ref{conditiondr}) , so that the limit $E_y\rightarrow 0$ automatically implies $1/\tau_s\rightarrow 0$, resulting in zero electric field in the central stripe. 
Formally the result (\ref{ind}) originates from the non-linear term $S_z\nabla_x S_z$ in the 
equation for the charge current, which includes the product of the two contributions in spin polarization given by Eq.(\ref{spinc}). The non-linear term itself follows from the dependence of diffusion coefficients for electrons with two spin projections on their corresponding densities,
as defined by Eq.(\ref{struct}).
 
Next, from the continuity of spin polarization we find the boundary $b_s$ of the central and the 
periphery stripes:
\begin{equation}
S_z^p(b_s)=S_z^c(b_s).
\end{equation}
Using Eqs.(\ref{spinp},\ref{spinc}) 
we obtain 
\begin{equation}
b_s=b\Big( 1-\frac{2a|b|}{L^2_s}\Big) \Big (1-\frac{4b^2}{L^2_s}\Big)^{-1}.
\label{bound}
\end{equation}
In the limit of strong electric fields $b_s$ and $b$ coinside, because $b\propto 1/E_y$. We note that the second bracket 
formally gives divergence of $b_s$ when the width of the central stripe is equal to spin relaxation length. This is precizely 
the condition when our expansion at small spin relaxation rate becomes invalid, as expected. 

For the set of parameters discussed in Sec VI, the spin polarization, the electric field and the potential energy in the 
presence of spin relaxation are shown in Fig.~\ref{Fig17}, Fig.~\ref{Fig18} and Fig.~\ref{Fig19}, correspondingly.
We observe that the stripe state manifests itself indeed even in the presence of spin relaxation.
There are three distinct regions, the center stripe and the two periphery stripes, with spin polarization reaching maximal possible value (positive or negative) at the edges of 
the sample. In the initial approximation to charge-field distribution, the value of the electric field in the presence of spin relaxation experiences jumps 
from the central stripe to the periphery stripes. Applying considerations of electrostatics 
as in Sec. IB and Appendix A, we find that the derivative of the electric field 
is not infinite (albeit large) as it changes from the center to periphery.  The corresponding field in the leading approximation is given by Eq.(\ref{trans}) with small correstions linear in $1/\tau_s$. Similarly to the case of absence of spin relaxation, the change occurs in a very narrow range of  $x$  close  to $b_s$.
The boundary of the central stripe and the periphery stripes at finite $\tau_s$ is marked by an abrupt change of the $\nabla_x S_z(x)$, as it occurs in the absence of spin relaxation. 

In our consideration of the stripe structure in the presence of spin relaxation, we introduced several 
simplifications. Like in the case of absence of spin relaxation, we assumed that there are two independent channels of conductivity, diffusive currents and spin currents, and the only effect of spin 
relaxation taken into account has been the modification of "continuity" equation for spin current 
via $1/\tau_s$ term. In general, there are contributions to conductivity and diffusion, 
which are off-diagonal in spin, caused by spin relaxation, but the leading terms in diffusion and conductivity are those we took into account. These off-diagonal contributions can change the details of corrections to spin polarization and electric field in the stripes, but not the principal 
features. Furthermore, spin relaxation can result in additional contributions to spin currents (or charge currents of 
spin-polarized carriers). For strong spin relaxation these contributions can be significant \cite{ALGP}, \cite{shytov}. 
We analyse several possible contribution to spin and AHE currents in system under consideration in Sec. IIIB.7. 
The upshot is that tn the case we consider, the condition for existence of stripes 
Eq.(\ref{conditiondr}) requires that the constant responsible for spin current is significantly 
bigger than the constant leading to spin relaxation. It then follows that contributions to spin current due to spin relaxation in the relevant regime will always be small. Therefore our approach is justified. 

In our illustration of domains in the presence of spin relaxation we discussed how the picture of domains caused by spin current defined by $\eta_1$ changes when $1/\tau_s$ is finite. This allowed direct comparison with traditional spin accumulation picture. If we take Sec. IA stripes in the absence of spin relaxation defined by both $\eta_1$ and $\eta_2$ (side-jump-like and skew scattering terms),  then, although the details of change itroduced by finite $1/\tau_s$ will differ, the principal conclusion is still that in spin relaxation gisve only a minor perturbance of the stripe picture.  
The next important task that needs to be accomplished is to prove that it is indeed possible to make the spin relaxation small but the spin current big. 

 \section{Strength of spin relaxation in various electronic systems and feasibility of electric field domains}

\subsection{Spin relaxation and spin current in the 3D bulk systems}

\subsubsection{Spin relaxation in the 3D bulk systems}
In bulk systems, the condition
Eq.(\ref{conditiongen}) requires very strong applied electric fields that are comparable with or larger
than internal confining fields of a heterostructure in semiconductors, or, in metals, give energies acquired by electrons 
in electric field on the scale of mean free path significantly exceeding the Fermi energy, which 
would lead to melting of the sample. As we are going to show, the reason is that the 3D spin-orbit 
interaction Hamiltonian in a metallic film
\begin{equation}
{\cal H}^{so}=\alpha {\mathbf \sigma}\cdot [{\mathbf k}\times \nabla_{\mathbf r} V({\mathbf r})]
\label{bulkso}
\end{equation} 
includes the spin-flipping terms containing $\sigma_x$, $\sigma_y$, which result in the relaxation of the 
$z$-projection of the electron spin, and this relaxation is defined by the same spin-orbit constant $\alpha$ that defines the spin current.  

We first discuss the effect of spin relaxation caused by spin flip in 3D samples. General expressions applied for 3D case 
will also be needed for quasi-two-dimensional samples. The latter are relevant for
for the spin relaxation in quantum wells, where  
it is important to
take into account a random distribution of impurities along the growth direction of the potential well. If spin-dependent scattering 
becomes a factor in spin relaxation in the case when usually more dominant effects in 2D samples are suppressed,  it is this  
effect of the thickness of the quantum well that 
leads to the existence of the spin-flip scattering, in contrast to the ideal 2D case described 
by Eq. (\ref{H}). 
To address these questions rigorously, we consider the evolution 
of the 2x2 spin density matrix $\rho$ in the presence of spin-dependent scattering. In 3D samples, the spin relaxation 
is defined by the Hamiltonian (\ref{bulkso}).
The quantum kinetic (rate) equation describing the evolution of the spin density matrix $\rho_{\mathbf k}$ 
in the presence of spin-flip
scattering can be derived by the standard Keldysh diagrammatic technique, see, e.g., \cite{ILGP,ALGP}, 
and is given by
\begin{eqnarray}
i\hbar \frac{\partial \rho}{\partial t}= \frac{2\pi}{\hbar}\sum_{{\mathbf k}^{\prime}}\delta (E_{\mathbf k}-
E_{{\mathbf k}^{\prime}}) \nonumber \\
\big ( {\cal H}^{so}_{{\mathbf k}{\mathbf k}^{\prime}}\rho_{\mathbf k}^{\prime}
{\cal H}^{so}_{{\mathbf k}^{\prime}{\mathbf k}}-
[ {\cal H}^{so}_{{\mathbf k}{\mathbf k}^{\prime}}{\cal H}^{so}_{{\mathbf k}^{\prime}{\mathbf k}},\rho_{\mathbf k}]\big ), 
\end{eqnarray}
 where $[A,B]=(AB+BA)/2$.
Scattering of electrons with average spin projection $S_i$ (i.e., electrons described by the component 
of the density matrix proportional to $S_i\sigma_i$ with $S_i$ independent of the direction 
${\mathbf o}$ of ${\mathbf k}$) 
in the presence of spin-orbit interactions (\ref{bulkso}) results in the two effects:
(i) spin relaxation of electrons described by spin relaxation time $\tau_{so}$, and (ii) the appearence of the 
correlation between the electron spin and momentum described by the spherical harmonic $(1-3o_i^2)$.
The effect (ii) is described by the component of the density matrix proportional to 
$ S_i\sigma_i(1-3o_i^2)$. 
Here we are interested in the spin relaxation only. The spin relaxation time for i-th projection 
of spin given by 
\begin{equation}
\frac{1}{\tau_{so,k,i}}=\frac{4\pi}{\hbar}\langle\sum_{{\mathbf k}^{\prime}}
\Big( |{\cal H}^{so}_{{\mathbf k},{\mathbf k}^{\prime}}|^2-
|{\cal H}^{so}_{{\mathbf k},{\mathbf k}^{\prime},i}|^2\Big)
\delta(E_{\mathbf k}-E_{{\mathbf k}^{\prime}})\rangle,
\label{spinrelax}
\end{equation}
where ${\cal H}^{so}_{{\mathbf k},{\mathbf k}^{\prime},i}$ is the part of matrix element 
(\ref{bulkso}) proportional to the i-th component of the Pauli operator, and angular brackets denote averaging over 
the directions of ${\mathbf o}$. The spin relaxation time $\tau_{so}$ is isotropic in directions of ${\mathbf o}$, and, in general, is
energy-dependent.  
%while anisotropy in momentum direction is important only for the
%correlation between spin and momentum distribution described by density matrix component defined by %$S_i(1-3o_i^2)$.
For point-like impurity scattering, with scattering amplitude defined by Eq.(\ref{potential}) and modified Eq.(\ref{matrix}) that includes terms proportional to $\sigma_i\cdot [{\mathbf k}\times {\mathbf k}^{\prime}]_i$, 
$i=x,y,z$, we have
\begin{equation}
\frac{1}{\tau_{so}}=\frac{8}{9}\alpha^2 k_F^4 \frac{1}{\tau},
\label{srelax}
\end{equation}
where the scattering time $\tau$ for 3D case is determined by
\begin{equation}
\frac{1}{\tau^{3D}}= \frac{V_0^2 n_i}{\hbar}\frac{m}{\hbar^2}\frac{k_F}{\pi}.
\label{tau3d}
\end{equation}
Then the spin-orbit legnth $L_{so}$
\begin{equation}
L_{so}=\sqrt{D\tau_{so}}=\frac{3}{4}\frac{v_F\tau}{\alpha k_F^2},
\label{solength} 
\end{equation}
where $v_F\tau=l$ is the mean free path. As clear from Eqs.(\ref{srelax},\ref{solength}), the spin relaxation rate for this mechanism grows as a square of the Fermi
energy and is proportional to the rate of electron-impurity collisions. In bulk semiconductors, this mechanism is 
called Elliott-Yafet mechanism, while in metals the analogius effect was discussed by Overhauser as early as 
\cite{awo}. In the case of 3D metals with inversion symmetry, Eqs.(\ref{tau3d}, \ref{solength}) fully define spin relaxation properties. If the symmetry of the crystall is lower, in heterostructures or in confined geometries, the Dyakonov-Perel mechanism of spin relaxation \cite{DP}
in the presence of intrinsic spin-orbit interactions is often the dominant spin relaxation 
mechanism in semiconductors. In terms of electron spin resonance line width, the spin relaxation rate 
given by Eqs.(\ref{srelax},\ref{tau3d}), characterizes the effects of collisional broadening of the line.
The Dyakonov-Perel spin relaxation mechanism is an example of motional narrowing effects \cite{PWA}.  
What is important, the collisional broadening spin relaxation mechanism in a bulk 3D metal cannot be suppressed once electrons contribute to the conductivity and scatter of impurities. In Sec.IIIA.2 we will see that this renders 
spin-electric domains (which are bulk analogs of spin-electric stripes) impossible. However, in quasi 2D samples, the situation is different, and we shall see that 
spin relaxation caused by spin-orbit scattering can be suppressed. We shall also see that intrinsic spin-orbit coupling due to low symmetry and the Dyakonov-Perel spin relaxation can be nearly eliminated. 
In order to discuss the spin-flip scattering in the 2D case, we need to derive the corresponding Hamiltonian.
The corresponding spin relaxation time will be discussed in Sec. IIIB.5.

\subsubsection{Spin current and external electric fields required for appearance of domains in bulk 3D systems}
We now calculate the spin current for bulk systems. 
Using (\ref{bulkso}), for the contribution of 
skew scattering effect 
we obtain
\begin{equation}
\sigma^{3D}_{\perp ,skew}=-\sigma\frac{\alpha k_F^2}{3} \sqrt{\frac{mk_F}{\pi\hbar^2 n_i}\frac{\hbar}{\tau}},
\label{bulkskew}
\end{equation}
where $\sigma$ refers to contribution of one projectrion of spin to conductivity.
The side jump contribution is given by
\begin{equation}
\sigma^{3D}_{\perp ,sj}=\sigma\frac{\alpha k_F^2}{2\pi}\frac{\hbar}{\epsilon_F\tau}.
\end{equation}
We note that for 3D metals, when the carrier density considerably exceeds the concentration of impurities,
the square root in (\ref{bulkskew}) is of order of unity, and at $\epsilon_F\tau/\hbar\gg 1$ 
currents of side jump group are much smaller than skew scattering contribution.

Using Eqs (\ref{conditiongen}, \ref{bulkskew}, \ref{solength}), we calculate the electric field that 
is needed to generate electric field 
domains induced by spin current in the bulk system. In order to discuss feasibility of an experimental 
observation of the effect, we multiply this electric field by the mean free path and the electron charge:
\begin{equation}
eEv_F\tau=\frac{4}{3}\frac{mv_F^2}{\sqrt{\frac{mk_F}{\pi\hbar^2 n_i}\frac{\hbar}{\tau}}}.
\end{equation}
With the denominator in this expression reaching minimum when the square root $\approx 1$, \cite{Born}
it follows that energy acquired by charge carriers in electric field on the scale of the mean free path 
exceeds the 
Fermi energy. These are unrealistic conditions, certainly impossible in metals. Note that if the side jump group 
contributions in some metallic conductors (metallic conductivity implicitly means $\epsilon_F\tau/\hbar >1$) 
exceed skew scattering conductivity, the electric fields needed for domains result in energies 
acquired by electrons on the scale of the mean free path is $\epsilon_F\tau/\hbar$ times bigger than 
the Fermi energy. The current induced by such external fields would melt the sample. Thus, 
the spin relaxation due to 
ordinary bulk 
spin orbit interactions effectively prohibits appearence of the electric domains caused by the spin current.

We observe that in order 
for electric fields to be experimentally feasible, energies acquired by charge carriers on 
the scale of the mean free 
path needs to be parametrically smaller than the Fermi energy. We
now demonstrate that it is possible to find
quasi-2D systems, in which applied electric field needed for observation of this effect is reasonable. 
These systems are characterized by a sizable spin current but the spin relaxation there is eliminated 
altogether, or at least
suppressed by several
orders of magnitude.

\subsection{Constructing quantum wells with no z-spin projection relaxation due to spin-orbit interactions.}

The example of the setting, in which it is feasible to observe electric field domains 
is the n-type semiconductor double heterostructures (quantum wells)
grown along [110] crystallographic direction.
Spin relaxation in these systems has been already experimentally shown to be much
smaller than for structures grown along
[001]\cite{flatte,Ganichev}. Let us now discuss if there is the ultimate limit for spin relaxation rate 
in such setting. 

\subsubsection{\it Intrinsic spin-orbit effects that need to be suppressed}
Spin-orbit interactions in quantum well structures, and semiconductors with low symmetry, apart from spin-orbit scattering given by 
Eqs.(\ref{H},\ref{bulkso}), generally includes intrinsic spin-orbit effects, i.e. terms in the Hamiltonian
\begin{equation} 
{\cal H}_{intr}= \hbar{\mathbf \sigma}\cdot ({\mathbf \Omega}^{(1)}+{\mathbf \Omega}^{(3)}).
\label{intrinsic}
\end{equation}
Here ${\mathbf \Omega}^{(1)}_i=\sum_{j} \beta_{ij} k_j$\cite{rashba1,rashba,altshuler1,dyakonovkachor},  ${\mathbf \Omega}^{(3)}_i=\sum_{jut} 
\gamma_{ijkl} k_jk_u k_t$, where $i,j,u,t$ denote projections on the principal axis of the crystal 
$x\parallel [100], y\parallel [010]$ and $z \parallel [001]$.
In bulk III-V systems, cubic in ${\mathbf k}$ terms are present with $\gamma_{xxyy}=-\gamma{xxzz}=
\gamma_{yyzz}=-\gamma_{yyxx}=\gamma_{zzxx}=-\gamma_{zzyy}$ and all other $\gamma$ components vanishing.  
This is the Dresselhaus term \cite{Dresselhaus}.  
Linear in ${\mathbf k}$ terms can arise in bulk crystals such with wurtzite or tellurium symmetry \cite{rashba1,ILGP}. 
Linear in ${\mathbf k}$ terms in bulk III-V systems arise only in the presence of uniaxial 
strain described by strain tensor $\epsilon_{ij}$ 
\cite{pikus,aronov_lyandageller}:
\begin{eqnarray}
&{\cal H}^{\epsilon}_{intr}= {\mathbf \Omega_{\epsilon}}\cdot  {\mathbf \sigma}= \nonumber\\
& C_4 (\sigma_x k_x(\epsilon_{yy}-\epsilon_{zz})+
\sigma_y k_y(\epsilon_{zz}-\epsilon_{xx})\nonumber\\
& +\sigma_z k_z(\epsilon_{xx}-\epsilon_{yy}))+
 C_2 (\epsilon_{xy}(\sigma_x k_y- \sigma_y k_x) \nonumber\\ 
&+\epsilon_{yz}(\sigma_y k_z- \sigma_z k_y) + \epsilon_{zx}(\sigma_z k_x-\sigma_x k_z))
\label{strain}
\end{eqnarray}
We note that non-zero diagonal components of the strain tensor naturally arise in the context of lattice-mismatched 
heterostructures. 
In the absence of strain, linear in ${\mathbf k}$ terms in III-V systems, such as GaAs, arise in 2D samples. 
In particular, the bulk Dresselhaus terms 
upon quantization along [001] direction to 2D become 2D Dresselhaus terms with $\beta^{D}_{xx}=-\beta^{D}_{yy}$ 
(all other $\beta^D$ are zero). In quantum wells with $x,y$ conducting plane perpendicular to growth 
direction z, subject to asymmetric potential confinement there arises 
Rashba term with $\beta_{xy}=-\beta_{yx}$\cite{rashba}. All these terms potentially result in spin relaxation of the z-component of spin. As we shall see below, in 110 grown III-V crystals, 
the situation is different, and intrinsic spin-orbit interactions leading to such spin relaxation are almost non-existent.

\subsubsection{\it  Intrinsic spin-orbit interactions for 110 orientation of the quantum well}
 
As we have seen above, rather moderate spin relaxation precludes spin-electric domains or stripes due to spin current.
Our task is therefore to find systems with spin-orbit interactions such that the terms responsible for 
spin current are sufficiently strong, but the terms responsible for spin 
relaxation of the z-projection of spin are suppressed or vanish altogether.
Let us consider heterostructures grown in [110] direction. 
In this case, our $z^{\prime}$-axis is along [110] crystallographic direction, $x^{\prime}$-axis is along $[{\bar 1}10]$ 
crystallographic direction, and $y^{\prime}\parallel [001]$. 
In these systems, several spin-orbit effects vanish \cite{vignale1}. In particular, the bulk Dresselhaus cubic 
in momentum term \cite{Dresselhaus} 
results only in the terms proportional to the combination of spin operators 
describing the spin projection along 
the normal to the 2D gas $z^{\prime}$:
\begin{equation}
{\cal H}_{D3}=\frac{\gamma_c}{2\sqrt{2}}\sigma_{z^{\prime}}(k_{x^{\prime}}k^2_{y^{\prime}}-k^3_{x^{\prime}}/2).
\label{cubic}
\end{equation}
The Dresselhaus 2D spin-orbit interaction arising 
from averaging over the quantized motion along the $z^{\prime}$-direction 
of the bulk cubic    
Dresselhaus term\cite{altshuler1,dyakonovkachor} also gives only terms containing spin operator 
projection along $z^{\prime}$: for electrons of the ground level of size quantization, assuming for 
simplicity an infinitely deep rectangular quantum well of width $d$,
\begin{equation}
{\cal H}_{D2}=\frac{\gamma_c\pi^2}{2\sqrt{2}d^2}\sigma_{z^{\prime}}k_{x^{\prime}}
\label{linear}
\end{equation}
 These terms give neither spin relaxation of the $z^{\prime}$-projection of spin, nor the spin current\cite{Vignale}. 
If the quantum wells have
symmetric heterostructure confinement, another source of spin-orbit interactions,
the Rashba term\cite{rashba}
vanishes. Furthermore, if neigboring heterostructure layers have common
atoms at the interface, yet another
source of the intrinsic spin-orbit interactions, the interface asymmetry, does not
arise \cite{interface,flatte}. Moreover, for lattice-matched systems, a strain does
not result in
extra spin-flipping interactions\cite{pikus} as well. Indeed, similar to the case of Dresselhaus 
terms in [110] grown structures, strain-induced terms Eq.(\ref{strain}) in the presence of the relevant strain tensor component 
$\epsilon_{z^{\prime}z^{\prime}}$
contain only a single Pauli matrix $\sigma_{z^{\prime}}$:
\begin{equation}
{\cal H}_{D2\epsilon}=C_4^{\prime}\sigma_{z^{\prime}}k_{x^{\prime}}\epsilon_{z^{\prime}z^{\prime}}.
\label{epsilon110}
\end{equation}
These terms also do not result in spin relaxation, 
and do not induce the spin current.
Thus, the only possible source of
spin-orbit effects in this system is
the impurity-induced or fluctuation-induced spin-orbit interactions. 

\subsubsection{\it Impurity and fluctuation-induced spin-orbit interactions}

These interactions are of
two types. Interactions of the first type, described
by Eq.(\ref{H}), conserve the spin projection perpendicular to the 2D layer. 
However, because of the finite
thickness of the 2D layer,
there is a possibility of a second type of interactions, where the impurity-induced terms lead to
spin relaxation. Impurities are randomly located with respect to centerplane $z=0$ of the 
quantum well, and the potential gradient of the impurity potential at the impurity location is sizable. 
Then, in much the same way as in the 3D conductors, at the first glance the spin-flip scattering is feasible, and 
can lead to spin relaxation.
Furthermore,
it has been suggested by Sherman\cite{sherman}
that random distribution of impurities in doping layers, even if it is
symmetric with
respect to the centerline of the quantum well, results in some minimal
spin-orbit
interaction of a "random Rashba" type that leads to spin relaxation. 

In the Appendices B and C we derive the effective spin-orbit Hamiltonian in systems 
confined to two dimensions in the 
presence of 3D random potential. The dependence on the third dimension turns out to be important. 
 Nevertheless, we demonstrate that
the effect of spin-orbit scattering by impurities or the spin-orbit effects in the presence 
of the long-range random potential on 
electrons in the quantum well can be greatly suppressed compared to similar 
effects in the 3D systems. Moreover, we show that these sources 
of spin relaxation can be eliminated altogether. 

The results of appendices B and C are as follows:
In the presence of 3D potential $V(x,y,z)$ affecting conduction electrons \cite{difference}, 
there are three principal contributions to the spin-orbit interactions. 
\newline $\bullet$ Spin-conserving spin-orbit scattering, i.e. the ideal 2D spin-orbit Eq.(\ref{H}).
\newline $\bullet$ Spin flipping spin-orbit scattering governed by the values of band offsets
\newline$ \bullet$ Spin flipping spin-orbit scattering governed by the difference of semiconductor parameters in the 
quantum well and the barriers.

The division of parameters that control spin-orbint interactions and the related relaxation of the 
$z$-projection of electron spin in quantum wells into two groups is a matter of convenience. 
The first group includes band offsets in the valence 
and the conduction band; the second group includes all the parameters that could differ in the barriers and the quantum well: 
the effective mass, the Kane parameter of the Kane model for III-V semiconductors \cite{Kane} (and potentially parameters of band coupling in other multiband models), and the spin-orbit splitting of the valence band. 

\paragraph{Band offsets and disorder-induced spin-orbit interactions}

In this paragraph we formulate the problem and present the results for the band-offset controlled spin-orbit constants.
Details of this consideration, including construction of the effective 2D Hamiltonian starting from 3D interactions, and the 
role of boundary conditions are discussed in Appendix B.

We derive a specific expression for spin-orbit constants for electrons in symmetric quantum wells that are of primary interest in this work. Our general conclusions apply to arbitrary confinement.  For simplicity we use the notation $x,y,z$ for 
Cartesian coordinates, with $z$ being the growth axis. We keep in mind the symmetry restrictions coming from the 
[110]
growth direction of the quantum well structure.  However, our results can be applied to spin phenomena in quantum structures with an arbitrary crystallographic orientation. 

Electrons in the conduction band of the double heterostructure are affected by confinement and scattering potentials 
\begin{equation}
U(z)={\tilde U}_c(1-\theta(z+d/2)+\theta(z-d/2))+V(x,y,z),
\label{c}
\end{equation}
where $d$ is the width of the well, ${\tilde U}_c$ is the conduction band offset. The valence electrons are subject to the potential (whose relevance is 
discussed in Appendix B)
\begin{equation}
U_v(z)= -E^w_g + {\tilde U}_v(1-\theta(z+d/2)+\theta(z-d/2))+V(x,y,z).
\label{v} 
\end{equation}
Here $E^w_g$ is the band gap in the quantum well, and ${\tilde U}_v$ is the valence band offset. We note that the only requirement is that 
conduction electrons are confined to form a 2D setting, 
while valence electrons for positive offset can experience a barrier potential and no confinement.
This latter setting realizes a so called type II heterostructure\cite{Danan}. 

\begin{figure}[t]
\vspace{-9mm}
\includegraphics[scale=0.35]{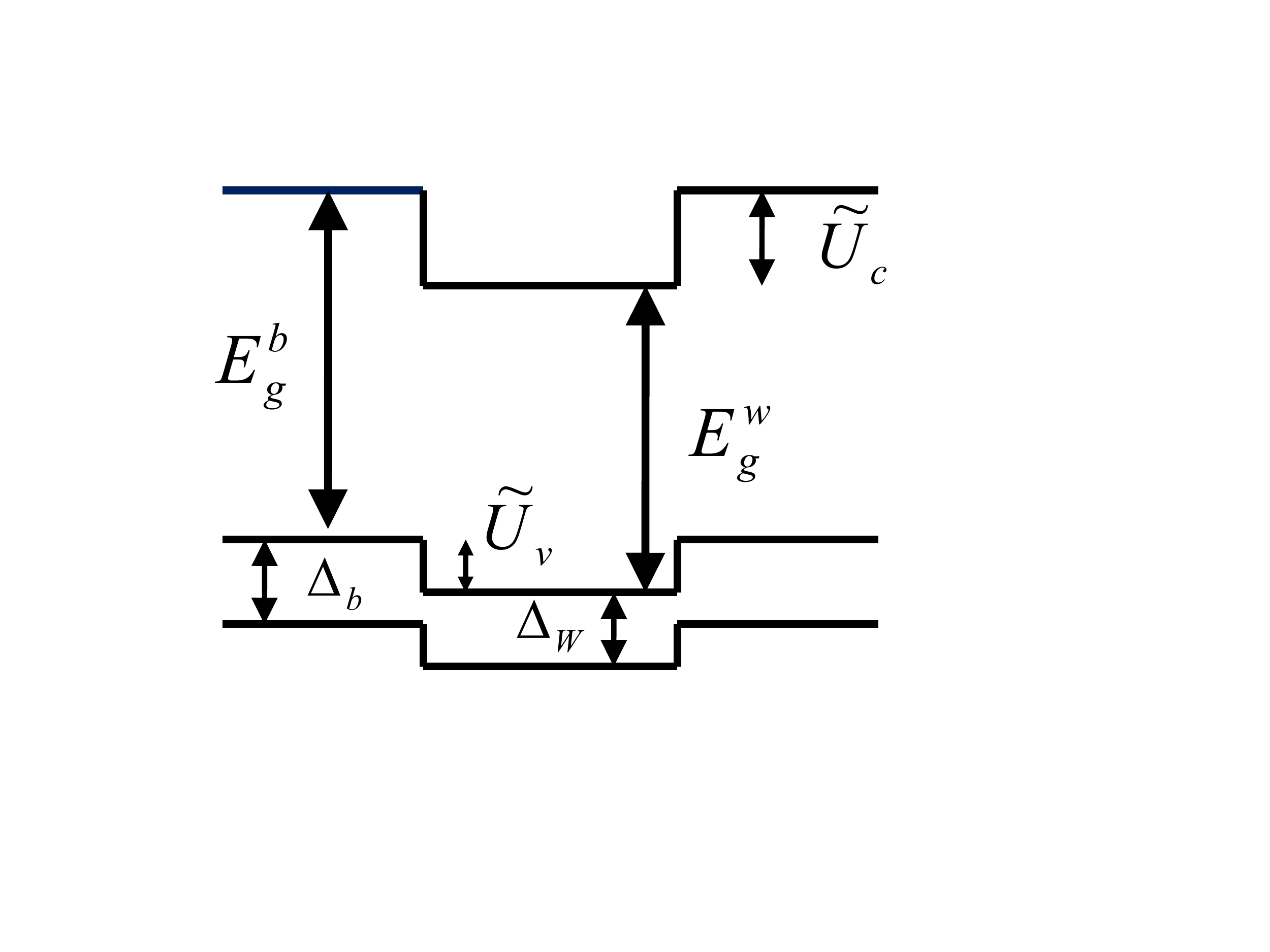}
\vspace{-19mm}
\caption{Potential profile of the heterostructure of the second type, in which electrons are quantized to a quantum well, but holes are not. Different features of the potential in barriers (subscript b) and the quantum well (w) are shown.}
\vspace{-2mm}
\label{structure}
\end{figure}

For direct gap semiconductor III-V structures, the structure of valence band includes two bands:
$\Gamma_8$ band of heavy and light holes,
whose confinement is described by Eq. (\ref{v}), and the split-off $\Gamma_7$ valence band, 
in which the potential acting on valence electrons is described by $U_{v7}=U_v-\Delta(z)$, where the coordinate-dependent 
spin-orbit splitting is given by
\begin{eqnarray}
&\Delta(z)= \Delta_w +\delta(\Delta (z)),\\
&\delta\Delta(z)= (\Delta_b-\Delta_w)(1-\theta(z+d/2)+\theta(z-d/2)), 
\label{7}
\end{eqnarray}
where $\Delta_b$, $\Delta_w$ are energy separations of the split-off band in the barrier and the quantum well, correspondingly. 
The heterostructure potential profile is shown in Fig.~\ref{structure}.

Taking into account the difference of offsets in the condution and $\Gamma_8$ valence band, and therefore the difference of the
bandgaps in the barrier and the quantum well ($E_g^w-E_g^b={\tilde U}_v-{\tilde U}_c$), but neglecting the difference of all other parameters in the well 
and the barriers, we obtain the following effective Hamiltonian 
(including both spin-conserving and spin-flipping  terms)
\begin{eqnarray}
{\cal H}_{so}^{2D}=\alpha
\Big (\sigma_z \cdot [ \big ( \nabla_{{\mathbf r}}V_{scat}({\mathbf r})\big )_{00}\times {\mathbf k}]_z \Big . \nonumber\\
+\Big (1-\frac{U_v}{U_c} \Big) \Big [\big ( \nabla_z V_{scat} \big )_{00}\big[
{\mathbf k}\times {\mathbf \sigma}\big ]_z + \nonumber\\
\Big.\Big.\big ( [{\mathbf \sigma}\times  \nabla_{\perp} V_{scat}]_z k_z \big )_{00}
\Big ]\Big ),
\label{fqwso}
\end{eqnarray}
where $(A)_{00}= 
[\int_{-\infty}^{\infty}dz \Psi^*(z)A(z)\Psi(z)$ is the expectation value of the operator 
${\hat A} ({\mathbf r})$ in the electron ground state in the confining potential $U(z)$. The constant $\alpha$ describes 
generic spin-orbit interactions in the Kane model, and is
given by Eq.(\ref{constso}). With our choice of ordering of operators in Eq.(\ref{fqwso}), $\alpha$ is negative.

The first term 
in Eq.(\ref{fqwso}) has the same symmetry as spin-orbit interaction (\ref{H}) in an ideal 2D system.
The two other terms provide channels for relaxation of the $z-$component of spin. Assuming wavefunctions $\Psi(z)$ to be real,  
the spin flipping terms take a simple form
\begin{equation}
{\cal H}_{so}^{2Dsf}=\alpha\kappa [(F(x,y)[{\mathbf p}\times {\mathbf \sigma}]_z +
[{\mathbf p}\times {\mathbf \sigma}]_zF(x,y)]/2,
\label{hermit}
\end{equation}
where the parameter $\kappa=1-{\tilde U}_v/{\tilde U}_c$, and the scattering potential gradient averaged over the ground state of the quantum well is $F(x,y)=\big ( \nabla_z V^{scat} \big )_{00}$. The symmetrization here gives a Hermitian Hamiltonian, arising 
from an accurate averaging over the growth direction (Appendix B). The term coming from the differential operator 
${\mathbf k}_{\perp}=-i\partial/\partial {\mathbf r}_{\perp}$ acting on $F(x,y)$ is especially important 
in the case of scattering potential $U(x,y,z)$ due to quantum well impurities. 

We now make several observations about these spin-orbit interactions. The spin-flip 
scattering terms in Eq.(\ref{fqwso}) vanish if potential offsets 
in the conduction band and in the valence band are equal. We underscore that in majority of heterostructures offsets differ significantly and have  
opposite signs.  Then $\kappa$ enhances the spin-flip scattering rate, as opposed to suppressing it at close offsets.

We now discuss the origin of the dependence of the coefficient $\kappa$ in spin-flip 
spin-orbit scattering on potential offsets. Part of this spin-orbit term comes from averaging of $\partial U_v/\partial z$ over the ground state in the quantum well. 
If this averaging would only include averaging of the  $\partial V_{scat}/\partial z$ over the wavefunctions calculated in the absence of scattering, then $\kappa=1$.  
However, scattering potential disturbes the electron wavefunctions in the quantum well. 
Then it is necessary to include the contribution of the gradient of symmetric confinement potential 
averaged over scattering-disturbed wavefunctions. That changes $\kappa$. Because the gradient of symmetric potential in rectangular quantum well is defined by offset 
potentials, $\kappa$ depends on offsets. Furthermore, a specific combination of offsets in $\kappa$ comes from the structure of perturbation expansion in the Kane model, 
in which electron spin-orbit effects are related to admixture of the valence band potential. Moreover, because $V_{scat}$ is essentially a part of the overall confining potential, 
vanishing $\kappa$ for ${\tilde U}_v={\tilde U}_c$ is related to the
Ehrenfest theorem. Indeed, according this theorem, the average gradient of the confining potential vanishes in any confined state. For equal offsets, the spin-orbit 
term under consideration is exactly such an average. We note  that because Hamiltonian describing spin-orbit interaction must be Hermitian, its part containing 
 ${\hat k}_z\partial U_v/\partial {\mathbf r}_{\perp}$ is also proportional to $\kappa$. 

In discussion of the 2D spin-related effects, such as weak antilocalization, or 2D spin relaxation 
in scientific literature, the term containing $-i\partial F(x,y)/\partial {\mathbf r}_{\perp}$ 
has not played any role. This has been a consequence of a common neglect of the dependence of scattering potential 
on the coordinate normal to the 2D gas, giving a vanishing diagonal matrix element of $p_z$ in the ground state 
in Eq.(\ref{fqwso}). Also, it has been usually asumed that 
the Dyakonov-Perel spin relaxation due to the Dresselhaus or Rashba term is dominant. 
However, when intrinsic spin-orbit interactions are significantly suppressed like in the growth axis [110] configuration, the scattering terms are important. 

\paragraph{Spin-orbit interactions due to a difference of parameters in the quantum well and barriers}

We now discuss the contributions to spin-orbit interactions associated with differences in 
materials parameters in the barriers and the quantum well.
As follows from Appendix C, the spin-orbit interaction Eq.(\ref{fqwso}) arising in the model of constant 
effective electron mass in the double heterostructure has to be modified if we take into account 
that simultaneously with potential offsets, the effect of variation of the value of materials parameters 
takes place across the structure. In particular, there are different masses 
in the barrier, $m_B$, and in the quantum well, $m_W$. 
For different masses in the barriers and the well, one obvious 
modification of (\ref{fqwso}) is simply to take 
into account the effect of mass difference on the wavefunction of the confined state 
$\Psi(z)$. However, the principal result is that the additional contribution to spin-orbit 
interaction due to mass difference arises, which is 
entirely independent from Eq.(\ref{fqwso}). This contribution does not contain the parameter 
$\kappa$ that vanishes at equal offsets, but instead contains the factor $m_B-m_W$:
\begin{eqnarray}
& \delta{\cal H}_{so}^{2D}= \alpha \hbar^2(m_B-m_W) \frac{{\tilde U_v}}{\tilde U_c} \nonumber\\
& \Big (\frac{1}{m}\nabla_z \Psi_{xy}^{*}(d/2)[{\mathbf k}\times 
{\mathbf \sigma}]_z \frac{1}{m}\nabla_z \Psi_{xy} (d/2)- \Big.\nonumber\\
& \Big.\frac{1}{m}\nabla_z \Psi_{xy}^{*}( -d/2)[ {\mathbf k}\times 
{\mathbf \sigma}]_z\frac{1}{m}\nabla_z \Psi_{xy} (-d/2)\Big ),
\label{diffmass}
\end{eqnarray}
where fluxes $\frac{1}{m}\nabla_z \Psi_{xy}(z)$ are conserved across single heterojunctions.
 The expression in brackets in Eq.(\ref{diffmass}) is nonzero because the scattering potential 
$V(x,y,z)$ or asymmetric potential $U_{asym}(z)$ admixes odd and even 
states of the Hamiltonian (\ref{noimp}), and makes  $\Psi_{xy}(z)$ a linear combination of these 
odd and even states. We note that that the wavefunction $\Psi_{xy}(z)$ has the dimensionality $1/\sqrt{d}$, 
so that  $\delta{\cal H}_{so}^{2D}$  has the dimensionality of energy.

In Appendices B and C we have also considered the variation of the Kane model 
parameter $P$  and spin-orbit splitting $\Delta$ between the $\Gamma_8$ band of heavy and light holes and 
the split-off $\Gamma_7$ across 
the double heterojunction, and derived the corresponding contributions to spin-orbit interactions caused by 
difference of these parameters in the barriers and the quantum well. 
These contributions to spin-orbit interaction that arise due to 3D asymmetric potential, 
and scattering effects are given by Eq.(\ref{dif}).

\paragraph{Remarks on Rashba interaction due to asymmetry of quantum well confinement}

We now discuss the consequences of our consideration for Rashba effect due to asymmetric potential confinement or applied bias. 
Although in order to observe spin-electric stripes one has to
get rid of asymmetry of confining potential and design the doping layers symmetrically with respect to the quantum well, so that intrinsic spin-orbit interactions of the Rashba origin are removed, our results are useful for consideration of other phenomena due to spin-orbit 
interactions.
When only $U_{asym}(z)$ is present, in the absence of random impurity potential, the Eq.(\ref{diffmass}) simplifies, so that different electron masses in the barriers and the 
quantum well result in an additional term of the Rashba symmetry
\begin{eqnarray}
& \delta{\cal H}_{so}^{2D}= \alpha \hbar^2 (m_B-m_W) \frac{{\tilde U_v}}{\tilde U_c}\nonumber\\
& \Big (\frac{1}{m}\nabla_z \Psi^{*}(L/2) \frac{1}{m}\nabla_z \Psi (L/2)- \Big .\nonumber\\
&\Big.\frac{1}{m}\nabla_z \Psi^{*}( -L/2) \frac{1}{m}\nabla_z \Psi (-L/2)\Big )  [{\mathbf k}\times 
{\mathbf \sigma}]_z,
\label{diffmassRashba}
\end{eqnarray} 
where the ground state wavefunction of an electron in asymmetric potential $\Psi(z)$  is independent of $x$ and $y$ coordinates. Similarly, for other contributions to spin-flipping spin orbit terms discussed in Appendix C, such as terms due to different spin-orbit splittings of the valence band  
or different Kane parameters in the barriers and the well,  for 1D asymmetric potential along the growth axis 
the Eq.(\ref{dif}) simplifies
and becomes a term of the Rashba symmetry given 
by Eq.(\ref{diffRashba}).
Indeed, all of our Eqs. (\ref{fqwso}, \ref{hermit}, \ref{diffmass}, \ref{dif}) describe not only 
the effect of random 
potential $V(x,y,z)$ on electrons confined 
to a symmetric heterostructure by the 
potential $U(z)$, but also the case when there is weak 
asymmetry of confining potential or externally applied voltage, leading to Rashba terms. 
The Rashba term arising from Eq.(\ref{hermit}) was discussed in\cite{Winkler}. The additional terms (\ref{diffmassRashba}), (\ref{diffRashba}) found here in the 
presence of asymmetric potential   
lead to a significant contribution to the Rashba 
interaction.

We note that in a recent work by Chalaev and Vignale \cite{Chalaev}, Rashba spin-orbit interactions 
arising from Coulomb interactions have  
been discussed. The authors assumed that confinement 
potentials acting on conduction electrons and valence electrons are 
proportional to each other, and deduced that Coulomb interactions is 
the only source of the Rashba term. While in general this conclusion is in 
line with our result that scattering of electrons by impurities 
or by potential fluctuations can be the dominant source of spin-flipping terms, 
we would like to offer a few clarifications to \cite{Chalaev}. First, it was assumed that confinement 
potentials for valence and conduction electrons are proportional to 
each other (AlGaAs/GaAs heterostructure was considered as an example). This is 
certainly true for the offsets part of potentials in double heterostructure quantum wells. 
However, for the asymmetric potential, whether it is related to the
external gate voltage, or to some smooth variation of the bandgap, this is generally not the case. 
Therefore, sources of asymmetry of the overall potential acting on valence electrons,
be it an arbitrary position of an impurity with respect to the centerplane $z=0$ of 
the quantum well, external voltage, asymmetric confinement, asymmetry 
in the location of delta-doping 
layers, or fluctuations of long-range potential for symmetric doping, will all result in the spin-flip terms. 
These effects will still all vanish once the difference of materials 
parameters in the barrier and the quantum well is ignored. 
Incidentally we note that consideration \cite{Chalaev} did not include the dependence of the 
Coulomb potential on the in-plane 
coordinates, which might be of interest in certain effects due to spin-orbit interactions, such as spin 
dephasing in quantum dots\cite{Brouwer}.

\paragraph{Feasibility of spin-orbit interactions with a forbidden flip of the $z$-projection of spin}

 It is important to note, that while simple Rashba symmetry terms can be removed by 
making the confinement potential symmetric, the terms caused by the spin-orbit scattering due to impurities and monolayer 
fluctuations derived here are always present when the bandgap or other materials parameters change 
across the heterostructure. 

\begin{figure}[t]
\vspace{-15mm}
\includegraphics[scale=0.35]{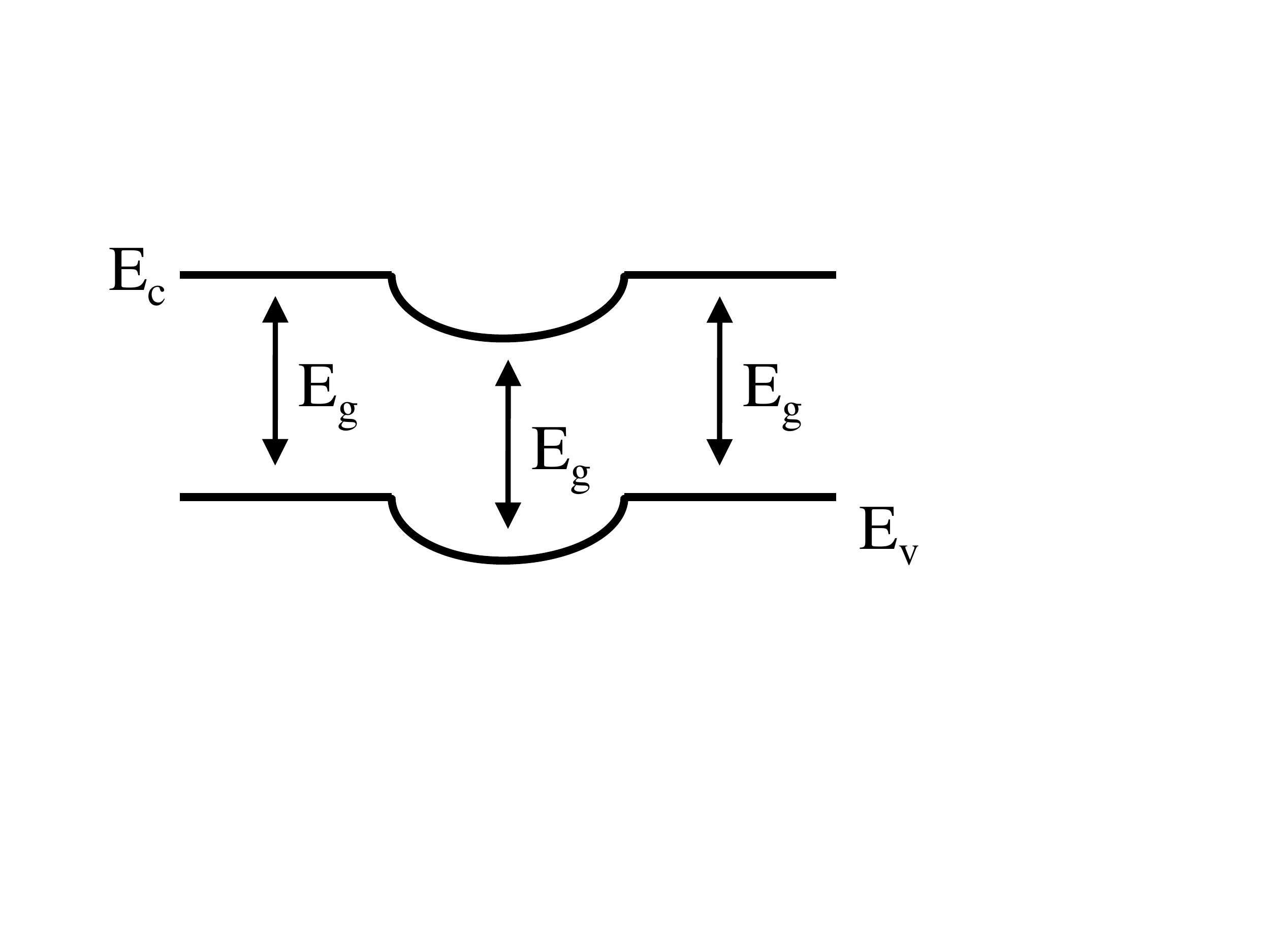}
\vspace{-28mm}
\caption{Ideal quantum structure confining conduction electrons but not holes with constant bandap and identical parameters in the 
well and barriers, in which the effective Hamiltonian of spin-orbit interactions is given by Eq.(\ref{H})}
\vspace{-3mm}
\label{gradual}
\end{figure}

These contributions can be significant especially
when materials in the barrier and the well differ substantially 
in their $E_g$, $\Delta$ and $P$. Thus, it is clear that for 
suppressing the spin relaxation one needs 
to choose a structure, in which the parameters in the well and the barrier(s), 
particularly $E_g$, are somewhat close. 
This latter condition is in line with the condition that $\kappa$ almost vanishes.
Indeed, if one engineers a heterostructure in which $E_g$ is constant, the spin relaxation
 would occur because of a minor possible difference of the Kane parameters in 
the barriers and the quantum well, while the spin-flipping terms in Eq.(\ref{fqwso}) would vanish precisely. 
Such confinement of electrons (in the absence of hole confinement) can  
be achieved in particular in systems with no abrupt hetero-boundaries but rather in situations when the confining potential for electrons, and a gradually varying potential in the valence band that repells holes are identical.   
In this case, a smooth variation of the band edges identical in the 
valence and the conduction bands, as shown in Fig.\ref{gradual},
leads only to existence of spin-orbit terms (\ref{H}) but not to spin-flipping terms\cite{interlevel}. This is a particular example of realizing symmetry that gives spin current but no z-projection spin flips.

 \subsubsection{\it Spin-orbit interactions due to large-scale potential fluctuations.}

Two types of disorder effects can contribute to spin-orbit interactions (\ref{fqwso}).
A random potential $V(x,y,z)$ comes from not only a short range disorder and 
charged impurities in the quantum well, but also from
a long-range disorder, particularly large-scale fluctuations 
of impurity density resulting in a smooth random potential. We now consider spin-orbit effects 
associated with the large-scale fluctuations. 

The large 
scale fluctuations were discussed in \cite{Melnikov} for 3D spin-orbit effects and in \cite{sherman} in the 
context of minimal spin-orbit interactions in 2D gas 
confined by a symmetric potential.  Our Eq.(\ref{fqwso}), adds the following important features to this spin-orbit
effect
(i) the factor $\kappa=1-U_v/U_c$ appears 
as a consequence of the rigorous procedure for averaging over the ground state in the present work;
(ii) the third term of (\ref{fqwso}) which makes the Hamiltonian Hermitian is missed in \cite{sherman}. 
 An accurate averaging over the $z-$direction is important because we need here a precize 
cap on spin-orbit interaction strength and spin-relaxation. We show that such averaging gives 
additional suppression of spin-orbit effects due to long-range potential fluctuations.

We consider the spin-orbit interactions in the quantum well located in the range of coordinates 
$ -d/2 \leq z \leq d/2$ caused by impurity Coulomb potentials of 
the two narrow ($\delta$) doping layers of width $w$ located 
in the range of coordinates $ -l-w/2 \leq z \leq -l+w/2$ and $ -l-w/2 \leq z \leq -l+w/2$. The 3D 
ionised impurity density in the layers is $n_i({\mathbf R})$, $i=1,2$. Average over the realizations of impurity 
potential, denoted by angular brackets, gives $ \langle n_1({\mathbf R}) \rangle =  \langle n_2({\mathbf R}) \rangle=  
\langle n \rangle$,
 so that the 2D density $n_{2D}$ 
in the quantum well induced by the doping layers is $n_{2D}= 2w \langle n({\mathbf R}) \rangle$. We assume 
the correlation function of the impurity densities in the doping layers to be of the "white noise" type:
\begin{eqnarray}
& \langle (n_i({\mathbf R})-\langle n\rangle) (n_i({\mathbf R}^{\prime})-\langle n\rangle) = 
\langle n\rangle \delta ({\mathbf R}-{\mathbf R}^{\prime} ) \nonumber \\
& \langle (n_2({\mathbf R})-\langle n\rangle) (n_1({\mathbf R}^{\prime})-\langle n\rangle) = 0
\end{eqnarray} 

The $z-$component of the electric field due to impurity density fluctuations acting on electron on the first level of size quantization (with coordinate ${\mathbf r}=(x,y,z)$)
 is given by
\begin{eqnarray} 
\delta E_z(x,y) = \sum_i \int dz \int d^3 R\Psi^2(z) \nonumber\\
 \Big ({\mathbf \nabla}_z \frac{e^2}{\epsilon |{\mathbf r}-{\mathbf R}|} \Big )
(n_i({\mathbf R})-\langle n\rangle).
\label{rfield}
\end{eqnarray}
This fluctuation-induced electric field results in a random Rashba term defined by Eq.(\ref{hermit}) with
$F(x,y)=\delta E_z(x,y)$. 
The average field $E_z(x,y)$ itself vanishes because of the symmetrically located doping layers.

The correlation function of the fluctuating electric fields $\langle \delta E(x,y)\delta E(0,0)\rangle$  
is expressed in terms of average square fluctuation electric field $\langle \delta E(0,0)^2\rangle$ 
and the correlator C(x,y)\cite{Shklovskii} :
\begin{equation}
 \langle \delta E_z(x,y)\delta E_z(0,0)\rangle= \langle \delta E_z(0,0)^2\rangle C(x,y).
\label{correlator}
\end{equation}
 Assuming here for simplicity the quantum well with infinite potential confinement with the wavefunction 
$\Psi(z)= \sqrt{\frac{2}{d}}\cos{\frac{\pi z}{d}}$, 
for $\langle \delta E_z(0,0)^2\rangle$ 
we obtain 
\begin{eqnarray}
\langle \delta E_z(0,0)^2\rangle= \frac{\pi n_{2D}e^4\beta^2}{\epsilon^2} \times \nonumber\\ 
(1-2l\beta[ci(2l\beta)\sin{(2l\beta)}-si(2l\beta)\cos{(2l\beta)},
\end{eqnarray}
where $\beta= 2\pi/d$. At large $4\pi l/d$, i.e., for large setbacks, this expression reduces to
\begin{equation}
\langle \delta E_z(0,0)^2\rangle= \frac{\pi n_{2D}e^4}{2l^2\epsilon^2}.
\label{largesetback}
\end{equation}
Thus, a precise averaging procedure over growth direction reduces $\langle \delta E(0,0)^2\rangle$ compared 
to \cite{sherman}. For the correlation function $C(x,y)$, 
at setbacks large compared to the width of the quantum well,
we obtain
\begin{equation}
C(x,y)=\frac{1}{\left ( 1 + \left ( \frac{\rho}{2l}\right )^2\right )^{3/2} },
\label{evalcor}
\end{equation}
where $\rho= \sqrt{x^2 +y^2}$. 

If the average Rashba interactions are suppressed by symmetric configuration of the two doping layers that are sufficiently close to the quantum well,
non-uniform fluctuation-induced Rashba interactions may become a dominant spin-orbit coupling in high mobility materials, with the possibility to realize experimental settings suggested in 
\cite{Tserkovnyak}. Here we are interested in suppressing these interactions, which is possible via large setbacks of doping layers. 
  
\subsubsection{Spin-orbit interactions in 2D systems. Suppression and enhancement of 
spin relaxation due to spin-orbit interactions.}

Once several intrinsic spin-orbit interactions in the structure are eliminated, the spin-orbit scattering in a quantum well with the scattering rate given by Eq.(\ref{spinrelax}) becomes an important z- projection of spin relaxation mechanism. For spin-orbit interactions defined by Eq.(\ref{fqwso}) , the matrix element of spin-flip scattering due to short-range interactions with impurities (\ref{potential}) in the ground state 
in the quantum well is given by
\begin{eqnarray}
&\Big({\cal H}^{2D}_{so}\Big)^{\sigma,\sigma^{\prime}}_{{\mathbf k}, {\mathbf k}^{\prime}}=
\sum_{i}\alpha\Big (1-\frac{{\tilde U}_v}{{\tilde U}_c} \Big)\frac{V_0}{2d}\sin{\frac{2\pi Z_i}{d}}\times\nonumber\\
&\Big( (k_y+k^{\prime}_y)\sigma^{\sigma,\sigma^{\prime}}_x-(k_x+k^{\prime}_x)\sigma^{\sigma,\sigma^{\prime}}_y\Big) e^{iR^{\perp}_i
\cdot ( {\mathbf k}-{\mathbf k}^{\prime})}.
\label{sfa}
\end{eqnarray}
Here we included for brevity only the spin-orbit interaction related to offsets, defined by $\kappa=1-\frac{{\tilde U}_v}{{\tilde U}_c}$.
Spin-orbit terms considered in Appendix C associated with a difference of parameters in the quantum well and barriers result in a contribution to the total spin-orbit constant that is added to $\alpha\kappa$.

We note that it is accurate inclusion of both spin-fipping terms in Eqs.(\ref{fqwso},\ref{hermit}) that makes this scattering amplitude Hermitian, 
\begin{equation}
\Big({\cal H}^{2D}_{so}\Big)^{\sigma,\sigma^{\prime}}_{{\mathbf k}, {\mathbf k}^{\prime}}=
\Big(\Big({\cal H}^{2D}_{so}\Big)^{\sigma^{\prime}\sigma}_{{\mathbf k}^{\prime}, {\mathbf k}}\Big)^*.
\end{equation}
Inserting the matrix element (\ref{sfa}) into Eq.(\ref{spinrelax}), and averaging over impurity locations across the quantum well $Z-i$, we obtain the spin relaxation time 
\begin{equation}
\frac{1}{\tau^z_{so}}=\frac{\alpha^2\kappa^2\pi k_F^2}{2d^2}\frac{1}{\tau}.
\label{2dsr}
\end{equation}
The corresponding spin-orbit length is given by $L_{so}=(dk_F) (p_F\tau/m)/\alpha p_F^2 \kappa$.  

The presence of the factor $\kappa$, or factors describing small 
variation of semiconductor 
parameters from barriers to the quantum well provides an important method for technological control of 
the strength of scattering-related 
spin-orbit effects,  
average Rashba spin-orbit interactions, and fluctuation-induced spin-orbit interactions.

Let us now discuss whether structures with strongly suppressed spin relaxation readily exist.
Indeed, there is a system which is very close to the ideal case. 
In $In_{0.52}Al_{0.48}As/InP/In_{0.52}Al_{0.48}As$ double heterostructure, 
for which  we obtain $\kappa\simeq 0.33$ using known band offsets and band-gaps in the barriers and the quantum well\cite{ruan,IP}. Such value of $\kappa$ decreases 
$z-$projection spin relaxation rate associated with (\ref{fqwso}) order of 
magnitude (assuming $k_F\sim 1/d$) compared to the spin relaxation of transverse $(x,y)$ spin components.
Furhermore, the effective electron mass in InP differs from that in $In_{0.52}Al_{0.48}As$ 
by just 5\%, which makes 
the contribution of Eq.(\ref{diffmass}) to spin-orbit interactions very small. 
Other heterostruture systems can be hopefully engineered 
with even smaller $\kappa$. We note, however, that in several other III-V compounds, $z-$projection spin relaxation in 2D case can be stronger than the relaxation of the transverse components of spin, due to 
the parameter $\kappa$, or because of a sizable difference between the quantum well and 
the barrier parameters. In particular, 
the  $z-projection$ spin relaxation rate associated with terms 
(\ref{sfa}) is the order of magnitude stronger than that for $x-$ and $y-$components in 
$Ga_{0.47}In_{0.55}As$ quantum well. 

We note that in GaAs/AlGaAs system, where a considerable 
decrease in the spin relaxation rate has been observed recently \cite{awschalom2009} 
in the case of the Dresselhaus and average Rashba terms of equal magnitudes, the random Rashba contribution 
and spin-orbit scattering induced relaxation of the $z-$projection of spin cannot be completely 
eliminated using the mechanism of reduction of $\kappa$ discussed here. 
This is because the experimental setting\cite{awschalom2009} 
implies a sizable Rashba term equal to the Dresselhaus term,
and thus nonvanishing 
$\kappa$. However, as follows from Eq.(\ref{largesetback}), random Rashba term due to fluctuations can always be suppressed by increasing a
setback (the distance between the quantum well and each of the two doping 
layers equally separated from it.) 

\subsubsection{\it  The Dyakonov-Perel spin relaxation for non-uniform intrinsic spin-orbit interactions}

The spin-dependent terms that contain $\sigma_x$ 
and $\sigma_y$ result in spin relaxation of the z-component 
of the spin. If such terms are independent of the electron coordinate or their coordinate dependence is smooth, 
the probability of a spin filp in a single scattering event is small, and 
the effective mechanism of the spin relaxation takes its origin in the Dyakonov-Perel 
spin diffusion \cite{DP}. Terms in (\ref{intrinsic}), (\ref{strain}) constitute an effective 
momentum-dependent field that makes electron spin precess. In the absence of scattering, this precession gives two spin-orbit "chiral" quantum states defined by on the projection of the electron spin on the effective field. Ordinary spin-independent scattering due to
random impurity potential  $U({\mathbf r})$ leads to a change of the electron wave vector, and thus randomly 
changes the direction of the effective momentum-dependent magnetic field.  If the angle of electron spin precession between two consecutive scattering events is small, the resultant effect is a random spin walk, diffusion,
leading to spin relaxation. For the coordinate-independent effective field, this phenomenon is very well discussed in the literature from different vantage points. In particular, it can be described in terms of a phase breaking of the electron wavefunction due to effective spin-dependent vector potential, which makes transparent the role of intrinsic spin-orbit interactions in weak localization and related phenomena \cite{iordanski1994,ylg1998,marcus2003}. 

In the context of spin-electric stripes, when several spin-orbit effects are suppressed, the coordinate-dependent intrinsic spin-orbit interactions can play an important role.
In order to compare the strength of various contributions to spin relaxation, we present an account of 
Dyakonov-Perel spin relaxation taking this coordinate dependence into account. 

In the presence of intrisic spin-orbit interactions, and spin-independent impurity scattering, the electron spin relaxation 
is described by the quantum kinetic equation for the spin density matrix
\begin{eqnarray}
i\hbar \frac{\partial \rho}{\partial t}+\frac{i}{\hbar}
\left\{ {\cal H}_{intr}({\mathbf r})+{\cal H}^{\epsilon}_{intr}({\mathbf r}),\rho_{{\mathbf k}}\right\}= \nonumber \\
\frac{2\pi}{\hbar}\sum_{{\mathbf k}^{\prime}}\delta (\epsilon_{\mathbf k}-\epsilon_{{\mathbf k}^{\prime}})
|U_{{\mathbf k}{\mathbf k}^{\prime}}|^2(\rho_{{\mathbf k}}-\rho_{{\mathbf k}^{\prime}}).
\label{DPrho}
\end{eqnarray}

In order to consider spin relaxation caused by residual intrinsic spin-orbit interactions due to long-range fluctuations of impurity doping, we now discuss the Dyakonov-Perel spin relaxation due to the
position-dependent Rashba interaction. We describe this interaction
in terms of the effective spin-orbit vector potential ${\mathbf A}_{so}({\mathbf r}_{\perp})$ :
\begin{equation}
{\cal H}_{intr}= -\frac{{\mathbf p}{\mathbf A}_{so}({\mathbf r}_{\perp})+{\mathbf A}_{so}({\mathbf r}_{\perp}){\mathbf p}}{2m},
\label{Aso}
\end{equation}
where for brevity we dropped a subscript $c$ in our notation for the effective mass $m$, ${\mathbf r}_{\perp}$ is the in-plane coordinate of the quantum well, 
${\mathbf A}_{so}({\mathbf r}_{\perp})= \alpha_2m \delta E_z(x,y) (\sigma_y, \sigma_x, 0)$, $\alpha_2=\alpha\kappa +\delta\alpha$, where $\delta\alpha$ is 
the spin-orbit constant defined by differences of parameters of materials in the quantum well and barriers, $E_z(x,y)$ is described by 
the Eqs.(\ref{rfield},\ref{correlator},\ref{largesetback},\ref{evalcor}). We note that the coordinate- and spin-dependent vector potential also describes the spin relaxation in mesoscopic ferromagnets with domain walls and 
other magnetic textures\cite{LGAG}.

In the relaxation time approximation for a collision integral due to impurity momentum scattering, 
the quantum kinetic equation (\ref{DPrho}) describing electron spin relaxation and spin diffusion can be written 
as
\begin{equation} 
\frac{\partial \rho}{\partial t}+\frac{{\mathbf p}}{m}\cdot\nabla_{\mathbf r}\rho - \frac{{\mathbf p}}{m}
[{\mathbf A}_{so}({\mathbf r}), \rho ({\mathbf r})]=- \frac{\rho -\rho_0}{\tau},
\label{srtau}
\end{equation} 
where square brackets give the commutator of operators that depend on Pauli matrices, and $\rho_0$ is the 
momentum-independent component of the density matrix. Here we drop the index ${\perp}$ for coordinates. 

As was shown in \cite{Glazov}, the electron-electron scattering,
which does not contribute to relaxation of momenta in semiconductors (that happens only in metals due to umklapp 
processes\cite{abrikosov}), contributes to the effective relaxation time that defines spin relaxation of electrons via the
Dyakonov-Perel mechanism.  The effect of electron-electron scattering also affects the electron spin relaxation for non-uniform intrinsic spin-orbit interactions. Therefore $\tau$ in Eq.(\ref{srtau}) includes a contribution from the electron-electron scattering defined by $\tau^{-1}=\tau_{imp}^{-1}+\tau_{ee}^{-1}$. For degenerate semiconductors at low temperatures $\tau_{ee}^{-1}= c_2 \epsilon_F/\hbar (T/ \epsilon_F)^2$, with $c_2=3.4$ \cite{Glazov}.
Defining Fourier-components $\rho({\mathbf r}) = \int d{\mathbf q}\rho({\mathbf q})\exp{i{\mathbf q}\cdot{\mathbf r}}$,
${\mathbf A}_{so}({\mathbf r}) = \int d{\mathbf q}{\mathbf A}_{so} ({\mathbf q})\exp{i{\mathbf q}\cdot{\mathbf r}}$,
we obtain
\begin{eqnarray}
\frac{\partial \rho({\mathbf q})}{\partial t}+\frac{1}{m}{\mathbf p}\cdot{\mathbf q}\rho({\mathbf q}) - \nonumber\\
 \frac{{\mathbf p}}{m}
\int d {\mathbf q}_1[{\mathbf A}_{so}({\mathbf q}-{\mathbf q}_1), 
\rho ({\mathbf q}_1)]= -\frac{\rho({\mathbf q}) -\rho({\mathbf q})_{0}}{\tau}.
\label{dm}
\end{eqnarray}
In order to describe spin relaxation, we consider an expansion of the spin-dependent part of $\rho({\mathbf q})$ 
in harmonics of ${\mathbf p}$. In the presence of Rashba-like interactions, it is sufficient to restrict this expansion 
to terms independent of $p_i$ and terms linear in $p_i$: 
\begin{equation}
\rho({\mathbf q})=  {\mathbf \sigma}\cdot {\mathbf S}^{(0)}({\mathbf q}) + \sigma_i S^{(1)}_{ij}({\mathbf q})p_j. 
\label{expand}
\end{equation}
Then, inserting Eq.(\ref{expand}) into Eq.(\ref{dm}), multiplying by $\sigma_z/2$ and taking the trace over spins, 
we express ${\mathbf S}^{(0)}_z({\mathbf q})$ in terms of $S^{(1)}_{ij}$:
\begin{eqnarray}
& \frac{\partial {\mathbf S}^{(0)}_z({\mathbf q})}{\partial t} + {\bar \frac{{\mathbf p}}{m}\cdot{\mathbf q}S^{(1)}_{zj}p_j} \nonumber\\
& - {\bar Tr \frac{\sigma_z}{2}\int d {\mathbf q}_1\frac{\mathbf p}{m}[{\mathbf A}_{so}({\mathbf q}-{\mathbf q}_1), 
\sigma_i S^{(1)}_{ij}({\mathbf q}_1)p_j]}
\label{S0}
\end{eqnarray}
In turn, $S^{(1)}_{ij}({\mathbf q})$ is expressed in terms of ${\mathbf S}^{(0)}_z({\mathbf q})$:
\begin{eqnarray}
&\frac{\partial {\mathbf S}^{(1)}_{ij}({\mathbf q})}{\partial t} + 
 \frac{q_j}{m}S^{(1)}_{ij}({\mathbf q})\nonumber\\
& -  Tr \sigma_i\int d {\mathbf q}_1\frac{1}{m}[ A_{so,j}({\mathbf q}-{\mathbf q}_1), 
\sigma_z S^{(0)}_{z}({\mathbf q}_1)]
\label{S1}
\end{eqnarray}
In the semiclassical limit, the term containing a commutator of the effective spin-dependent vector potential 
${\mathbf A}_{so}$ and the spin components 
describes the spin presession with frequency $\Omega= \alpha_{2}k_F E_z({\mathbf r})/\hbar$.
We assume that the classical angle of rotation is small, $\Omega\tau \ll 1$ during the time interval equal to  
the momentum relaxation time.
Using this condition and inserting the solution of Eq.(\ref{S1}) into Eq.(\ref{S0}), we obtain the equation 
of the evolution the average spin ${\mathbf S}^{(0)}_z({\mathbf q})$.  In order to obtain the magnitude of the spin relaxation time, it is sufficient to examine the resulting equation in the stationary state, which reads
\begin{equation}
Dq^2 S^{0}= -\frac{S^{0}}{\tau_{so}},
\end{equation} 
where
\begin{equation}
\frac{1}{\tau_{so}}= \alpha_2^2 k_F^2 \langle\delta E_z(0,0)^2 \rangle \tau,
\label{nDP}	
\end{equation}
 $D=v_F^2\tau/d$ is the diffusion coefficient, and $d$ is the dimensionality of the system. 
In the model that we use for a calculation of the spin relaxation caused by the fluctuations-induced Rashba term, the spin diffusion coefficient D turns out to be the same as the electron diffusion coefficient. This is because $\tau$  is defined by the spin-independent scattering,  and therefore, $S^{(1)}p_j$ relaxes in the same way as the electron momentum. Once the spin-dependence of scattering 
is taken into account, this will no longer be the case. However, in systems with potential spin-independent scattering being the dominant scattering mechanism, the difference between the spin 
diffusion coefficient and the diffusion coefficient is negligible. 
  
It is worth noticing that as seen from Eq.(\ref{nDP})  
for the fluctuation-induced Rashba term, the spin relaxation rate is expressed in 
terms of average squared of the electric field caused by fluctuations, $\langle\delta E_z(0,0)^2 \rangle$, 
and does not depend on the correlation function $C(x,y)$. 

In much the same way we obtain the spin relaxation time in cases when impurity scattering 
depends on scattering angle. The procedure is also easily takes into account
the smoothly varying intinsic spin-orbit interactions of various symmetries. 
If  $|{\mathbf \Omega}_1|$, $|{\mathbf \Omega}_1|$ given by Eqs.(\ref{intrinsic},\ref{strain}) are much less 
than the frequency of impurity scattering, the 
solution of equation for spin density matrix in the second order in the effective precession field gives the spin relaxation time
\begin{eqnarray}
\frac{1}{\tau_s^{z}}= \langle ({\mathbf \Omega}^{(1)2}_x+{\mathbf \Omega}^{(1)2}_y) \rangle \tau_1 +\nonumber\\
 \langle ({\mathbf \Omega}_{\epsilon x}^2+{\mathbf \Omega}_{\epsilon y}^2) \rangle \tau_1 +\nonumber\\
\langle ({\mathbf \Omega}^{(3)2}_x+{\mathbf \Omega}^{(3)2}_y) \rangle \tau_3,
\label{DP}
\end{eqnarray}
where $\frac{1}{\tau_n}=  \sum_{{\mathbf k}^{\prime}}(W^(0)_{{\mathbf k}, {\mathbf k}^{\prime}} 
(1 -\cos{n\theta})$, $\theta$ is the angle between ${\mathbf k}$ and  ${\mathbf k}^{\prime}$. 
For point-like scattering, $\tau_1=\tau_3=\tau$.

\subsubsection{Analysis of additional contributions to the spin and anomalous Hall currents}

The spin and AHE currents discussed in Sec. IE are due to spin-conserving spin-orbit interaction given by Eq.(\ref{H}).
In our discussion of systems with reduced relaxation of the $z-$component of spin, we have seen that several other spin-orbit terms are present. These are: (i) spin-flip scattering described by Eq.(\ref{sfa}), (ii) the random Rashba interaction given by Eqs. (\ref{hermit},\ref{Aso}), and (iii) intrinsic spin-orbit interactions described by 
Eqs. (\ref{linear},\ref{cubic},\ref{epsilon110}) containing only $z-$Pauli operator. The question arises whether these 
effects can result in a sizable contribution to the transverse current.

We begin with the discussion of the spin-flip scattering. Using Eq.(\ref{sfa}), it is easy to see that the spin-flip processes result 
in both skew scattering [Eq.(\ref{skr})] and side-jump-like [(Eq.(\ref{sj})] currents. However, these contributions are quadratic in the spin-orbit constant responsible for spin relaxation, $\alpha\kappa$. An example of the structure of such contribution, e.g., for skew scattering of electrons with a spin projection $\sigma$, is as follows: one of the matrix elements in Eq.(\ref{asym}), which now includes the summation over spins of intermediate states, is proportional to $k_x(\sigma_y)^{\sigma,\sigma^{\prime}}$, where the spin projection $\sigma^{\prime}$ is opposite to $\sigma$. Of the remaining two matrix elements in  Eq.(\ref{asym}) one includes a term $k_y^{\prime}(\sigma_x)^{\sigma^{\prime},\sigma}$, and the other is spin-independent. The resultant sum over $\sigma^{\prime}$ contains $k_xk_y^{\prime}(\sigma_z)^{\sigma,\sigma}$. Such a term leads to generation of the spin current in $x-$direction in the presence of non-equilibrium distribution function $f^E\propto k_y^{\prime}E_y$. Because this effect is quadratic in $\alpha\kappa$, its ratio to
the transverse current due to spin-conserving interaction Eq.(\ref{H}) is described by the parameter
\begin{equation}
\frac{\alpha\kappa }{d^2}\simeq 0.7\times 10^{-3},
\label{param}
\end{equation}
where the numerical estimate is made for the InAlAs/InP/InAlAs quantum well with 300\AA~ width. As we seek a system with weak relaxation of the $z-$ component of spin, every extra order of spin orbit constant results in additional smallness of the corresponding contribution. Thus, we can neglect these effects.

Discussing the role of random Rashba interactions, we first note that for uniform linear in momentum spin-orbit interactions, no 
intrinsic contributions to non-equilibrium spin conductivity is generated. This fact 
was derived in numerous contributions over the past decade, see e.g.\cite{shytov,inoue,Raimondi,Rashba}, as well 
as is apparent from the calculation of the component of the nonequilibrium spin density matrix  proportional 
to the Pauli operator and momentum in early work \cite{ALGP}. Because of smooth non-uniformity 
of the random Rashba term, a small spin current arises, which at weak spin-orbit interactions is 
quadratic in $\alpha\kappa$. Therefore, its smallness is characterized by Eq.(\ref{param}), and it can be neglected. Second, random Rashba interactions can serve as a sourse of asymmetry that leads to skew scattering or to side-jump-like contributions, instead of spin-orbit term in Eq.(\ref{H}).  Such contribution was discussed, e.g., 
 for skew scattering AHE current in \cite{borunda,Nunner} for strong uniform Rashba interaction, and it was shown that in certain conditions this effect vanish.  Besides this, for small $\alpha\kappa$, any such skew scattering or side-jump like current, even if present for non-uniform Rashba coupling, must again be quadratic in  $\alpha\kappa$, and is negligibly small.

We now analyse the role of intrinsic spin-orbit interactions that contain only $\sigma_z$ Pauli operator for the spin current. The 
Hamiltonian for a single current-contributing conduction band reads
\begin{equation}
{\cal H}_{so}^z= \alpha_z\sigma_z A(k_x,k_y),
\label{alphaz}
\end{equation}
where $A(k_x,k_y)$ is odd with respect to the time reversal symmetry.  In the case of interest for us, $A(k_x,k_y)$  is defined by 
terms in Eqs.(\ref{cubic},\ref{linear},\ref{epsilon110}). This Hamiltonian results in the spin-orbit term in electron energy:
\begin{equation}
E_{\mathbf k}=\frac{\hbar^2 k^2}{2m}\pm \alpha_zA(k_x,k_y)
\label{en}
\end{equation}

We begin with a question whether the terms of this symmetry can contribute to the side-jump group 
currents. These currents originate from two sourses: the "spin dipole" $\Omega$  given by Eq.(\ref{Omega}), and the phase of a scattering matrix element. It is easy to see that the spinors entering the electron wavefunctions and defining $\Omega$  in this case are just $[0,1]$ or $[1,0]$, and are momentum-independent. Then the cooridnate contribution associated with $\Omega$ in the side jump given by Eq.(\ref{jump}) vanishes. Furthermore, the "intrinsic" current related to ${\mathbf curl}\Omega$\cite{Luttinger1954} vanishes as well. Moreover, if a scattering potential itself does not 
contain a spin-orbit term, the phase of the matrix element of scattering is spin-independent (and is due to different momenta in the initial and final
states of a scattering process). 
Then, the phase part of the side-jump given by Eq.(\ref{jump}) also vanishes. 

We next check if any contribution may appear as a result of renormalization of the velocity operator (and the current operator) containing the spin-orbit correction
\begin{equation}
{\mathbf  v}_{\pm}=\frac{\hbar {\mathbf k}}{m}\pm \frac{\alpha_z}{\hbar} \frac{\partial A(k_x,k_y)}{\partial {\mathbf k}}.
\label{vel}
\end{equation}
Such spin-orbit correction is capable to produce the spin current, which is defined by terms odd in $\alpha_z$ in the expression
\begin{equation}
{\mathbf  j}_{\pm}= -e\sum_{\mathbf k} {\mathbf  v}_{\pm} f_{\pm}({\mathbf k}).
\label{isc}
\end{equation}
Here  $f_{\pm}({\mathbf k})$ is the nonequilibrium density matrix in the presence of externally applied electric field, given by Eq.(\ref{flow}), which is linear in velocity.  It is easy to see that when the expression under the sign of sum is odd in $\alpha_z$, it is also odd in ${\mathbf k}$, and the sum vanishes. Similarly, the contributions in which the 
spin-independent component of the velocity operator (\ref{vel}) defines the current (\ref{isc}),
and the spin-dependence comes from energy (\ref{en}), vanish in summation over ${\mathbf k}$.

We analyse now the possibility of the skew scattering asymmetry caused by the spin-orbit interaction (\ref{alphaz}). 
First, due to momentum-independent and real spinors, the standard skew scattering probability given by Eq.(\ref{asym})  vanishes.
Second, higher-order in scattering amplitude terms in the scattering probabilities also vanish. Indeed, an asymmetric distribution function in the presence of electric field is linear in the current (velocity) operator, and it contains odd in ${\mathbf k}$ term independent of $\alpha_z$ and even in ${\mathbf k}$ term linear in $\alpha_z$. The current operator that is averaged over the distribution function has exactly the same property, while the spin-dependent part of energy is linear in $\alpha_z$ and odd in 
${\mathbf k}$. Then if an expression for the current is odd in $\alpha_z$, which is required for terms (\ref{alphaz}) in order to get currents 
of opposite spin projections to be opposite in sign,  it is also odd in ${\mathbf k}$,  so that summation over the directions of ${\mathbf k}$ gives zero spin current.  

Thus we see that nonequilibrium spin currents in the case of Hamiltonian (\ref{alphaz}) all vanish. However, there is quite an interesting twist here related to the spin currents in equilibrium. This effect was pointed out by E.I. Rashba in connection with the spin-flipping Dresselhaus and Rashba terms\cite{equilibrium}. For such contribution, the accompanying spin relaxation of the $z-$component of spin is strong, and therefore the condition (\ref{conditiongen}) cannot be satisfied, so that no spin-electric stripes occur. However, 
 the "equilibrium spin current" 
caused by Hamiltonian (\ref{alphaz}) is not accompanied by spin-relaxation. This effect is given by the expression of Eq.(\ref{isc}) with the function $f_{\pm}({\mathbf k})$ being 
the Fermi function. The linear in $\alpha_z$ spin current 
vanishes as an integral of the full differential, but the cubic or higher power in $\alpha_z$ currents are present. These currents
are very small, so that normally the residual spin relaxation, in particular due to nuclear spins, will preclude spin-electricity. Still there exists a theoretical possibility that in a setting with the spin-orbit Hamiltonian (\ref{alphaz}), parameter $\kappa\sim 0$, and nuclei with spin 0, spin-electric stripes can arise. It would  be curious to develop a material with these characteristics. However, we have no suitable candidate at present, and in what follows we will discuss only spin-electric stripes caused by the external electric field, i.e., the electric current flowing through the sample. 

As follows from the discussion above, the contributions of all spin-orbit interactions 
except the principal term of Eq.(\ref{H}) to non-equilibrium spin currents are either vanishingly
 small or vanish altogether. Therefore, our consideration of spin-relaxation as the only effect of 
the spin-flipping terms in a microscopic Hamiltonian on the constituitive 
equations of the system Eqs. (\ref{spindensity},\ref{chargecurrent}) is justified.

\subsection{Residual spin relaxation: electron-nuclear interactions}

In our quest to eliminate as much spin relaxation as possible we inevitably arrive to the point when the spin-orbit source of $z-$projection spin 
relaxation is reduced significantly, and even eliminated in an ideal structure with equal potential 
confinement of conduction and valence electrons in $z$-direction, with no change of materials parameters accross the structure, while the spin-orbit  terms giving electric field domains are intact. However, as we are 
discussing III-V semiconductors as the most probable experimental setting, we must remember that these materials contain 
rather heavy nuclei in the host crystalline lattice, which are the source of hyperfine contact interactions of 
electron and nuclear spins. The contact interaction is usually the strongest electron spin- nuclear spin interactions, and potentially can result in electron spin relaxation. 

In the problem of spin Hall current, we are dealing with delocalized (free) electrons. Hyperfine coupling has never been 
considered as a viable spin relaxation mechanism for free charge carriers (as opposed to spin relaxation of 
localized electrons). For free electrons, the relaxation mechanisms associated with spin-orbit interactions are usually by far the most efficient, with relaxation rates 5-6 orders of magnitude higher than for spin relaxation due to 
hyperfine coupling. However, because we aim at reducing (eliminating) spin-orbit interactions spin relaxation mechanisms, 
we have to analyse the electron spin relaxation induced by hyperfine coupling.

The hyperfine interaction of electron and nuclear spins is described by the Hamiltonian, \cite{abragam}
\begin{equation}
{\cal H}^{(hp)}=\frac{8\pi}{3}\nu_0\nu_N g_0\sum_{i,n} g_i\delta ({\mathbf r}- {\mathbf R}_{i,n})
{\mathbf \sigma}\cdot{\mathbf I}_{i,n},
\label{hpi}
\end{equation}
where $\nu_0= e\hbar/2m_0 c$ and $\nu_N=e\hbar/2M_p c$ are the electron and nuclear magnetons, $m_0$ and $M_p$ being the free electron and proton mass correspondingly, $g_0\sim 2$ is the free electron g-factor, index $i$ enumerates 
the types of species of nuclei present in a crystal unit cell and $n$ enumerates all nuclei of a 
given type in a sample, $g_i$ is the nuclear Lande-factor 
and $I_{i}$ is the 
nuclear spin operator for the nuclei of $i$-type, $R_{i,n}$ is the nuclear coordinate.

Spin relaxation of electrons due to hyperfine interactions comes from the following two mechanisms. One is the direct mutual electron-nuclear 
spin-flip scattering. For the interaction given by Eq.(\ref{hpi}), it is important that electron density is non-uniform over the 
crystal unit cell. For the conduction electron wavefunction in the ground state of an infinite quantum well 
$\Psi_k({\mathbf r}_{\perp})=(1/\sqrt{S})u({\mathbf r}_{\perp})
\exp{i{\mathbf k}_{\perp}\cdot{\mathbf r}_{\perp}}(\sqrt{2}/d)\cos{\pi z/d}$, where $u({\mathbf r}_{\perp})$ is the Bloch amplitude periodic with the 
lattice constant, $d$ is the width of the quantum well, and $S$ is the 2D sample area, we obtain scattering matrix element
\begin{equation}
{\cal H}^{(hp)}_{{\mathbf p}\sigma,{\mathbf p}^{\prime}\sigma^{\prime}}=
\frac{V_c}{S} \sum_{i}A_i\sum_n C_{i,n}
\Big [{\mathbf \sigma}\Big ]_{\sigma,\sigma^{\prime}}\cdot{\mathbf I}_{i,n} 
e^{i{\mathbf R}_{i,n}\cdot({\mathbf k}-{\mathbf k}^{\prime})},
\label{mel}
\end{equation}
where $C_{i,n}=(2/d) \cos^2 (\pi Z_{i,n}/d)$ is the coefficient arising from the quantum well confinement, $V_c$ is the unit cell volume,  
$Z_{i,n}$ is the nuclear coordinate, and the hyperfine constant $A_i$ is defined by
\begin{equation}
A_i=\frac{8\pi}{3}\nu_0\nu_N g_0 g_i |u(0)|^2_i.
\end{equation} 
The spin-flip scattering probability is given by Eq.(\ref{spinrelax}) with the substitution of the spin-orbit scattering matrix elements by 
Eq.(\ref{mel}), with inclusion of averaging over initial states of the nuclei and summation over final nuclear spin states. Averaging over coordinates $Z_{i,n}$, gives the probability of the flip of the $z-$projection of spin of the quantum well electrons due to hyperfine interaction and the corresponding spin relaxation rate $1/\tau_{NS}$:
\begin{equation}
\frac{1}{\tau_{NS}}=\sum_{{\mathbf k}^{\prime}}{\cal W}^{(hp)}_{{\mathbf k}\sigma,{\mathbf k}^{\prime}\sigma^{\prime}}=\sum_i \frac{1}{2\hbar}(V_c A_i)^2 n_i \frac{m}{\hbar^2d} I_i(I_i+1),
\label{ncb}
\end{equation}  
where $n_i$ is the density of nuclei species of i-type. 

The second mechanism of electron spin relaxation comes from the effective magnetic field $B_N$ 
acting on electrons spins due to hyperfine interaction (\ref{hpi}):
\begin{equation}
B_N=\frac{V_c}{\nu_B  g*}\langle\sum_{i,n}A_iC_{i,n} {\mathbf I}_{i,n} \rangle_{N}
\label{BN}
\end{equation}
 where angular brackets denote averaging over all nuclear wavefunctions and $g*$ is the effective 
electron g-factor in a crystal. If this field were static, it would induce no electron spin relaxation.
To analyze its dynamics, we first observe that there is also an effective magnetic field $B_e^i$ 
due to hyperfine 
interactions that acts on each nuclei of i-type due to electrons:
\begin{equation}
B_e^i= \frac{V_c}{\nu_N  g_i} A_iC_{i,n} {\mathbf I}_{i,n}.
\end{equation}
The effective field $B_e^i\ll B_N$,  in particular because the nuclear magneton and spin splitting is 3 orders of magnitude smaller than that for electron.
We now observe that in the absence of the external magnetic field, and other sources of nuclear spin polarization, the effective field $B_N$ averaged over times $t> \hbar/\nu_N  g_iB_e^i$ must be zero on symmetry grounds. On a smaller time scale, nuclear spins retain some polarization caused by electron spins, and, in turn, the electron spin is subject to a frozen fluctuation of magnetic field arising from nuclear spins. These 
fluctuations of hyperfine-induced effective magnetic field acting on electron spins are described by the
dispersion of the nuclear field distribution, see, e.g., \cite{paget}. Assuming the nuclear spin directions independent, for nuclei acting on quantum well electrons, this
dispersion is given by
\begin{equation}
\Delta^2_B=\frac{2}{3}\sum_{jn} (I_j(I_j+1) A_j^2 C_j^2 \Big ( \frac{V_c}{\mu_B g^* }\Big )^2.
\end{equation}
Substituting the sum over $n$ by an integration, we obtain
\begin{equation}
\Delta^2_B=\frac{1}{(\mu_B  g^*)^2}\sum_{j}A_j^2 (I_j(I_j+1)\frac{ V_c}{Sd},
\end{equation} 
where the summation runs over different types of nuclear species in one unit cell. 
This dispersion is very small for delocalized electrons. The corresponding time scale becomes of order of 
the characteristic times for nuclear dipole-dipole relaxation time, $\tau_{dd}\sim 10^{-4}$s\cite{abragam}. Nuclear dipole-dipole interactions, in contrast to hyperfine interactions, do not conserve the total spin of electrons and nuclei. Then, the nuclear dipole-dipole interactions, as a result of flip-flop of the nuclear spins, randomise the effective magnetic fields acting on electronic spins due to hyperfine interactions, so that on the scale of $\tau_{dd}$, the nuclear fields seen by electronic spins are rapidly changing. If the characteristic precession frequency of electrons spins is  $\mu_B g^*\sqrt{\Delta_B^2}\ll \hbar\tau^{-1}_{dd} $, the electron spin undergoes random rotations by an angle of characteristic magnitude $  \mu_B g^*\sqrt{\Delta_B^2}\tau_{dd}/\hbar$. Such diffusive motion of the electron 
spin results in spin relaxation characterized by the relaxation rate
\begin{equation}
\frac{1}{\tau_{s,N}^z}= \frac { (\mu_B g^*)^2 \Delta_B^2\tau_{dd}}{\hbar^2}.
\end{equation}
The shorter $\tau_{dd}$, the faster the electron spin relaxation.
Therefore in contrast to direct scatttering of electron spins by nuclear spins due to hyperfine interactions,
i.e., collisional broadening, the dipole-dipole nuclear interactions result in motional narrowing. 
In an experimental setting with almost completely suppressed spin-orbit induced spin relaxation,
both collisional broadening and motional narrowing due to nuclear spins can be important factors in 
weak relaxation of $z-$projection of the electron spin. 

\section{Spin-electric stripes in 110 grown quantum wells with limited spin flips}

We now calculate the magnitude of the applied electric field 
needed for the observation of the spin-electric domains, and the magnitude of 
the transverse electric field in domains. We consider 
[110] grown 
$In_{0.52}Al_{0.48}As/InP/In_{0.52}Al_{0.48}As$ heterostructure, and take $\kappa=0.33$ \cite{ruan}. This is the lowest 
$\kappa$ in known III-V systems. 
The Kane parameter in InP is $1.36\times 10^{-19} erg\cdot cm$, the bandgap is $E_g=1.35 eV$, the effective mass 
is $m=0.073m_0$\cite{IP} and the 
spin-orbit splitting of the valence bands $\Gamma_8$ and $\Gamma_7$ ia $\Delta=0.11 eV$. This gives spin-orbit 
constant $\alpha =1.92\times 10^{-16} cm^2$. 

We condider a rectangular quantum well with 300\AA width, in sample with the planar dimensions $94 \mu$ m $\times 1$ cm. 
We take a strongly disordered quantum well with well doping by donors, and compensating doping by acceptors, at a total level of
donor and acceptor impurity density $n_t=2\times 10^{12}$ cm$^{-2}$. In addition, two symmetrically positioned remote doping layers provide 
charge carriers, defining the total carrier density at a level $n=5\times 10^{11}$ cm$^{-2}$ in all settings we are going to consider. When donor density $n_d$ exceed 
acceptor density $n_a$, donors provide additional electrons in the quantum well, and weaker remote doping is needed to get 
this level of total density. In the setting with acceptors more numerous than donors, remote doping layers need both to compensate the action of acceptors that take electrons from donors and to provide the total carrier density. This can be achieved by taking different separations of $\delta$-doping layers (setbacks) from the centerplane $z=0$ of the quantum well.
 
At large setbacks, the dominant scattering mechanism are quantum well impurities. We take scattering to be short-range but assume that it mimics shallow donors and acceptors capable of providing or accepting one electron. Scattering amplitudes for such donor and acceptor scattering have the same magnitudes and opposite signs.  We consider three different settings corresponding to cases I, II and III of the Sec. IA:
Case I corresponds to total compensation, $n_d=10^{12}$ cm$^{-2}$, $n_a=10^{12}$ cm$^{-2}$, 
where $n_d$ and $n_a$ are the donor and acceptor densities correspondingly.  In the case II, the donor doping level is $n_d=1.1\times 10^{12}$ cm$^{-2}$, and the acceptor doping level is  $n_a=0.9\times 10^{12}$ cm$^{-2}$. 
In the case III, quantum well acceptors overcompensate quantum well donors: $n_d=0.9\times 10^{12}$ cm$^{-2}$,  $n_a=1.1\times 10^{12}$ cm$^{-2}$. Remote doping 
provides charge carrier electrons in all three cases.

For case I, we consider the stripes first neglecting the spin relaxation of the $z-$component of spin, and then see how spin relaxation modifies the picture. The $\delta$-doping layers setbacks are $l=1100\AA$, 
providing $n=5\times 10^{11}$ cm$^{-2}$ electron charge density \cite{Eisenstein}, which results in 
Fermi energy $\epsilon_F=16.5 meV$. The used setbacks suppress both scattering by smooth random potential and the contribution of long range-fluctuations into spin-orbit interactions. The impurity momentum relaxation time in this setting is 
$\tau=0.4\times 10^{-13}$, and the mobility is $\mu = 10^3 cm^2/V\cdot s$. In this case of total compensation, the 
the spin current is defined by side-jump-like contributions of Eq. (\ref{sj}) only. The parameter characterizing the strength of disorder in this setting is $k_F\ell=2$, and the transverse  (spin Hall) mobility 
$\eta_1=0.6 cm^2/(V\cdot s) $. A periphery stripes of $15\mu m$ width and transverse electric field $E_x=2.4 V/cm$, and a central stripe of the width $2b=64 \mu m$ are induced in such low-mobility sample by an electric field $E_y=4kV/cm$. This gives energy acquired by an electron in electric field on the scale of the mean free path equal to $eE\ell=4.2 meV$ or $\sim 0.25 E_F$
This would lead to certain heating, but nevertheless, the observation of such stripes in InAlAs/InP/InAlAs heterostructures is feasible. Designing heterostructures with lower $\kappa$ will allow to decrease electric fields required for the observation of the effect and thus to reduce heating substantially. The spin polarization, electric field and potential distribution neglecting spin relaxation are plotted in Fig.~\ref{Fig5}, Fig.~\ref{Fig6} and Fig.~\ref{Fig2} correspondingly. 

The choice of parameters made above is consistent with Eq.(\ref{conditiongen}) and has a comfortable edge. Indeed, the calculation of the spin relaxation rate for the $z-$component of spin shows that two mechanisms are dominant: First, weak residual spin-orbit interactions due to spin-orbit constants (\ref{fqwso}), 
(\ref{diffmass}) result in spin-orbit scattering off quantum well impurities described by Eq. (\ref{2dsr}), which leads to $1/\tau_{ss}=5.47 \times 10^{4}s^{-1}$.  
Second, the biggest spin $z-$component relaxation comes from direct scattering of electron spin by nuclear spins, due to hyperfine interactions. For InP, In nuclei has spin 9/2,  $g_{In}=1.11$, and 
the corresponding density of the Bloch electron at the nuclei is $|u_c(0)|^2=7.63\times 10^{25}$ cm$^{-3}$\cite{Denninger}. For P nuclei spin is 1/2, $g_P=2$, and  $|u_c(0)|^2=3.26\times 10^{25}$ cm$^{-3}$.  The relaxation rate given by
Eq.(\ref{ncb}) leads to $1/\tau_{NS}=1.297 \times 10^{5}s^{-1}$. The spin relaxation due to random Rasbha terms is $\sim 200$ times smaller than that due to spin-orbit scattering, because of the low mobility and large setbacks. The spin relaxation due to fluctuations of the hyperfine field in the presence of nuclear spin dipole-dipole scattering is very small due to massive sample.
The total spin relaxation rate is  $1/\tau_{s}=1.75 \times 10^{5}s^{-1}$ results in spin relaxation length $L_s=95.3 \mu m$.
Thus, in the chosen structure the strip is narrower than $L_s$, which is more than 3 times bigger than the half-width of the central stripe, and more than 6 times exceeds the width of the periphery stripe. 

The big spin relaxation length and long times are the consequences of weak spin-orbit constant in InP, structure of offsets 
in the InAlAs/InP/InAlAs heterostructure, "spin-orbit engineering", particularly, $[110]$ crystallographic orientation of the quantum well, symmetric confinement, large setbacks for $\delta$-doping layers, and a wide quantum well. For comparison, the spin relaxation rate due to spin-orbit scattering in the [110] GaAs quantum well, in an identical symmetrical structure,
 is $1/\tau_{ss}= 0.65 \times 10^{7}s^{-1}$. In a 100\AA width quantum well, $1/\tau_{ss}= 0.6 \times 10^{8}s^{-1}$, and the corresponding spin relaxation length is $L_s=4.7 \mu m$.

\begin{figure}[t]
\vspace{-5mm}
\includegraphics[scale=0.35]{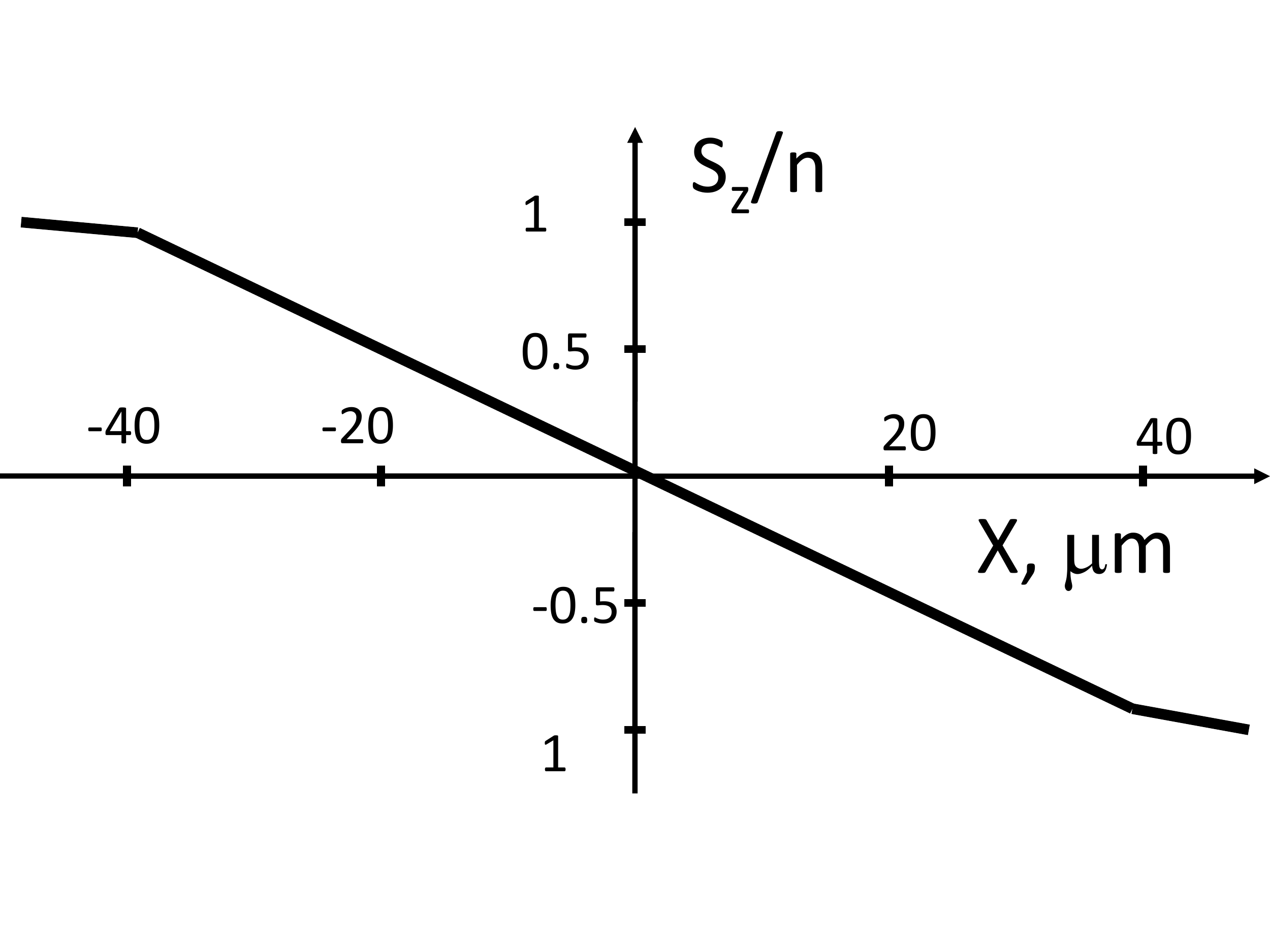}
\vspace{-8mm}
\caption{Spatial dependence of spin polarization on the coordinate transverse to the flowing electric current in the presence of spin relaxation of the $z-$component of spin, case I. All parameters are discussed in the text.}
\vspace{0mm}
\label{Fig17}
\end{figure}

\begin{figure}[t]
\vspace{-3mm}
\includegraphics[scale=0.35]{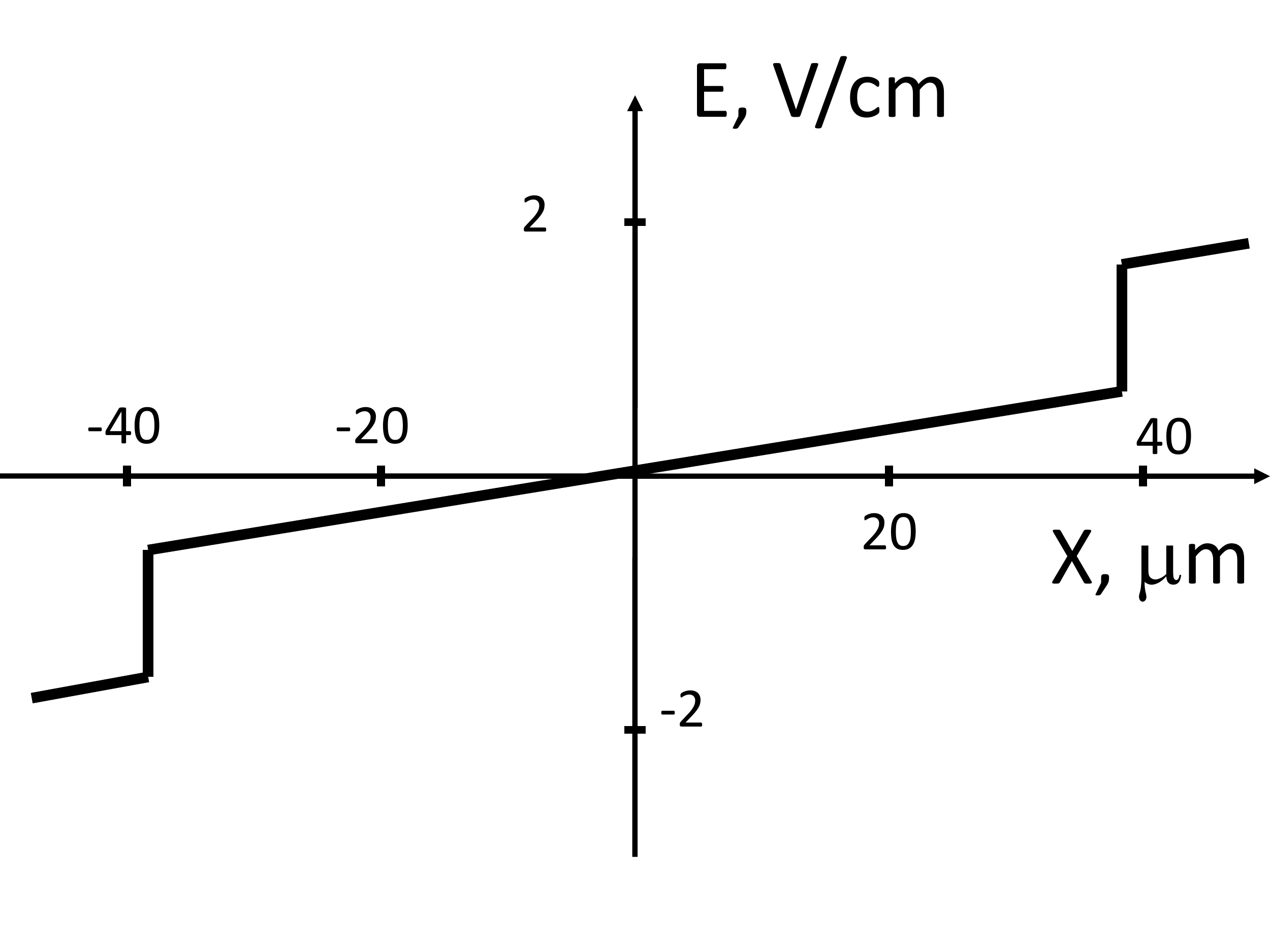}
\vspace{-5mm}
\caption{Spatial distribution of transverse electric field in the presence of spin relaxation, case I.}
\vspace{-3mm}
\label{Fig18}
\end{figure}

\begin{figure}[t]
\vspace{3mm}
\includegraphics[scale=0.7]{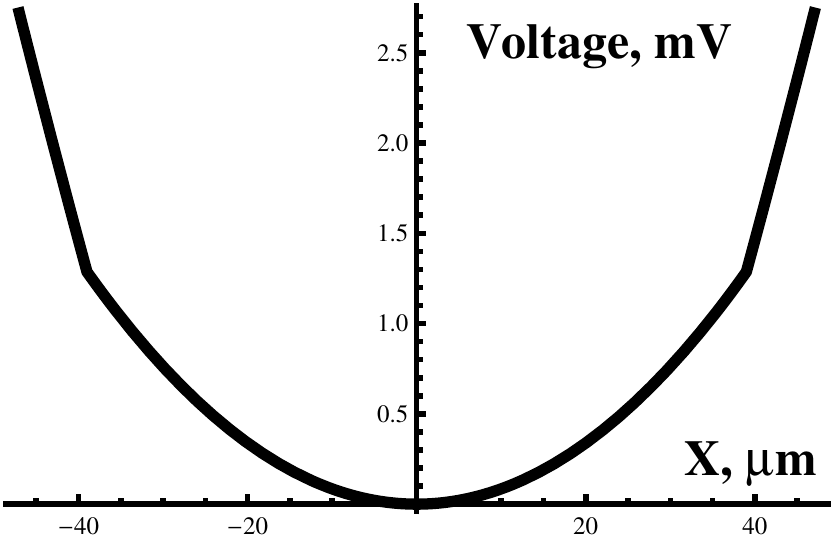}
\vspace{-2mm}
\caption{Potential profile in the presence of spin relaxation, case I.}
\vspace{-5mm}
\label{Fig19}
\end{figure}

For case I, we calculate how the spin relaxation changes the picture of the periphery and central stripes.
The half-width of the central stripe becomes $b_s= 39 \mu m$, and the width of the periphery stripe is now $8\mu m$. 
The spin polarization, electric field and potential distribution across the sample are shown in 
Fig.~\ref{Fig17}, Fig.~\ref{Fig18} and Fig.~\ref{Fig19} correspondingy.
A small electric field linear with coordinate is induced in the center reigion (Sec. IIC). This electric field, however, results 
in sizable build-up of the potential in the central region over the increased half-width of the central stripe. Still, placing the inside-sample contact for voltage probe within 8 microns from the edge, one would observe almost a linear increase of voltage with 
the increase of external electric field, characterizing the periphery stripe, as opposed to markedly quadratic dependence in the 
central stripe.

We now discuss parameters of the system for cases II and III. Here the electron mobility is the same as in case I. Because the acceptors do not fully compensate donors, a skew scattering current is present.
In the low density samples that we consider, the side jump and the skew scattering mobilities have equal magnitudes in the absence of compensation, $n_a=0$. For $|n_d-n_a|/  (n_d+n_a)=0.1$, as in cases II and III, the skew scattering is just 10\% of the side jump-like currents. We remark that the spin drag effect that decreases the skew scattering contribution\cite{Vignale},
 for small density of charge carriers (4 times less than in \cite{Vignale}) and temperature T=10 K , in our setting reduces 
the skew scattering by just $\sim 2\%$ of its value. The electric field and potential distribution neglecting spin relaxation are plotted $E_y=4kV/cm$ for case II in Fig.~\ref{Fig7} and Fig.~\ref{Fig8}, and for case III in Fig.~\ref{Fig9} and Fig.~\ref{Fig10} correspondingly. The transverse electric field in the periphery domain is $E_x=2.16 V/cm$ in case II, and $E_x=2.64 V/cm$ in case III. In these cases, a linear in coordinate electric field in the central stripe arises already in the absence of spin relaxation. That means that at in the presently available setting it would be rather difficult to distinguish side-jump-like and skew scattering contributions by measuring the distribution of electric field and potential. However, if  a system is engineered with bigger $\alpha$ and smaller $\kappa$, or weaker constants of the hyperfine interactions, such an experiment would be feasible. We also note that case IV with dominant skew scattering or case V with skew scattering opposite in sign to side jump and exceeding it $\sim 1.5$ times in value requires the values of $k_F\ell\sim 10$ that result in 
significant of $eE_y\ell\sim E_F$. These cases may become feasible in systems with stronger spin current and weaker spin relaxation. 

The electric field in both in stripes can be observed in experiments similar to \cite{willett}.  A voltage 
between internal contacts in the center region and contacts on a sample periphery should be measured. 
A 100\% spin polarization at the edges, opposite in two pperiphery stripes, can be measured in 
optical experiments similar to \cite{Sih1}. Furthemore, as Fig.~\ref{Fig19} indicates, a build-up of electric potential 
and a significant spin polarization can be observed even if the periphery stripes are not yet developed, in a narrower sample.

\section{Future work. Conclusion}

Considering the charge and spin currents, we restricted ourselves to  
currents linear in the gradient of density and electric fields. We considered zeroth approximation in charge-electric field distribution and calculated the first order modification of charge distribution, but did not consider the corresponding changes in accordance with the consituitive equations of the system. However, the approximations made capture the principal features of the spin-electric state. The current manusript is purposedly analytic, and all figures represent the analytic equations presented in the paper.  Future work will address numerical analysis of time-dependent picture of the phenomena, self-consistent calculation of
 charge-electric field distribution together with the constitutive equations, and hydrodynamic approach. 
Also, in this paper we restricted our consideration to conduction band electrons. The account of spin-electric stripes in systems with charge-carrier holes will be presented shortly.

In conclusion, we the reiterate the principal findings of our work.
When the electric current is flowing through a two-dimensional (2D) uniform conductor, the spin-orbit scattering results in the two spin-electric stripes near the two opposite boundaries of the conductor parallel to the direction of the current. In each periphery stripe there is an electric field transverse to the flowing current. The directions of the electric fields in the two stripes are opposite to each other, and their magnitude is the same. If the relaxation of the spin component perpendicular to the 2D plane is negligible, the magnitude of the electric field is constant throughout the stripes.  In the stripes, electrons are fully spin polarized, with opposite spin orientations in the two stripes. The electric fields accompanied by a 100\% spin polarizations in the periphery stripes are the consequences of the inability of the spin relaxation to limit the spin accumulation when a spin current flows. Spin polarization is limited only by its maximal (minimal) value. This leads to Anomalous Hall effect in the periphery stripes, so that charge carriers with the two opposite spins stream to the respective boundaries. It is then the Coulomb interactions that limit the streams and give rise to the electric fields.

The two periphery stripes are separated by the center region (the third stripe) of the 2D plane. On the center-line, the electron spin polarization is zero. The picture of the spin polarization in the central stripe is almost like in the conventional spin accumulation, except that when the spin polarization grows from the center-line to the
boundary between the central and the periphery stripe (or falls for the opposite spin direction) in absence of spin relaxation, it reaches the maximal $\pm 1$ value at such boundaries and cannot grow any further into the periphery stripes.  The magnitude and direction of the electric field in the central stripe in the case of negligible spin relaxation depends on the relation between the microscopic mechanisms of the bulk spin current. Experimental measurement of the voltage caused by the electric fields in the periphery stripes would
constitute a direct electric measurement of the spin and AHE currents. In the case of reaching the limit of negligible spin relaxation, measuring the spatial distribution of the electric fields can provide a useful tool for experimental separation of the skew scattering and side-jump like effects. 

When a weak relaxation of the spin component normal to the 2D plane is present, the spin polarization and the electric field in the periphery stripes experience a small change from the periphery towards the center region. At strong spin relaxation stripes do not arise.The division of the sample into three stripes remains meaningful in the presence of weak spin relaxation: at the boundaries between the preriphery stripes and the central stripe the electric field is a steep function of the transverse coordinate. Also a 100\% spin polarization is present at the edges of the sample. 
If a single periphery stripe becomes a part of a separate circuit with the electric current flowing, it is possible to transfer this spin polarization for the purpose of applications. 

Favorable experimental settings, in which an electron spin relaxation of the spin component transverse to the 2D plane is suppressed but the spin current is not, are discussed. The role of the remote doping by donors and doping in the quantum well by donors and acceptors is studied. In a consideration of the spin-orbit effects for 2D electrons, a 3D short-range and long-range random potentials are taken into account. The role of the boundary conditions and the role of the difference of parameters in the barriers and the quantum well are investigated. All the spin-orbit spin-flipping terms can be eliminated from the 2D Hamiltonian by engineering a confinement potential for conduction and valence electrons. In the proposed experimental setting of electrons confined to $InAlAs/InP/InAlAs$ symmetric double heterostructure, it is possible to suppress the spin-orbit induced relaxation of the normal component of spin to such an extent that hyperfine interactions of spins of delocalized conduction electrons and nuclear spins play a significant role in electron spin relaxation.

\section{Acknowledgements}
I am grateful to I.L. Aleiner for help in solving the electrostatic problem (section 1.B. and Appendix A) 
and useful discussions. I am grateful to J.J. Eisenstein for the discussion of remote doping in 2D samples. Support from National Science Foundation under the grant ECCS-0901754 is gratefully acknowledged. 

\section*{Appendix A}

Here we present the details of a solution of the electrostatic problem, and find the electric field in the whole range of $x$ and the distribution of charge density across the 2D sample consistent with the solution of constituitive equations (\ref{j}) for the electric field in domains. Rather than deriving the function of complex variable $E(t)$ of section I.B from the complex potential for 2D electrostatic problem found by using the relevant Schwartz-Kristoffel transformation, we will derive  the Kramers-Kronig like dispersion relations for some auxillary analytic function, which is 
directly related to $E(t)$.  

\begin{figure}[t]
\vspace{-8mm} 
\includegraphics[scale=0.35]{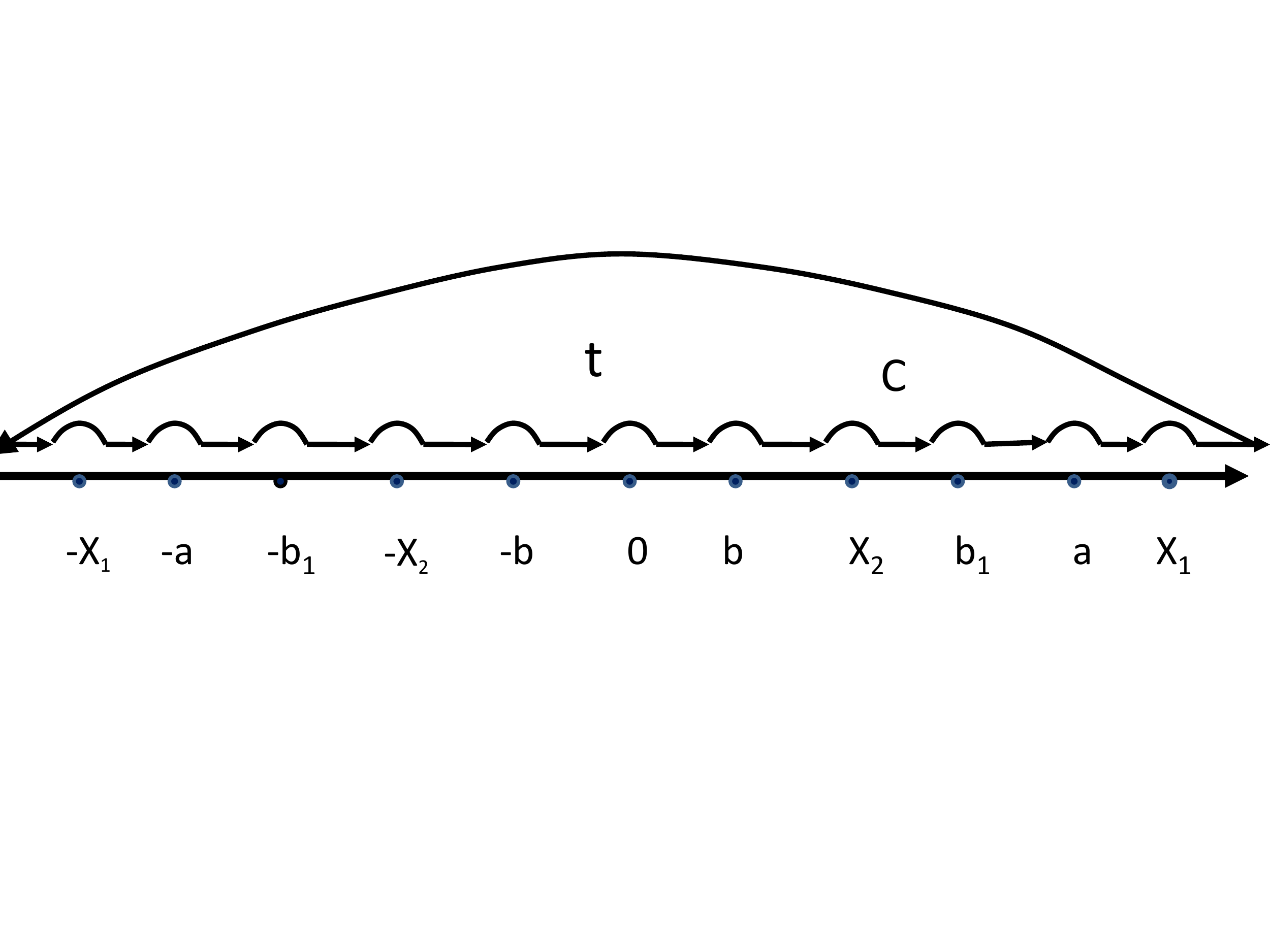}
\vspace{-30mm}
\caption{Contour in the complex plain t. The functions $f(t)$ and $1/f(t)$, are analytic inside and on the contour formed by real axis, semicircular detours of infinitesimal radius around singular points $\pm a$, $\pm b_1$, $\pm b$, around points of extrema of  $f(t)$ at $\pm X_1$, $\pm X_2$ and 0, and around infinite-radius semicircle.}
\vspace{-5mm}
\label{Contour-t}
\end{figure}

We consider a contour $C$ in the complex plane $t$ shown in Fig. \ref{Contour-t}, and a conformal 
mapping descibed by the function 
\begin{equation}
f(t)=\sqrt{\frac{(t^2-b_1^2)(t^2-b^2)}{t^2-a^2}}.
\label{mapping}
\end{equation} 
The function $f(t)$ has three extrema with the derivative $f^{\prime}_t(t)=0$ at 
\begin{eqnarray}
& t_1=(a^2 +\sqrt{(a^2-b_1^2)(a^2-b^2)})^{1/2} & |t|> a\nonumber\\
& t_2=(a^2 -\sqrt{(a^2-b_1^2)(a^2-b^2)})^{1/2} & b<|t|<b_1\nonumber \\
& t=0
\end{eqnarray} 

Thus, the functions $f(t)$ and $1/f(t)$ are analytic in the upper half plane of the complex variable $t$ and on the real axis (with the exception of points 
$t=\pm t_1, \pm a, \pm b_1, \pm t_2, \pm b, 0$). The complex function $E(t)$ is analytic in the same region of the complex plane as well, and 
so is the auxillary function $E(t)/f(t)$. Furthermore, from boundary conditions $E(t)$ given by Eqs.(\ref{a},\ref{b},\ref{c1},\ref{d}) and from Eq.(\ref{mapping})
\begin{equation}
\lim_{|t|\rightarrow\infty} E(t)/f(t)=0.
\label{semicircle}
\end{equation}
Then, from the integral Cauchy formula, analytic properties of $E(t)/f(t)$ and Eq.(\ref{semicircle}), we have
\begin{equation}
E(t)/f(t)= \frac{1}{\pi i} P\int_{-\infty}^{\infty}\frac{E(u)/f(u)}{u-t}du,
\label{C}
\end{equation}
where $P$ indicates the Cauchy principal value. 

The real and imainary parts of $E(x)/f(x)$ obey the dispersion relations
\begin{eqnarray}
& Re[E(t)/f(t)] &=\frac{1}{\pi } P\int_{-\infty}^{\infty}\frac{Im[E(u)/f(u)]}{u-t}du\\
& Im[E(t)/f(t)] &=-\frac{1}{\pi } P\int_{-\infty}^{\infty}\frac{Re[E(u)/f(u)]}{u-t}du.
\label{dispersion}
\end{eqnarray}
The function $f(t)$ with the branching points $\pm a$, $\pm b_1$ and $\pm b$ is alternatively real and imaginary in the intervals $|x|>a$, $b_1 <|x|< a$, $b <|x|<b1$ and $|x|<b$,  ($x$ is the real value) :
\begin{eqnarray}
& f(x)=sign(x) \sqrt{\frac{(x^2-b_1^2)(x^2-b^2)}{x^2-a^2}} \hspace{5mm} \mbox{if $|x|>a$} \nonumber\\
& f(x\pm i0)=\mp i \sqrt{\frac{(x^2-b_1^2)(x^2-b^2)}{a^2-x^2}} \hspace{5mm} 
\mbox{ if $b_1 < |x|< a$} \nonumber\\
& f(x)=sign(x) \sqrt{\frac{(b_1^2-x^2)(x^2-b^2)}{a^2-x^2}} \hspace{5mm} 
\mbox{if  $b < |x|< b_1$}\nonumber\\
& f(x\pm i0)=\pm i \sqrt{\frac{(b_1-x^2)(b-x^2)}{a^2-x^2}}\hspace{5mm} \mbox{ if $|x|<b$}
\label{f}
\end{eqnarray}
Using dispersion relations (\ref{dispersion}), boundary conditions Eqs. (\ref{a},\ref{b},\ref{c1}), and Eq.(\ref{f}) with complex values taken in the upper half plane, we find unknown electric fields $E(|x|>a)$ and $E(b <|x|<b1) $ and charge densities $n(b_1 <|x|< a) $ and $n(|x|<b)$. For an unknown electric field,  i.e., the electric field  in the intervals of $x$, where it is not set by the boundary conditions, we have
\begin{eqnarray}
&  E(x, |x|>a) = sign(x)  \sqrt{\frac{(x^2-b_1^2)(x^2-b^2)}{(x^2-a^2)}}\times\nonumber\\
& \left[
\frac{2E_0}{\pi}\int_{b_1}^a\frac{sds}{s^2-t^2} \sqrt{\frac{a^2-s^2}
{(s^2-b_1^2)(s^2-b^2)}}\right. \nonumber \\
 & \left. -4N_0 \int_{b_2}^{b_1}\frac{sds}{s^2-t^2} \sqrt{\frac{(b_1^2-s^2)(s^2-b^2)}{a^2-s^2}} \right].
\label{E1}\\
&  E(x, b <|x|<b1) = sign(x)  \sqrt{\frac{(b_1^2-^x2)(x^2-b^2)}{(a^2-x^2)}}\times\nonumber\\
& \left[
\frac{2E_0}{\pi}\int_{b_1}^a\frac{sds}{s^2-t^2} \sqrt{\frac{a^2-s^2}
{(s^2-b_1^2)(s^2-b^2)}}\right. \nonumber \\
 & \left. -4N_0 \int_{b_2}^{b_1}\frac{sds}{s^2-t^2} \sqrt{\frac{(b_1^2-s^2)(s^2-b^2)}{a^2-b^2}} \right].
\label{E2}
\end{eqnarray}

The charge density in the regions, where it is not defined by the constraint, is given by
\begin{eqnarray}
&  n(x, b_1<|x|<a) = -  \sqrt{\frac{(x^2-b_1^2)(x^2-b^2)}{(a^2-x^2)}}\times\nonumber\\
& \left[
\frac{E_0}{\pi^2}\int_{b_1}^a\frac{sds}{s^2-t^2} \sqrt{\frac{a^2-s^2}
{(s^2-b_1^2)(s^2-b^2)}}\right. \nonumber \\
 & \left. -2\frac{N_0}{\pi} \int_{b_2}^{b_1}\frac{sds}{s^2-t^2} \sqrt{\frac{(b_1^2-s^2)(s^2-b^2)}{a^2-s^2}} \right].
\label{N1}\\
&  n(x, b <|x|<b1) =- \sqrt{\frac{(b_1^2-^x2)(x^2-b^2)}{(a^2-x^2)}}\times\nonumber\\
& \left[
\frac{E_0}{\pi^2}\int_{b_1}^a\frac{sds}{s^2-t^2} \sqrt{\frac{a^2-s^2}
{(s^2-b_1^2)(s^2-b^2)}}\right. \nonumber \\
 & \left. -2\frac{N_0}{\pi} \int_{b_2}^{b_1}\frac{sds}{s^2-t^2} \sqrt{\frac{(b_1^2-s^2)(s^2-b^2)}{a^2-b^2}} \right].
\label{N2}
\end{eqnarray}

Using Eqs.(\ref{dispersion}),\ref{E1}),(\ref{E2}),(\ref{N1}) and (\ref{N2}), one can check that 
the electric field and the charge density in the regions where they are set by Eqs.(\ref{a}), (\ref{b}) 
and (\ref{c1}) are given precisely by these values. Also, the complex function $E(t)$  is indeed analytic and satisfies the condition ({\ref{d}). 
Evaluation of integrals in (\ref{E1}),(\ref{E2}) and Eqs. (\ref{a},\ref{b}), 
yields the resulting electric field profile given by 
\begin{eqnarray}
& E_x=-\left[\frac{2E_0}{\pi}
\left(\frac{\sqrt{a^2-b_1^2}}{\sqrt{a^2-x^2}}\ln{\frac{y}{4}}+ \arctan{\frac{\sqrt{a^2-b_1^2}}{\sqrt{a^2-x^2}}}\right)\right.\nonumber \\
& +\left.2\pi N_0\sqrt{\frac{a^2-b_1^2}{x^2-a^2}}\right]sign(x)\hspace{3mm}\mbox{if $|x|>a$}\\
& E_x=E_0 sign(x)\hspace{5mm} \mbox{if  $b_1<|x|<a$}\\
& E_x=\frac{2E_0}{\pi}\left[\sqrt{\frac{a^2-b_1^2}{a^2-x^2}}\left(\frac{\pi}{2}-\arctan{\frac{b_1^2-x^2}{x^2-b^2}}\right)\right.+\nonumber\\
& \frac{x^2-b^2}
{b_1^2-x^2}\left(\ln{\frac{\sqrt{a^2-x^2}}{\sqrt{a^2-x^2}+\sqrt{a^2-b_1^2}}}+\right.
\nonumber\\
& \left.\left.\sqrt{\frac{a^2-b_1^2}{a^2-x^2}}\ln{2}\right)\right]sign(x) \hspace{5mm} \mbox{if $ b<|x|<b_1$}
\label{trans}\\
& E_x=0 \hspace{5mm} \mbox{if $|x|<b$},
\label{electricfield}
\end{eqnarray}
where 
\begin{equation}
y=\sqrt{(b_1^2-b^2)/(a^2-b^2)}.
\label{y}
\end{equation}

At large $|x|\rightarrow\infty$, the asymptotic expansion of the electric field gives, on the one hand,
\begin{equation}
|E_x|(|x|\rightarrow\infty)=\frac{\sqrt{a^2-b_1^2}}{|x|}\left(  \frac{2E_0}{\pi}\ln{\frac{ye}{4}}+2\pi N_0\right).
\label{asymp}
\end{equation}
On the other hand, the magnitude of electric field at $|x|\rightarrow\infty$ is related the electric charge in the sample (see e.g.,\cite{smythe}),
\begin{equation}
|E_x|(|x|\rightarrow\infty)=\frac{2q}{|x|}=0, 
\label{relation}
\end{equation}
where  $q$ is the linear charge density $q$, i.e., the charge per unit length in $y-$direction, which vanishes for an electrically neutral sample.

From Eqs.(\ref{asymp}, \ref{relation}), we determine $b_1$:
\begin{equation}
b_1=b+\frac{8(a^2-b^2)}{e^2b}\exp{-\left[\frac{2\pi^2 N_0}{E_0}\right]}.
\label{b1}
\end{equation}
We see indeed that the constraint on charge density turns out to be important almost at a single point $b_1$, which merges with $b$ with an exponential accuracy. However, 
neglecting the constraint would result in logarithmic divergency of the charge density. 
Evaluating the integrals in Eqs.(\ref{N1}) and (\ref{N2}) taking the constraint into account gives 
the charge density profile:
\begin{eqnarray}
& n=\frac{E_0}{\pi^2}\left(\frac{1}{2}\ln{\frac{x^2-b_1^2}{a^2-x^2}}+\right.
\sqrt{\frac{(a^2-b_1^2)(x^2-b^2)}{(a^2-x^2)(x^2-b_1^2)}}\ln{2}\nonumber\\
& \frac{1}{2}\sqrt{\frac{x^2-b^2}{x^2-b_1^2}}\ln{\frac{a^2-x^2}{(\sqrt{a^2-x^2}+\sqrt{a^2-b_1^2})^2}}-\nonumber\\
& \left. \sqrt{\frac{a^2-b_1^2}{a^2-x^2}}
\ln
{
\frac
{\sqrt{x^2-b_1^2}\sqrt{b_1^2-b^2}}
{(\sqrt{x^2-b^2}+\sqrt{x^2-b_1^2})\sqrt{a^2-b_1^2}}
}
\right)
-N_0\sqrt{\frac{a^2-b_1^2}{a^2-x^2}} \nonumber \\
& if \hspace{4mm} b_1<x<a
\label{n1}
\end{eqnarray}
\begin{eqnarray}
n=-N_0 & if \hspace{4mm} b_2< x< b_1
\label{n0}
\end{eqnarray}
\begin{eqnarray}
& n=\frac{E_0}{\pi^2}
\left(\frac{1}{2}\ln{\frac{b_1^2-x^2}{a^2-x^2}}+
\sqrt{\frac{(a^2-b_1^2)(b^2-x^2)}{(b_1^2-x^2)(a^2-x^2)}}\ln{2}\right.\nonumber\\
& +\frac{1}{2}\sqrt{\frac{b^2-x^2}{b_1^2-x^2}}\ln{\frac{a^2-x^2}{(\sqrt{a^2-x^2}+
\sqrt{a^2-b_1^2})^2}}+ \nonumber\\
&- \left. \sqrt{\frac{a^2-b_1^2}{a^2-x^2}}
\ln
{
\frac
{\sqrt{x^2-b_1^2}\sqrt{b_1^2-b^2}}
{(\sqrt{b^2-x^2}+\sqrt{x^2-b_1^2})\sqrt{a^2-b_1^2}}
}
\right)
-N_0\sqrt{\frac{a^2-b_1^2}{a^2-x^2}}\nonumber\\
& if \hspace{4mm} x<b_2
\label{n2}
\end{eqnarray}
Electric field and charge density profile are presented in Fig.~\ref{Fig14} and Fig.~\ref{Fig15} for case I parameters given in Sec.VI.

\section*{Appendix B \\
Spin-orbit constants and interactions in various systems}

In this appendix we give the values of spin-orbit scattering constants. We will present a discussion of 
spin-orbit effects in metals, bulk semiconductors and 2D electron gas in semiconductor quantum wells. 
Spin-orbit interactions have been treated in numerous monographs and textbooks. 
Nevertheless, the phenomenon we deal with in 
the present paper requires a special vantage point: how to suppress the spin-orbit effects resulting in spin relaxation of the $z-$component of spin while keeping 
the part of spin-orbit interactions 
responsible for the spin Hall current intact. This is our primary 
focus here, which requires development of theory of spin-orbit scattering in quantum wells, as well as re-consideration and clarification of several facts about spin-orbit interactions, including intrinsic spin-orbit effects, e.g., of the Rashba type.

\subsection*{\it Magnitudes of constants describing electron spin-orbit interactions in bulk systems}

\subsubsection*{\it Spin-orbit constants for relativistic electrons and electrons in metals} 

In relativistic quantum mechanics,  
the spin-orbit scattering potential in the Schroedinger-Pauli equation arises as a result of spin-dependent 
admixture of positron states to electron states by coupling proportional to the speed of light and 
the electron momentum in the Dirac Hamiltonian. The standard hand-waving explanation of the spin-orbit coupling 
is that a moving electron 
experiences the magnetic field $v{\mathbf \nabla} U/c$ that couples to its magnetic moment. This explanation results in 
almost the same 
magnitude of the relativistic spin-orbit constant as the rigorous derivation from the Dirac 
equation, except it is a factor of two higher\cite{Drell}. 

For ordinary metals with a broad conduction band and the Fermi energy $\sim 10eV$, the resulting spin-orbit constant 
for electrons has essentially a relativistic value $\alpha= \hbar/4m^2c^2$. Electron-electron 
interactions in a Fermi liquid may somewhat renormalize this constant, in much the same way they renormalize the effective mass 
of electrons in metals. Modification of the magnitude of the spin-orbit constant due to an admixture of states 
with different symmetry to electronic states (similar to the effect that we will discuss below for semiconductors) is 
very small for metals. Most importantly, 
in simple metals, like Al, Cu, or Au, the spin-orbit interaction is isotropic, i.e., the relaxation of spins $\sigma_x$, $\sigma_y$ and $\sigma_z$ is defined by the same time constant as in Eq.(\ref{srelax}). 
In order to be able to suppress spin relaxation for one component of spin but to keep the terms responsible for the spin current, significant anisotropy of spin-orbit interactions is neccessary, which is 
not the case for simple metals. We need the spin-orbit interaction given by Eq.(\ref{H}) be dominant, while similar terms containing $\sigma_x$ and $\sigma_y$ be small. This  might be possible in a two-dimensional metal, but such structures similar to the 2D semiconductor systems have not been realized. The apparent reason for difficulties in designing metallic quantum wells, in which only one ground level of size quantization is occupied, is a very high electron density in metals. 

\subsubsection*{\it Spin-orbit interactions in semiconductors, including heterogeneous media}

In semiconductors, the spin-orbit constants are significantly different from those in metals. 
The solution of the Schroedinger-Pauli equation in a periodic potential $V({\mathbf r})$
results in a group of bands, which are energetically close to each other. In particular, in the Kane model 
\cite{Kane} describing III-V systems,
the conduction band $| c\rangle$ (group theory notation $\Gamma_6$) is separated from the top of 
the valence band 
$|\Gamma_8\rangle$ by  $E_g$ just $\sim$1eV. The band $\Gamma_8$ with the total angular momentum 3/2 is separated 
from electron states in the split-off valence 
band $|\Gamma_7\rangle$ with total angular momentum 1/2 by the spin-orbit splitting $\Delta$
\begin{equation}
\Delta=-\frac{3}{4}\frac{i\hbar}{4m^2c^2}\langle x|({\mathbf \nabla}V({\mathbf r})\times {\mathbf p})_y|z\rangle,
\end{equation}
 where functions $\langle x|$, $\langle y|$, $\langle z|$ are orthonormalized functions that behave like 
vector components $x,y,z$ upon the symmetry transformations of the crystalline point group. The spin-orbit 
splitting in 
III-V systems can be smaller or bigger than the band gap. Furthermore, in much the same way as the Dirac electrons 
are coupled to the Dirac positrons by spin-dependent coupling, the conduction band electrons are coupled 
to the valence band electrons, so that the Kane Hamiltonian ${\cal H}^{K}$ matrix elements are given by:
\begin{eqnarray}
{\cal H}^{K}_{c, 1/2;\Gamma_8, 1/2 }= \sqrt{\frac{2}{3}}k_z P \nonumber \\
{\cal H}^{K}_{c, 1/2;\Gamma_8, -1/2 }= -\frac{1}{\sqrt{3}}k_- P\nonumber \\
{\cal H}^{K}_{c, 1/2;\Gamma_8, 3/2 }=  -k_+ \nonumber \\
{\cal H}^{K}_{c, 1/2;\Gamma_8, -3/2 }= 0\nonumber \\
{\cal H}^{K}_{c, -1/2;\Gamma_8, 1/2 }= -\frac{1}{\sqrt{3}}k_+ P\nonumber \\
{\cal H}^{K}_{c, -1/2;\Gamma_8, -1/2 }= \sqrt{\frac{2}{3}}k_z P\nonumber \\
{\cal H}^{K}_{c, -1/2;\Gamma_8, 3/2 }= 0\nonumber \\
{\cal H}^{K}_{c, -1/2;\Gamma_8, -3/2 }= k_- P\nonumber \\
{\cal H}^{K}_{c, 1/2;\Gamma_7, 1/2 }= -\frac{1}{\sqrt{3}}k_z P\nonumber \\
{\cal H}^{K}_{c, 1/2;\Gamma_7, -1/2 }= \sqrt{\frac{2}{3}}k_z P\nonumber \\
{\cal H}^{K}_{c, -1/2;\Gamma_7, 1/2 }= -\sqrt{\frac{2}{3}}k_-P\nonumber \\
{\cal H}^{K}_{c, -1/2;\Gamma_7, -1/2 }= \frac{1}{\sqrt{3}}k_+P,
\label{Kane}
\end{eqnarray}
where the Kane matrix element $P=\frac{i\hbar}{m}\langle S | p_x |X\rangle$, 
$ |S\rangle$ is the orthonormalized function 
that behaves like 
a spherical function upon the symmetry transformations of the crystalline point group, $ |X\rangle$ is the orthonormalized 
wavefunction that behaves like an $x$-component of the coordinate upon these symmetry transformations, and
$k_{\pm}=(k_x\pm ik_y)/\sqrt{2}$. 
When the Kane parameter changes with the coordinate across the heterostructure, 
the order of operators $k_z$ and $P(z)$ in Eq.(\ref{Kane}) is important.  Below we will discuss the choice made for the order of these operators in Eq. (\ref{Kane}) and its consequences for the form of the conduction band Hamiltonian, 
the effective mass, 
the spin-orbit constant and for the boundary conditions at heterostructure interfaces. At this stage,
it is important that such choice results in the corresponding form of the Hermitian conjugated terms 
of the Kane Hamiltonian, e.g.:
 \begin{equation} 
{\cal H}^{K}_{\Gamma_8, 1/2; c, 1/2 }= \sqrt{\frac{2}{3}}P k_z .
\end{equation}
The spin-orbit interaction in the conduction band of bulk semiconductors arises in the third order 
perturbation theory.  
We shall see here that the magnitude of 
spin-orbit interactions is affected by
a spatial dependence of the electron mass,  $P(z)$  and $\Delta (z)$. 
Then, both the second and third order perturbation theory terms are important for us. 
In the second order, the kinetic energy operator associated with the $z-$direction motion becomes:
\begin{equation}
{\cal H}^{kin}_z=-\frac{1}{3}\frac{\partial}{\partial z}P(z)\Big[\frac{2}{E_g(z)}+
\frac{1}{E_g(z)+\Delta (z)}\Big]P(z)\frac{\partial}{\partial z}.
\label{z-kinetic}
\end{equation}
We see that our choice of ordering of the operators in the Kane model leads to a form of kinetic energy that is described 
exclusively in terms of the position-dependent Kane effective mass given by
\begin{equation}
\frac{m}{m_c(z)}= \frac{2m}{3\hbar^2}P(z)\Big[\frac{2}{E_g(z)}+
\frac{1}{E_g(z)+\Delta (z)}\Big]P(z).
\label{pdmass}
\end{equation} 
The total kinetic energy, assuming that the effective electron mass changes along the growth drection only, is
\begin{equation}
{\cal H}^{kin}= \frac{\hbar^2}{2m_c(z)}(k_x^2+k_y^2) - 
\hbar^2 \frac{\partial}{\partial z}\frac{1}{m_c(z)}\frac{\partial}{\partial z}.
\end{equation}
A choice of any other ways  of ordering of $P(z)$ and $\frac{\partial}{\partial z}$ in the Kane operator 
would result in description of motion of electrons in $z-$direction 
by three varying parameters ($P(z)$, $E_g(z)$, and $\Delta (z)$), instead of a single parameter $m_c(z)$.
It is noteworthy that electrons in the conduction band with varying effective 
mass have been often described in the literature by the Hamiltonian, see, e.g., \cite{Morrow}
\begin{equation}
{\cal H}^{kin}_z= m_c^{\alpha}\frac{\partial}{\partial z}\frac{1}{(m_c(z))^{1+2\alpha}}\frac{\partial}{\partial z}m_c^{\alpha}.
\end{equation}  
Therefore we see that the Kane model leads to such description only in the case $\alpha=0$, which corresponds to 
our choice of ordering operators. 
We now consider the spin-dependent part of the electron Hamiltonian.
The admixture of the valence band to conduction band electrons in the Kane model gives
\begin{eqnarray}
{\cal H}_{so}= -\frac{1}{3}\big [ \nabla_{\mathbf r} \frac{P^2(z)}{E-U_v({\mathbf r})}
\times {\mathbf k}\big ] \cdot {\mathbf \sigma} +\nonumber\\
\frac{1}{3}\big [ \nabla_{\mathbf r} \frac{P^2(z)}{E-U_v({\mathbf r})-\Delta (z)}\times {\mathbf k}\big ]
\cdot {\mathbf \sigma},
\label{generalso} 
\end{eqnarray}
where $E$ is the electron energy, the first term comes from $\Gamma_8$ bands and the second term originates from the split-off band. 
The spin-orbit coupling (\ref{generalso}) is defined by such non-relativistic parameters such as the Kane matrix 
element $P$ and the band gap $E_g$. It is related to relativistic spin-orbit coupling via spin-orbit 
splitting of the valence band $\Delta$ and vanishes in the limit of zero $\Delta$. 
The magnitude of spin-orbit interaction 
critically depends on the ratio of $E_g$ and $\Delta$. 
An important feature of (\ref{generalso}) is its dependence on the potential 
$U_v({\mathbf r})$ that differs from potential acting on conduction electrons. This feature attracted attention 
in \cite{Lassnig} and was discussed in \cite{Winkler} in connection with the consideration of 
the Rashba term in the 2D electron gas.
As we shall see, the story has much broader implications, and this feature is crucial for understanding the 
constant of spin-orbit scattering in quantum wells, spin-orbit coupling due to long-range fluctuations, and the possibility to suppress both spin-orbit scattering and various
Rashba effects.

\subsubsection*{Effective 3D Hamiltonian for abrupt heterojunctions. Boundary conditions.}

Our primary interest are spin-orbit effects in symmetric quantum well structures with 
abrupt heterojunctions. 
We will first obtain "3D" Hamiltonian describing such systems, and then derive the 
2D effective Hamiltonian by averaging over the wavefunction in the ground quantum well subband.  
The potential affecting conduction electrons is given by Eq.(\ref{c}), the potential for $\Gamma_8$ valence electrons 
is given by  Eq.(\ref{c}), and for $\Gamma_7$ electrons by Eq.({\ref{7}).  
Assuming the coordinate-dependent parts of the potential affecting valence electrons to  
be much smaller than $E_g^w$ and $E_g^w +\Delta_w$, we find
\begin{eqnarray}
& {\cal H}_{so}= -\frac{1}{3} \big [  \nabla_{\mathbf r}\big( P^2(z) U_v ({\mathbf r})\big )
\times {\mathbf k}\big ]
\cdot {\mathbf \sigma} [\frac{1}{(E_g^w)^2}-\frac{1}{(E_g^w+\Delta_w)^2}]\nonumber\\
& +\frac{1}{3(E_g^w+\Delta_w)^2}  \nabla_z\Big[ P^2(z)\delta(\Delta(z)) \Big ]\cdot
\big( {\mathbf k}\times {\mathbf \sigma}\big)_{z} \nonumber\\
&-\frac{1}{3} \nabla_z\big( P^2(z))[\frac{1}{E_g^w}-\frac{1}{E_g^w+\Delta_w}] \big )
\cdot\big [{\mathbf k}\times  {\mathbf \sigma}\big ]_z.
\label{qwso3}
\end{eqnarray}
We observe that the spin-orbit Hamiltonian includes three contributions.
The first term in (\ref{qwso3}) is a generalization of standard 3D spin-orbit interactions 
\begin{equation}
H^{so}=\alpha {\mathbf \sigma}\cdot [ \nabla_{{\mathbf r}}U_v({\mathbf r})\times {\mathbf k}],
\label{so}
\end{equation}
with the constant $\alpha$ given by
\begin{equation}
\alpha= -\frac{P^2}{3}
\Big ( \frac{1}{E_g^2}- \frac{1}{(E_g+\Delta)^2}\Big )
\label{constso}
\end{equation}
to a heterogeneous 
media with a varying Kane constant. We will use these equations to describe spin-orbit effects in the 
approximation neglecting all differences in materials constants in the well and the barriers, except for $E_g$. Then the materials parameters in Eqs.(\ref{so}, \ref{constso}) will correspond to their values in the quantum well. 
 
The spin-orbit coupling associated with the first term in Eq. (\ref{qwso3}) 
arises from coupling of the conduction and the valence bands 
in a manner similar to 
the appearance of the relativistic spin-orbit coupling from electron-positron coupling in the Dirac Hamiltonian. 
This is a third order perturbation theory term: the first stage of the process comes from coupling of the 
conduction and the valence bands given by Eq.(\ref{Kane}), the second stage is an action of the varying potential 
$U_v({\mathbf r})$ (including offsets)
acting on the valence electrons, and the final stage couples 
valence bands back to the conduction band. 
The specific form of the first term (\ref{qwso3}), i.e., the order of differential operators 
and $P(z)$ is defined by our 
choice of ordering of these operators in the Kane model. 
The second term in (\ref{qwso3}) comes from the difference of the energy separation between valence bands $\Gamma_8$ and $\Gamma_7$ in the quantum well and in the barriers. 
Finally, the third term arises in the second order perturbation theory and is associated with the difference 
of the Kane matrix element $P$ in the barriers and the well. The specific form of this term is also 
the consequence of our choice of ordering of the operators in the Kane model.

We now discuss how the ordering of operators in the Kane Hamiltonian affects the boundary conditions for the wavefunction and spin-orbit interactions.
As discussed in \cite{IP},\cite{foreman},\cite{ER}, 
the general way to order operators in the Kane model is to write down the matrix elements in the 
form,
\begin{equation}
{\cal H}^{K}_{c, 1/2;\Gamma_8, 1/2 }= \sqrt{\frac{2}{3}}P^{\alpha}k_z P^{1-\alpha},
\label{Kalpha}
\end{equation} 
with $0\le\alpha\le 1$. The multiband wavefunction in the Kane model is constructed, 
according to Suris \cite{suris}, in the form
\begin{equation}
\psi= u|S\rangle +{\mathbf v}\cdot |{\mathbf R}\rangle,
\end{equation}
with 
\begin{equation}
u=\left [ \begin{array}{c}
u_{1/2}\\u_{-1/2}
\end{array} \right ],
\end{equation}
and
\begin{equation}
{\mathbf v}=\left [ \begin{array}{c}
{\mathbf v}_{1/2}\\{\mathbf v}_{-1/2}
\end{array} \right ],
\end{equation}
where $u_l$ and $v_{il}$ $(l=\pm 1/2, i=x,y,z)$, ${\mathbf R}_x= X$, ${\mathbf R}_y= Y$, and ${\mathbf R}_z= Z$.
Then the continuity equation is given by
\begin{equation}
\frac{\partial}{\partial t}( |u|^2 + |{\mathbf v}|^2 ) + Pdiv (u^+{\mathbf v} + u{\mathbf v}^+)=0.
\end{equation}
The conserved flux is 
\begin{equation}
I=P(u^+{\mathbf v} + u{\mathbf v}^+),
\end{equation}
and $P^{\alpha} u$ and $P^{1-\alpha}v_z$ are continuous at the boundaries, where $z$ is the normal 
to the boundaries. Then, relating $u$ and ${\mathbf v}$ from the Kane equation 
${\cal H}^{K}\psi=E\psi$ , and using the 
continuity of $P^{1-\alpha}v_z$, we obtain continuity  of the quantity
\begin{eqnarray}
& P^{2-\alpha}\left [\left ( \frac{2}{E+E_g}+\frac{1}{E+E_g+\Delta}\right )\frac{\partial u}{\partial z}-\right.\nonumber\\
& \left.\frac{\Delta [{\mathbf \sigma} {\mathbf k}]_z u}{(E+E_g)(E+E_g+\Delta)}\right]
\label{sur}
\end{eqnarray}
at the boundaries. At $\alpha \ne 0$,  $u$ is not continuous. However, there is no contradiction here with general quantum-mechanical 
principles, because, at an abrupt heterojunction, the total wavefunction constitues a mixture 
of the wavefunction of many bands, and only a total wavefunction is subject to continuity and flux continuity, rather than
its individual components, such as $u$. An example of boundary condition in the Kane model, in which $u$ is continuous but 
 $(du/dz)/m_c$ is not, was devired in\cite{sham}. 
Remarkably, our choice of ordering in the Kane Hamiltonian with $\alpha = 0$ at $E=0$, as follows from Eq.(\ref{sur}),  results in the continuity 
of $u$ and $(\partial u/\partial z)/m_c$ themselves, i.e., in
the natural and traditional 
boundary conditions in the quantum well. We also note that the model considered here for coupling of 
conduction and valence electrons is a "pure" Kane model (a model with no bare diagonal terms quadratic in $k$  
with different masses in the conduction and valence bands). Therefore the choice $\alpha = 0$ is fully consistent,
and there is no reason for the asymmetric choice $\alpha = 1$, such as made in modified Kane model \cite{foreman}. 

We also note that for every choice 
of $\alpha$, our principal conclusions about spin-orbit coupling will hold. 
We see that in order 
to reduce spin-flipping terms, materials in the quantum well and the barrier should have close values of materials constants. 

Concluding this part we remark that in semiconductor physics, the five band and seven band models 
are sometimes used, which are more 
elaborate than the Kane model. 
In these cases,
it is possible to obtain contributions to spin-orbit coupling in which ${\mathbf \sigma}$ and 
$\nabla_{{\mathbf r}}U_v({\mathbf r})$ are coupled to an expression cubic (or higher power) 
in components of the momentum operator. 
These terms are much smaller than spin-flipping terms in (\ref{so}), 
and alter the symmetry of the Hamiltonian in a minimal way,
not essential for our purposes.
   
\subsection*{\it Spin-orbit effects in the ground conduction band subband of 2D electrons in quantum wells} 

Having derived the 3D Hamiltonian (\ref{generalso},\ref{qwso3}), we now proceed to describing the 
spin-orbit coupling for 2D electrons.  
We will temporarily ignore the effect of different masses, $P$ and $\Delta$ 
in the barrier 
and the quantum well, which will be taken into account in Appendix C, and assume that the only feature 
of a heterostructure 
is the potential confinement characterized by the offsets in the conduction and valence bands.

\subsubsection*{The effective Hamiltonian for quantum wells}
The most general potential describing charge carriers in quantum wells is given by
\begin{equation}
U_v({\mathbf r})= U(z) + V(x,y,z),
\label{potential1} 
\end{equation}
where 
\begin{equation}
U(z)=U_{sym}(z)+U_{asym}(z)
\label{zpotential}
\end{equation}
 describes the potential confinement of
charge carriers to two dimensions, and is uniform in $x,y$. The asymmetric potential $U_{asym}(z)$
can include an external electric field applied along $z$.
The potential (\ref{zpotential}) must result in a confined state or states, so that no free motion or  
escape into the delocalized state along the $z-$direction takes place.
The separation of $U(z)$ onto symmetric and  antisymmetric components is convenient in the case of the double 
heterostructure rectangular quantum well, in which components of $U(z)$ are symmetric or antisymmetric 
with respect to the centerline of the QW at $z=0$.
The second term in Eq.(\ref{potential}) can describe charge carrier scattering, 
fluctuating electric fields due to 
large-scale fluctuations in impurity density distribution, or the effects of fluctuations 
of the quantum well width by one or two monolayers. In two dimensions, we assume that 
$V(x,y,z)$ does not result in 
confinement in 
transverse ($x$ or $y$) directions, so that the in-plane motion of charge carriers is almost free, and is
subject to effects of scattering by $V(x,y,z)$.    

The closest situation to an ideal two dimensional case 
occurs when electrons occupy just one level
of size quantization along the $z$ direction. Actual transitions of charge carriers to higher size quantization 
levels at low densities, such that the Fermi energy 
lies below the bottom of the band originating from the first excited level of the quantum well, 
require energy conservation, i.e., phonons, 
and are negligible at low 
temperatures. Thus, scattering leaves electrons on the same ground size quantization level. However, as 
we shall see shortly, due to a dependence of scattering potential on $z$, an admixture of excited 
electron states in the confining potential $U_{sym}(z)+U_{asym}(z)$ to the ground state plays an important role.

The ideal 2D spin-orbit interactions described by Eq.(\ref{H}) correspond to the case of $V(x,y,z)\equiv V(x,y)$, 
i.e., when no 
dependence of random potential on the third dimension takes place. As we neglect for now the 
difference of masses,
$P$ and $\Delta$ in barriers and the quantum well, our starting point here is Eq. (\ref{so}). 
 
In general, the Hamiltonian (\ref{so}) is not reduced to the Hamiltonian (\ref{H}) in two dimensions. Each impurity has a certain location in a quantum well along $z$. From a symmetry standpoint, this
provides a vector lowering the 
symmetry of the system and results in symmetry allowed spin-dependent invariants in the 
Hamiltonian of the system. 
In a rectangular quantum well of a double heterostructure, 
the potential of impurities is neither symmetric nor antisymmetric with respect to $z=0$. 
As a result, the Hamiltonian responsible for electron scattering      
in general includes not only $\sigma_z$ but also $\sigma_x$ and $\sigma_y$ and can lead to relaxation of the $z$-component of spin. 
 
We take a realistic potential of Eq.(\ref{potential1}) and go 
beyond traditional approximations of a scalar potential being a sum of two functions, one depending on 
the growth axis coordinate and the other depending only on the in-plane coordinate. Furthemore, scattering potential in general is not even a product of such two functions. We begin with the derivation of the {\it spin-independent} scattering in two dimensions. 
The Schroedinger equation is 
\begin{equation}
[\frac{p_{\perp}^2}{2m}+\frac{p_z^2}{2m}+U(x,y,z)]\Psi(x,y,z)=E\Psi(x,y,z).
\label{Sch}
\end{equation}
We represent the wavefunction in a form 
\begin{equation}
\Psi(x,y,z)=\Psi_{xy}(z)\Phi(x,y)
\label{psi}, 
\end{equation}
where
\begin{equation}
[\frac{p_z^2}{2m}+U_{sym}(z)+U_{asym}(z)+ V(x,y,z)]\Psi_{xy}(z)=E_{xy}\Psi_{xy}(z)
\label{z}
\end{equation}
is the Schroedinger equation for $z$-direction at the fixed values of in-plane coordinates 
treated as parameters, $E(x,y)$ is an eigenenergy, and $\Psi_{xy}(z)$ is the wavefunction,
with the normalization condition $\int_{-\infty}^{+\infty}dz |\Psi_{xy}(z)|^2=1$. 
The 2D problem is then separated into the two parts. First we need to find $E(x,y)$ and $\Psi_{xy}(z)$.
Our interest is the ground state solution of (\ref{z}), 
so $E(x,y)$ and $\Psi_{xy}(z)$ from now on 
refer to the ground level.
We insert (\ref{psi}) into (\ref{Sch}), multiply its both parts by 
$\Psi_{xy}^*(z)$, and integrate over the coordinate $z$, 
thus obtaining the Schroedinger equation for 2D electrons
\begin{equation}
[\frac{p_{\perp}^2}{2m}+ E_{xy}+P(x,y)+ L_{xy}] \Phi(x,y)=E\Phi(x,y),
\label{inplane}
\end{equation}
where $P(x,y)=({\mathbf p}_{\perp}\cdot {\mathbf A}^{a}_{\perp}+
{\mathbf A}^{a}_{\perp}\cdot{\mathbf p}_{\perp})/2m$ describes an in-plane motion in the presence of 
an effective vector potential
$ {\mathbf A}^{a}_{\perp}=i\int_{-\infty}^{\infty} dz 
\Psi_{xy}^*(z)\partial/\partial {\mathbf r}_{\perp} 
\Psi_{xy}(z)$. This vector potential and the associated magnetic field 
${\mathbf B}^{a}=curl {\mathbf A}^{a}$ vanish if 
the wavefunction $\Psi_{xy}(z)$ is real. For localized states in the absence of 
magnetic field and absence of 
additional non-trivial dependence of $V(x,y,z)$ on spin (coming, e.g., 
from magnetic textures or magnetic impurities), 
this is indeed the case. Furthermore, in Eq.(\ref{inplane}),
 $L_{xy}=-\frac{1}{2m}\int_{-\infty}^{\infty} dz 
\Psi_{xy}^*(z)\partial^2/\partial {\mathbf r}_{\perp}^2 
\Psi_{xy}(z)$ is an effective scalar potential. Both $P(x,y)$ nd $L(x,y)$ are induced by scattering 
modulation [Eq.(\ref{z})] of the wavefunction 
of the confined state. These terms are small only if the modulation can be considered 
slow in adiabatic approximation. However here such adiabatic approximation is not essential. 
We note that it is easy to generalize Eq.(\ref{inplane}) to an effective matrix Hamiltonian that describes 
several excited states $\ell$ of the Hamiltonian (\ref{z}) in addition to the ground state, and 
takes into account both diagonal and off-diagonal terms of the 2D Hamiltonian in the 
basis of wavefunctions $\Psi^{\ell}_{xy}(z)$. 

\subsubsection*{Spin-orbit interactions in the quantum well. Offsets contribution}

The effect of spin-orbit interaction on the in-plane motion arises due to
an additional term, $S(x,y)$, 
in the the Hamiltonian Eq.(\ref{inplane}) when the 3D spin-orbit term,e.g., Eq.(\ref{so}), is averaged over the 
$z$-direction,
\begin{equation}
 S(x,y)= \int_{-\infty}^{\infty} dz 
\Psi_{xy}^*(z)H^{so} 
\Psi_{xy}(z)
\label{soperturb}
\end{equation}

If $H^{so}$ is strong by some reason, the spin-orbit term must be treated not as a  
perturbation (\ref{soperturb}), but rather included into $V(x,y,z)$. Then, $L(x,y)$, $P(x,y)$, and 
$E(x,y)$ will include the associated spin-orbit effects. Here we use the perturbative scheme that includes 
spin-orbit effects via Eq.(\ref{soperturb}).

We now consider the spin-orbit interactions in a particular case of scattering in 
a symmetric rectangular quantum well of a finite depth. This results in 
a single source of spin-orbit effects, the scattering potential. 
We will easily generalize our result to the case of an arbitrary potential confinement.
Because we are interested in a scattering problem, 
in order to average $H^{so}$ over the ground state of the Hamiltonian (\ref{z}), we will use the perturbation 
theory. In linear in $V(x,y,z)$ approximation, it is important to include two contributions: in (i) $V(x,y,z)$ enters $H_{so}$; in (ii)  $V(x,y,z)$ enters $\Psi_{xy}(z)$, while 
$H_{so}$ is defined by Eq.(\ref{zpotential}) or just by symmetric potential confinement. In the first order in scattering 
\begin{equation}
\Psi_{xy}(z)=\Psi^{0}(z)+\sum_n \frac{V_{0n}(x,y)}{E_0-E_n}\Psi^{n}(z),
\label{wavefunction}
\end{equation}
where $\Psi^{n}(z)$ is the solution to the Schroedinger equation describing the $z-$direction only
\begin{equation}
{\cal H}^{z}\Psi_n(z)=[\frac{p_z^2}{2m}+U(z)]\Psi_n(z)=E_n\Psi_n(z),
\label{noimp}
\end{equation}
where index $n$ includes both integers $n=0,1,2...$ corresponding to confined states of 
a discrete spectrum and
a continuum of wavevectors corresponding to the continuous spectrum. The summation over $n$ in Eq.(\ref{wavefunction}) includes integral 
over continuum of states and sum over the discrete spectrum. 
In the case when $U(x,y,z)= U(z) + V_{scat}(x,y,z)$ is the sum of the quantum well confinement potential and a scattering potential, for electrons with energy $E_1$ on the ground level of the size quantization, the effective 
2D potential $E_{xy}$ is
\begin{equation}
E_{xy}= E_1 + \int dz \Psi_{xy}^*(z) V_{scat}(x,y,z)\Psi_{xy}(z).
\end{equation}
Averaging over $\Psi_{xy}(z)$ here includes sum over all virtual excited states of the Hamiltonian 
(\ref{noimp}), in accord with Eq.(\ref{wavefunction}). 
 
Then the 2D spin-orbit term reads
\begin{eqnarray}
&{\cal H}_{so}^{2D}=\alpha \int_{-\infty}^{\infty}dz \Psi^*_{x,y}(z) 
[\sigma_z [ \nabla_{{\mathbf r}}U_v({\mathbf r})\times {\mathbf k}]_z + \nonumber \\
& \nabla_z U_v({\mathbf r})[ {\mathbf k}\times {\mathbf \sigma}]_z+
[{\mathbf \sigma} \times \nabla_{{\mathbf r}}U_v({\mathbf r})]_z k_z]]\Psi_{x,y}(z),
\label{2Dso} 
\end{eqnarray} 

The first term in brackets under the sign of integral in (\ref{2Dso}) results in the spin-orbit 
interaction of the same symmetry as the 
ideal 2D case 
spin-conserving interaction (\ref{H}).
 In the first order in $ V_{scat}$ this term is given by
\begin{equation}
{\cal H}_{\sigma_z}^{2D}=\alpha \int_{-\infty}^{\infty}dz
\Psi^{*}_{0}(z)[\sigma_z [ \nabla_{{\mathbf r}}V_{scat}(x,y,z)\times {\mathbf k}]_z\Psi_{0}(z)
\end{equation}

The two other terms in (\ref{2Dso}) potentially can lead to the relaxation 
of $z-$projection of spin. The second terms in brackets, in the case when a potential 
confining charge carriers to the quantum well is asymmetric, is known to result in the Rashba 2D 
term associated with 
this asymmetry. For  $U_v({\mathbf r})$ due to scattering potential, this term produces spin-flip scattering, and 
for $U_v({\mathbf r})$ associated with fluctuations in $\delta$ doping, a random Rashba term arises.
 Furthermore, the third term in brackets, which makes the Hamiltonian Hermitian also contributes to the
relaxation of the $z-$component of spin due to scattering or the random Rashba interaction. 

We now analyze the second term in Eq.(\ref{2Dso}) containiing $\nabla_z U_v({\mathbf r})$. 
Inserting Eq.(\ref{c}) into Eq.(\ref{potential}), 
and Eqs.(\ref{potential},\ref{wavefunction},\ref{v}) 
into Eq.(\ref{2Dso}), we obtain the following contribution to the 2D spin-orbit interaction:
\begin{eqnarray}
& {\cal H}_{\nabla_z}^{2D}=\alpha \Big( \int_{-\infty}^{\infty}dz
 [\Psi^{*}_{0}(z)\nabla_z V_{scat}(x,y,z)\Psi_{0}(z) \Big.+ \nonumber\\ 
& \sum_n \frac{V_{n0}(x,y)}{E_0-E_n}\Psi^{*}_{n}(z)\nabla_z U_v(z)\Psi_{0}(z)+ \nonumber \\
& \Big.\Psi^{*}_{0}(z)\nabla_z U(z)\frac{V_{0n}(x,y)}{E_0-E_n}\Psi_{n}(z)] \Big)
 [{\mathbf k}\times {\mathbf \sigma}]_z  \nonumber \\
&+ T(x,y)
\label{so1}  
\end{eqnarray}
 Here the first term in the integral originates from the gradient of the scattering potential, and the other two terms originate 
from the gradient of the potential confining the valence electrons (even if this potential is symmetric), 
but include the scattering 
potential via the wavefunction (\ref{wavefunction}). In these terms, ${\hat {\mathbf k}}_{\perp}$ is an operator acting on wavefunctions 
$\Phi(x,y)$ in Eqs.(\ref{psi}, \ref{inplane})  The last contribution $T(x,y)$ reads:
\begin{eqnarray}
& T(x,y)= \alpha \int_{-\infty}^{\infty}dz
 \sum_n\Big[ \Big( [{\mathbf k}\times {\mathbf \sigma}]_z \frac{V_{n0}(x,y)}
{E_0-E_n} \Big) \times \Big.\nonumber \\
&\Psi^{*}_{n}(z) \nabla_z U_v(z)\Psi_{0}(z) +
\Psi^{*}_{0}(z)\Big([{\mathbf k}\times {\mathbf \sigma}]_z\frac{  
V_{0n}(x,y)}{E_0-E_n}\Big)\times \nonumber \\
&\Big. \nabla_z U_v(z)\Psi_{n}(z)\Big]. 
\label{so2}
\end{eqnarray}
In $T(x,y)$, the differential operators ${\hat {\mathbf k}}_{\perp}$ in $[{\mathbf k}\times {\mathbf \sigma}]_z$
act only on $V_{n0}(x,y)$, which originates from the effect of scattering 
on the wavefunction of confined electrons. Correspondingly, due to $T(x,y)$, the wavefunction $\Phi(x,y)$ of the 
transverse motion is affected by the operators $i \partial V_{n0}(x,y)/\partial x$ and 
$i \partial V_{n0}(x,y)/\partial y$. 

We now transform the first three terms of Eq.(\ref{so1}).
From Eq.(\ref{z}), we have an identity
\begin{equation}
\nabla_z U(z)= -i[{\cal H}^{z},k_z],
\label{comc}
\end{equation}
where $[A,B]=AB-BA$.
Therefore, from Eq.(\ref{c},\ref{v})
\begin{equation}
\nabla_z U_v(z)= \frac{{\tilde U}_v}{{\tilde U}_c}[{\cal H}^{z},k_z].
\label{comv}
\end{equation}
Then, e.g., the matrix element  $\big (\nabla_z U(z)\big )_{0n}$, in Eq. (\ref{so1}) is given by
\begin{equation} 
\big (\nabla_z U_v(z)\big )_{0n}=\frac{{\tilde U}_v}{{\tilde U}_c}  \big ( k_z \big )_{0n} (E_0-E_n).
\label{me}
\end{equation}
Inserting Eq.(\ref{me}) into Eq.(\ref{so1}), and using the identity
\begin{equation}
\sum_n- [V_{0n} \big ( k_z \big )_{n0}- \big ( k_z \big )_{0n}V^{scat}_{n0} ]=
 \big ( \nabla_z V^{scat} \big )_{00},
\end{equation}
we arrive to the result 
\begin{equation}
{\cal H}_{\nabla_z}^{2D}=\alpha \Big (1-\frac{{\tilde U}_v}{{\tilde U}_c}\Big )\big ( \nabla_z V^{scat} \big )_{00}\cdot
[{\mathbf k}\times {\mathbf \sigma}]_z +T(x,y).
\label{term1+}
\end{equation}

Considering now the $T(x,y)$ term, and applying the identitities of Eq.(\ref{comv}) and the identity
\begin{equation}
 \sum_n [\big ( k_z \big )_{0n}\frac{\partial V^{scat}_{n0}}{\partial {\mathbf r}_{\perp}} 
 ]= 
  \big ( k_z \frac{\partial}{\partial {\mathbf r}_{\perp}} V^{scat} \big )_{00},
\end{equation}
we obtain
\begin{equation}
T(x,y)=\alpha \frac{{\tilde U}_v}{{\tilde U}_c}\big (\big( k_z 
\frac{\partial}{\partial x} V^{scat} \big )_{00} \sigma_y- 
\big ( k_z
\frac{\partial}{\partial y} V^{scat} \big )_{00} \sigma_x \big).
\label{termT}
\end{equation}

Finally, combining Eqs.(\ref{term1+}, \ref{termT}) with the term in Eq.(\ref{2Dso}) containing $k_z$, 
for the spin-flipping terms of the 2D spin-orbit interactions we obtain
\begin{eqnarray}
{\cal H}_{\nabla_z, p_z}^{2D}=\alpha \Big (1-\frac{{\tilde U}_v}{{\tilde U}_c}\Big )\big[ \big ( \nabla_z V^{scat} \big )_{00}
{\mathbf k}\times {\mathbf \sigma}]_z + \nonumber\\
\big ( [{\mathbf \sigma}\times \nabla_{\perp} V^{scat}]_z k_z \big )_{00}
\big ].
\label{spinflip2D} 
\end{eqnarray}
These are spin-orbit interaction terms that lead to relaxation of the $z-$projection of spin, due to 
asymmetric scattering potential $V(x,y,z)$ in the rectangular quantum well due to different offsets in the conduction and valence bands. 
More generally, for the arbitrary shape of the electron quantum confinement, the corresponding spin-orbit terms are defined by Eqs.(\ref{so1},\ref{so2}). From 
Eq.(\ref{comc}), it follows that such scattering-induced terms are non-zero only if the potentials acting on the conduction and valence electrons differ from each other. This generalizes a similar 
result for average Rashba spin-orbit interaction due to one-dimensional asymmetric potential\cite{Winkler} to spin-orbit interactions due to the 3D scattering potential 
or due to long-range fluctuations.

\section*{Appendix C. Effect of the difference of parameters in barriers and the quantum well on spin-orbit interactions in 2D systems}
We now address the contribution to spin-orbit effects associated with the change in the effective masses,
Kane matrix element $P$, and the spin-orbit splitting $\Delta$ at the heterostructure interfaces. 
We will begin with the effect of the kinetic energy term in Eq.(\ref{z-kinetic}). The presence of this term leads to two contributions. First, obviously
a decay of wavefunctions in 
barriers and oscillatory behaviour of the wavefunctions inside the quantum well are described by different
 masses, and this changes the magnitudes of matrix elements in Eq.(\ref{spinflip2D}). Second, an additional non-trivial contribution to spin-orbit terms arises due to modification of Eqs.(\ref{comc},\ref{comv}).
Indeed, in the presence of mass difference described by the Hamiltonian Eqs.(\ref{z-kinetic},\ref{pdmass}),
the gradient $\nabla_z U(\mathbf r)$ is given by
\begin{equation}
\nabla_z U(z)= -i[{\cal H}^{z},k_z]+
\frac{\partial}{\partial z}\frac{1}{m(z)}\frac{\partial m(z)}
{\partial z}\frac{1}{m(z)}\frac{\partial}{\partial z},
\label{grad}
\end{equation}
 and we now take into account the second term in this identity.
Using Eq.(\ref{c}, \ref{v}) we have 
\begin{equation}
\nabla_z U_v(\mathbf r)= \nabla_z U(\mathbf r)\frac{{\tilde U_v}}{\tilde U_c}.
\end{equation}  
Then from Eq.(\ref{so}) the 2D spin-orbit interaction due to mass difference in the quantum well and barriers is given by
\begin{eqnarray}
\delta{\cal H}_{so}^{2D} 
= 
 \alpha \hbar^2 (m_B-m_W) \frac{{\tilde U_v}}{\tilde U_c}\nonumber\\
\Big [\frac{1}{m}\nabla_z \Psi_{xy}^{*}(d/2) [{\mathbf k}\times 
{\mathbf \sigma}]_z \frac{1}{m}\nabla_z \Psi_{xy} (d/2)-\Big.\nonumber\\
\Big.\frac{1}{m}\nabla_z \Psi_{xy}^{*}( -d/2)[ {\mathbf k}\times 
{\mathbf \sigma}]_z\frac{1}{m}\nabla_z \Psi_{xy} (-d/2)\Big ].
\end{eqnarray}
We note that the flux $\frac{1}{m}\nabla_z \Psi_{xy} (z)$ is continuous at the 
interfaces for chosen boundary conditions. Importantly, as a result of scattering or due to the asymmetric potential,
the flux values are different  at $z=-d/2$ and at $z=d/2$.  
We also note that in the presence of scattering potential $V(x,y,z)$, the operator $[{\mathbf k}\times 
{\mathbf \sigma}]_z$ that includes spatial derivatives over in-plane coordinates acts both 
on $\Phi(x,y)$ and on $\Psi_{xy}(\pm d/2)$. The latter in the first order in scattering potential is defined 
by Eq.(\ref{wavefunction}). The sum of the corresponding terms gives a Hermitean spin-orbit 
Hamiltonian. 

When only the 1D asymmetric potential $U_{asym} (z)$ is present, the spin-orbit interaction due to mass difference in the well and barriers contributes to the Rashba term:
\begin{eqnarray}
& \delta{\cal H}_{so}^{2D}= \alpha (m_B-m_W) \frac{{\tilde U_v}}{\tilde U_c}\hbar^2 \nonumber\\
& [\frac{1}{m}\nabla_z \Psi^{*}(d/2) \frac{1}{m}\nabla_z \Psi (d/2)- \nonumber\\
& \frac{1}{m}\nabla_z \Psi^{*}( -d/2) \frac{1}{m}\nabla_z \Psi (-d/2)][{\mathbf k}\times 
{\mathbf \sigma}]_z.
\label{mRashba}
\end{eqnarray}
In a weak asymmetric potential, e.g., due to gate voltage applied to a symmetric quantum well or due to asymmetric doping of symmetric heterostructure, 
in the first order in the asymmetric potential we have
\begin{equation}
 \Psi(z)=\Psi^{0}(z)+\sum_n \frac{U^{asym}_{0n}}{E_0-E_n}\Psi^{n}(z),
\end{equation}
and inserting the second term into Eq.(\ref{mRashba}) results in a Rashba term associated with potential asymmetry. 

For strongly asymmetric quantum wells Eq.(\ref{mRashba})
will contain the difference of fluxes at the interfaces of the effective mass change $z=-d/2$ and $z=d/2$. 
We note that the contribution to spin-orbit terms due to 
different masses will also take place in inversion layers. Such contribution can be calculated 
using general expressions for the spin-orbit terms Eq.(\ref{2Dso}) and Eq.(\ref{z-kinetic}). Strong asymmetry in 
inversion layers will lead to sizable Rashba term as compared to spin-orbit terms that conserve 
the $z-$projection of spin.
  
We now consider the effects described by the second and the third term of Eq.(\ref{qwso3}),
 due to different Kane parameters $P$ and different splitting $\Delta$ of $\Gamma_8$ and 
$\Gamma_7$ 
valence bands.
Averaging these terms over the quantized motion in the ground state of the quantum well gives:  
\begin{eqnarray}
\delta {\cal H}_{so}= [ 
\beta P_b^2(\Delta_b-\Delta_w) + \gamma (P_b^2-P_w^2)
 ] \nonumber\\
 \Big ( (|\Psi_{xy}(d/2)|^2 -|\Psi_{xy}(-d/2)|^2)
[ {\mathbf k}\times{\mathbf \sigma}]_z - \Big.\nonumber\\
i[  \big ( \Psi_{xy}^*(d/2)
\frac{\partial}{\partial {\mathbf r}_{\perp}}\Psi_{xy}(d/2) -\big. \nonumber\\
\Big.\big.\Psi_{xy}^*(-d/2)\frac
{\partial}{\partial {\mathbf r}_{\perp}}\Psi_{xy}(-d/2)  \big ) \times {\mathbf \sigma}]_z\Big )
\label{dif}
\end{eqnarray}
where the constant $\beta$ is given by
\begin{equation}
\beta= \frac{1}{3(E_g^w + \Delta_W)^2},
\label{beta}
\end{equation}
and the constant $\gamma$ is given by
\begin{equation}
\gamma=\frac{1}{3} \Big ( \frac{1}{E_g^w}-\frac{1}{E_g^w + \Delta_w}\Big ).
\label{gamma}
\end{equation}
A difference of the probabilities at the left and right heterointerfaces, or a difference of 
products of the wavefunction and its derivative over in-plane coordinate at the left and 
right heterointerfaces entering Eq.(\ref{dif}) 
is due to scattering or due to an asymmetric potential. When only an asymmetric potential in the quantum well is taken into account, only the first term in brackets of Eq.(\ref{dif}) contributes to spin-orbit interactions.  
In order to calculate such term for small difference of materials parameters, we use  the identity
\begin{equation}
\langle \nabla_z U_c(z)\rangle =\int_{-\infty}^{\infty}dz \Psi^*(z) 
 \nabla_z U_c({\mathbf r})\Psi(z)=0,
\end{equation}
which constitutes the Ehrenfest theorem, 
and obtain
\begin{equation}
|\Psi(d/2)|^2-|\Psi(-d/2)|^2=  -\frac{1}{{\tilde U}_c}
\langle \nabla_z U_{asym}(z)\rangle.
\end{equation} 
We note that only the average of the asymmetric part of the confinement potential, not the full confinement potential, enters this expression. 
For gate voltage applied externally $\langle \nabla_z U_{asym}(z)\rangle= eE_z$, where $E_z$ is the applied electric field. From Eq.(\ref{dif}), we have
\begin{eqnarray}
&\delta {\cal H}_{so}= -\frac{1}{{\tilde U}_c}
[ \beta P_b^2(\Delta_b-\Delta_w) + \gamma (P_b^2-P_w^2)]\times& \nonumber\\
&\langle \nabla_z U_{asym}(z)\rangle
\big [ {\mathbf k}\times{\mathbf \sigma}\big ]_z. &
\label{diffRashba}
\end{eqnarray}
This term due to asymmetric potential defined by the mass and valence band splitting difference in the quantum well and barriers 
has the symmetry of Rashba coupling.


\begin{thebibliography}{999}
\bibitem{mott} N.F Mott and H.S.W. Massey, The Theory of atomic collisions, (Oxford
University press, New York, 1964)
\bibitem{dyakonovperel} M.I. Dyakonov and V. I. Perel, Phys. Lett A {\bf 35}, 459 (1971)
\bibitem{aronov_lyandageller} A.G. Aronov and Y.B. Lyanda-Geller JETP Letters {\bf 50} 489 (1989)
\bibitem{edelstein} V.M. Edelstein, Solid State Comm., {\bf 73} 233-235 (1990)  
\bibitem{awschalom}Y.K. Kato, R.C. Myers, A.C. Gossard and D.D. Awschalom, 
Science {\bf 306} 1910 (2004)
\bibitem{Wunderlich} J. Wunderlich, B Kaestner, J. Sinova and T. Jungwirth, Phys. Rev. Lett {\bf 94} 047204 (2005)
\bibitem{Sih1} V. Sih, R.C. Myers, Y.K. Kato, W.H. Lau, A.C. Gossard and D.D. Awschalom, Nature Physics, {\bf 1} 31-35  (2005)
\bibitem{Stern} N.P. Stern, S. Ghosh, G. Xiang, M. Zhu, N. Samarth and D.D. Awschalom,  Phys. Rev. Lett., {\bf 97} 126603 (2006) 
\bibitem{Sih2} V. Sih, W.H. Lau, R.C. Myers, V.R. Horowitz, A.C. Gossard and D.D. Awschalom, Phys. Rev. Lett., {\bf 97} 096605 (2006) 
\bibitem{Gossard} N.P. Stern, D.W. Steuerman, S. Mack, A.C. Gossard and D.D. Awschalom, Nature Physics, {\bf 4}, 843-846  (2008) 
\bibitem{nitta} T. Seki, Y. Hasegawa, S. Mitani, S. Takahashi, H. Imamura, S. Maekawa, J. Nitta and K. Takanashi,
Nature Materials,  {\bf 7} 125-129 (2008).   
\bibitem{Matsu} S. Matsuzaka, Y. Ohno and H. Ohno, Phys. Rev. B., {\bf 80}, 241305 (2009)      
\bibitem{Jung} J. Wunderlich, B.J. Park, A.C. Irvine, L.P. Zarbo, E. Rozkotova, P. Nemec, V. Novak, J. Sinova and T.  Jungwirth, 
Science, {\bf 330} 1801-1804  (2010)
\bibitem{hirsch} J.~E.~ Hirsch 
Phys. Rev. Lett. {\bf 83} 1834 - 1837 (1999)
\bibitem{Shuzhang} S.~F.~ Zhang, Phys. Rev. Lett., {\bf 85} 393  (2000)   
\bibitem{zhang} Murakami S, Nagaosa N, Zhang SC, Science {\bf 301} 1348-1351 (2003)
\bibitem{macdonald} Sinova J, Culcer D, Niu Q, N.A. Sinitsyn, T. Jungwirth and A.H. Macdonald 
Phys. Rev Lett, {\bf 92} 126603 (2004)
\bibitem{shytov} E.G. Mishchenko, A.V. Shytov and B.I. Halperin, Phys. Rev. Lett., {\bf 93} 226602 (2004)
\bibitem{Nikolic} B.K. Nikolic, S. Souma, L.P. Zarbo and J. Sinova, Phys. Rev. Lett., {\bf 95} 046601 (2005) 
\bibitem{halperin}
H.A. Engel, B.I. Halperin and E.I. Rashba, Phys.Rev. Lett.,
{\bf 95} 166605 (2005)
\bibitem{Adagideli} I. Adagideli and G.E.W. Bauer,
Phys.Rev. Lett.,  {\bf 95}   256602   (2005) 
\bibitem{inoue} J. I. Inoue, G.E.W. Bauer and L.W. Molenkamp, Phys. Rev. B. {\bf 70} 041303 (2004)
\bibitem{Rashba} E.I. Rashba, Phys. Rev. B., {\bf 70} 201309 (R) (2004).
\bibitem{Raimondi} R. Raimondi and P. Schwab, Phys. Rev. B. {\bf 71} 033311 (2005)
\bibitem{ALGP} A.G. Aronov, Y.B. Lyanda-Geller and G.E. Pikus, 
Zh. Exsp. Teor Fiz., {\bf 100} 973-981 (1991),
[JETP, 73, 537-541 (1991)]
\bibitem{altshuler} B.L. Altshuler, private communication.
\bibitem{awschalom1} Y.K. Kato, R.C. Myers, A.C. Gossard  and D.D. Awschalom, Phys. Rev. Lett, {\bf 93} 176601 (2004)   
\bibitem{silov} A.Y. Silov, P.A. Blajnov, J.H. Wolter, R. Hey, K.H. Ploog and N.S. Averkiev,
Appl. Phys. Lett., {\bf 85} 5929 (2004)
\bibitem{ganichev} S.D. Ganichev, Int. J. Mod. Phys. B   {\bf 22}   1-26   (2008) 
\bibitem{rokhinson}A. Chernyshov, M. Overby, X. Liu, J.K. Furdyna, Y. Lyanda-Geller and L.P. Rokhinson, Nature Physics, 5, 656-659  (2009). 
\bibitem{ferro}I.M. Miron, G. Gaudin, S. Auffret, B. Rodmacq, A. Schuhl, S. Pizzini, J. Vogel, P. Gambardella, Nature Materials, {\bf 9} 230-234 (2010) 
%\bibitem{silsbee} R.H. Silsbee, Journ. Phys. Cond. Matter, {\bf 16}  R179-R207 (2004).
\bibitem{Luttinger} J.M. Luttinger, Phys. Rev., {\bf 112} 739-751 (1958). 
\bibitem{Berger}L. Berger, Phys. Rev. B., {\bf 2}  4559-4566 (1970)  
\bibitem{willett} R.~L.~Willett, L.~N.~Pfeiffer, and K.W. West,
 Phys. Rev. Lett, {\bf 93}   026804 (2004) 
\bibitem{finkler} The field in domains is linear in the applied electric field. 
The effect does not require nonlinearities considered in 
I.G. Finkler, H.A. Engel, E.I. Rashba and B.I. Halperin, Phys. Rev. B {\bf 75}, 241202 (2007).
\bibitem{stanescu} B. Anderson, T.D Stanescu, and V. Galitski, {\bf 81}, 121304  (2010). 
\bibitem{pershin_diventra} Our effect does not require macroscopic 
inhomegeneity of
charge besides the applied longitudinal external electric field as 
in Y.V. Pershin and M. DiVentra, Journ. of Phys.Cond Matt, {\bf 20} 025204 (2008).
\bibitem{note2} Such spin-orbit term in the 2D Hamiltonian arises from the standard Dirac spin-orbit interaction Hamiltonian or Kane model 3D spin-orbit scattering
in semiconductors when only the in-plane momenta $p_x$ and $p_y$ are allowed and the $z-$direction is completely 
eliminated from the picture\cite{hikami}.
\bibitem{hikami} S. Hikami, A.I. Larkin and Y. Nagaoka, 
Progr. Theor. Phys. {\bf 63} 707-710 (1980)
\bibitem{lgm} Y.B. Lyanda-Geller and A.D. Mirlin, Phys. Rev. Lett., {\bf 72} 1894 (1994).
\bibitem{vignale1}
E.M. Hankiewicz, G. Vignale and M.E. Flatte,
Phys. Rev. Lett., {\bf 97} 266601 (2006).
\bibitem{knap} W. Knap, C. Skierbiszewski, A. Zduniak, E. Litwin-Staszewska, 
D. Bertho, F. Kobbi, J.L. Robert, G.E.Pikus, F.G. Pikus, S.V. 
Iordanskii, V. Mosser, K. Zekentes and Y.B. Lyanda-Geller, 
Phys. Rev B, 53 3912 (1996).
\bibitem{aleiner}
I.L. Aleiner and V.I. Fal'ko, Phys. Rev Lett., 87 256801 (2001).
\bibitem{zhang1} B.A. Bernevig, J. Orenstein, and S.C. Zhang, 
Phys. Rev. Lett. 97, 236601 (2006).
\bibitem{awschalom2009}J. D. Koralek, C. P. Weber, J. Orenstein, B. A. Bernevig, Shou-Cheng Zhang, S. Mack and 
D. D. Awschalom, Nature 458, 610-613 (2009). 
\bibitem{maslov} M. Duckheim, D.L. Maslov and D. Loss D, {\bf 80} 235327 (2009)  
\bibitem{Vignale}E.M. Hankiewicz and G. Vignale,  Phys. Rev. B. {\bf 73} 115339 (2006).
\bibitem{dasSarma} W.K.Tse and S. Das Sarma, Phys.Rev. Lett., {\bf 96} 056601  (2006).   
\bibitem{review} N. Nagaosa, J. Sinova J, S. Onoda, A.H. MacDonald and N.P.  Ong, Review of Modern Physics, {\bf 82}   1539-1592  (2010) 
\bibitem{YLG}Yu.B. Lyanda-Geller, JETP Letters {\bf 46} 489 (1987)
\bibitem{culcer} D. Culcer, E.M. Hankiewicz, G. Vignale and R. Winkler, Phys. Rev. B, {\bf 81} 125332 (2010) 
\bibitem{Chazalviel} J.N. Chazalviel and I. Solomon, Phys. Rev. Lett., {\bf 29} 1676 (1972)
\bibitem{sherman} E.Y. Sherman, Phys.Rev. B {\bf 67} 161303 (2003).
\bibitem{remark}
We note that the contribution to the conductivity from electron-electron interactions, that leads to the spin drug effect on the spin Hall current
\cite{Vignale}, can still be accounted  by these constituitive equations with minor modifications. However, 
in all cases we consider, the correction to the skew scattering current due to spin drug is estimated to be less than 1\% (section IV).
\bibitem{sign} Sign of the constant $\alpha$ for electrons in metals and semicondutors can be the same or opposite to 
the sign of this constant for 
relativistic free electrons \cite{Abakumov}. 
\bibitem{Abakumov} V.N. Abakumov and I.N. Yassievich, Zh. Exsp. Teor Fiz., 61 2571-2579 (1971).
[JETP, 34 1375-1378 1972]
\bibitem{Giuliani} G. F. Giuliani and G. Vignale, Quantum theory of the electron liquid, Cambridge University Press, Cambridge (2005)
\bibitem{rendell}R.W. Rendell and S.M. Girvin, Phys. Rev B {\bf 23} 6610 (1981)
\bibitem{Landau} L.D. Landau and E.M. Lifshitz, Electrodynamics of Continuous Media, Second Edition: Volume 8 (Course of Theoretical Physics), Elsevier: New York (1984).
\bibitem{halle} E. Hallen, Electromagnetic Theory, Wiley and Sons, New York (1962)
\bibitem{smythe} W.R. Smythe, Static and Dynamic Electricity, McGraw-Hill (1968)
\bibitem{but}However, if we solve the Poisson equation without the constraint
on positive
charge, i.e., without the introduction of $b_1$, and use
Eq.(\ref{b}) and Eq.(\ref{a}) at $b_1=b$, but not Eqs.(\ref{c},\ref{d}),
we obtain unphysical divergency of positive charge density in the vicinity
of $b$.
\bibitem{borunda}M.F. Borunda, T.S. Nunner, T. Luck, N.A. Sinitsyn, C. Timm, J.Wunderlich, T. Jungwirth, A.H. MacDonald and J. Sinova,
Phys. Rev. Lett., {\bf 99}  066604  (2007)    
\bibitem{Nunner} T.S. Nunner, N.A. Sinitsyn, M.F. Borunda, V.K. Dugaev, A.A. Kovalev, Ar. Abanov, C. Timm, T. Jungwirth, J. I. Inoue, A.H. MacDonald and J. Sinova,
Phys. Rev. B., {\bf 76}  235312  (2007) 
\bibitem{Born} When the square root parameter reaches the order of 1,  the scattering 
amplitude in Born approximation has to be substituted by the exact scattering amplitude in 
expressions for relaxation times and conductivities, similar to the description of the Kondo effect,
see discussion in \cite{abrikosov}.
\bibitem{abrikosov} A.A. Abrikosov, Fundamentals of the theory of metals,  North-Holland, Amsterdam (1988).
 \bibitem{ivchenko1} E.L. Ivchenko, Y.B. Lyanda-Geller, G.E.Pikus {\it et al}, Fiz. Tekhn. Poluprovodn. {\bf 18} 93 (1984); [Sov. Phys. Semicond. {\bf 18} 55 (1984)] 
\bibitem{Luttinger1954} R. Karplus and J.M. Luttinger, Phys. Rev., {\bf 95} 1154-1160  (1954) 
\bibitem{Nozieres} P. Nozieres and C. Lewiner, Journal de Physique, {\bf 34} 901-915 (1973). 
\bibitem{Zutic} W.K. Tse, J. Fabian, I. Zutic and S Das Sarma, Phys. Rev. B., {\bf 72}  241303  (2005) 
\bibitem{ILGP} E.L. Ivchenko, Y.B. Lyanda-Geller and G.E.Pikus, Zh. Eksp. Teor Fiz., {\bf 98} 989-1002 (1990),
[JETP, 73, 550-557 (1991)]. 
\bibitem{awo} A.W. Overhauser, Physical Review, {\bf 89} 689-700 (1953).
\bibitem{DP} M.I. Dyakonov and V.I. Perel, Sov. Phys. JETP-USSR, {\bf 33} 1053 (1971)
\bibitem{PWA} P.W. Anderson, Journ. Phys. Soc. Jpn., {\bf 9} 316-339 (1954)
\bibitem{flatte} K.C. Hall, K. Gundogdu, E. Altunkaya, W.H. Lau, M.E. Flatte, 
T.F. Boggess, J.J. Zinck, W.B. Barvosa-Carter and S.L. Skeith,
Phys. Rev B   {\bf 68} 115311 (2003). 
\bibitem{Ganichev}V.V. Bel'kov, P. Olbrich, S.A. Tarasenko, D. Schuh, W. Wegscheider, T. Korn, C. Schuller, 
D. Weiss, W. Prettl and S.D Ganichev, 
 Phys.Rev. Lett., {\bf 100} 176806 (2008)
\bibitem{Dresselhaus} G. Dresselhaus. Phys. Rev B, {\bf 100} 580 (1955)
\bibitem{rashba1} E.I. Rashba, Soviet Physics-Solid State, {\bf 2} 1109-1122 (1960)
\bibitem{pikus}G.L. Bir and G.E. Pikus, Sov Phys. Solid State, {\bf 3} 2221 (1961)
\bibitem{rashba} Y.A. Bychkov and E.I. Rashba, J.Phys. Solid State {\bf 17} 6093 (1984).
\bibitem{altshuler1}B.L. Altshuler, A.G. Aronov, A.I.Larkin and D.E. Khmelnitskii,
Sov Phys. JETP {\bf 81} 788 (1981).
\bibitem{dyakonovkachor} M.I. Dyakonov and V.Y. Kachorovskii, Soviet Physics Semiconductors-USSR, 
{\bf 20} 110-112 (1986).
\bibitem{interface} O. Krebs and P. Voisin, Phys. Rev. Lett. 77, 1829 (1996).
\bibitem{difference} This 3D potential affecting electrons in the conduction band 
(besides the potential confining to 2D) can be the same as or differ with the 3D potential 
affecting valence electrons. While, e.g., external voltage and potentials associated with 
impurities are the same for conduction and valence electrons, potential associated with islands of 
fluctuations of the quantum well width by a monolayer or few, differ for conduction and valence electrons. 
This difference is associated with the same difference in potential offsets as in confining potentials for 
conduction and valence electrons.
\bibitem{Kane} E.O. Kane, Journal of Physics and Chemistry of Solids, {\bf 1} 249 (1957)
\bibitem{Danan}G. Danan, B. Etienne, F. Mollot, R. Planel, A. M. Jean-Louis, 
F. Alexandre, B. Jusserand, G. Le Roux, J. Y. Marzin, H. Savary, and B. Sermage, Phys. Rev. B {\bf 35}, 6207 (1987). 
\bibitem{Winkler} R. Winkler, Spin–Orbit Coupling Effects in Two-Dimensional Electron and Hole Systems, Springer-Verlag Berlin Heidelberg (2003)
\bibitem{Chalaev} O. Chalaev and G. Vignale, Phys. Rev. Lett., 104 226601 (2010).
\bibitem{Brouwer} P.W. Brouwer, J.N.H.J. Cremers and B.I. Halperin, Phys. Rev. B, {\bf 65} 081302 (2002).  
\bibitem{interlevel}Spin-nonconserving terms due to scattering accompanied by 
transitions between 
levels of size quantization in the quantum well remain, 
but their effect is negligible for systems in which only the 
first level of size quantization is occupied.
\bibitem{Melnikov} V.I. Melnikov and E.I. Rashba, Zh. Eksp. Teor. Fiz, {\bf 61} 2530 (1971) [Sov. Phys. JETP, {\bf 34} 1353 (1972)]. 
\bibitem{Shklovskii} B.I. Shklovskii and A.L. Efros, Electronic properties of doped semiconductors (Springer-Verlag, Berlin, 1984).
\bibitem{Tserkovnyak} Y. Tserkovnyak and A. Brataas, Phys. Rev. B 76, 155326 (2007). 
\bibitem{ruan}Y.C. Ruan and W.Y. Ching, Journ. Appl. Phys. {\bf 62} 2885-2887 (1987).
\bibitem{IP} E.L. Ivchenko and G.E.Pikus, Superlattices and Other Heterostructures, Springer-Verlag, Berlin (1995)
\bibitem{iordanski1994} S.V. Iordanskii, Y.B. Lyanda-Geller and G.E. Pikus, JETP Letters, {\bf 60} 206-211 (1994)
\bibitem{ylg1998} Y. Lyanda-Geller, Phys. Rev. Lett, {\bf 80} 4273 (1998)
\bibitem{marcus2003}J.B. Miller, D.M. Zumbuhl, C.M. Marcus, Y.B. Lyanda-Geller, D. Goldhaber-Gordon, K. Campman and A.C. Gossard,
Phys. Rev. Lett., {\bf 90} 076807 (2003)
\bibitem{LGAG} Y.B. Lyanda-Geller, I.L. Aleiner and P.M. Goldbart, Phys. Rev. Lett, {\bf 81} 3215 (1998).
\bibitem{Glazov} M.M. Glazov and E.L. Ivchenko, JETP, {\bf 99} 1279-1290 (2004).
\bibitem{equilibrium} E.I. Rashba, Phys Rev. B {\bf 68} 241315 (2003)
\bibitem{abragam} A. Abragam, Principles of Nuclear Magnetism, Oxford University Press, 1983, c1961.  
\bibitem{paget} D. Paget, G. Lampel, B. Sapoval and V.I. Safarov, Phys. Rev. B, {\bf 15} 5780-5796 (1977)
\bibitem{Eisenstein} J.J. Eisenstein, Dependence of 2D carrier density on remote doping, private communication. 
\bibitem{Denninger} B. Gotschy, G. Denninger, H. Obloh, L. Wilkening and J. Schneider, Solid State Comm., {\bf 71} 629-632 (1989)
\bibitem{Drell}J. D. Bjorken and S. D. Drell, Relativistic quantum mechanics, New York, McGraw-Hill (1964)
\bibitem{Lassnig} R. Lassnig, Phys. Rev. B, {\bf 31} 8076 (1985)
\bibitem{Morrow} R.A. Morrow and K.R. Brownstein,  Phys. Rev. B, {\bf 30} 678  (1984)   
\bibitem{foreman}B.A. Foreman. Phys. Rev. B, {\bf 56} 12748 (1997)
\bibitem{ER} A.V. Rodina, A.Y. Alekseev, A.L. Efros, M. Rosen and B.K. Meyer,  Phys. Rev. B {\bf 65} 125302 (2002)   
\bibitem{suris} R.A. Suris, Fiz. Tekhn. Poluprovodn., {\bf 20} 2008 (1986) [Sov. Phys. -Semicond, {\bf 20} 1258 (1986) 
\bibitem{sham}S.R. White and L.J. Sham, Phys.Rev. Lett., {\bf 47} 879 (1981)   
\end{thebibliography}
\end{document}